\documentclass[12pt]{article}

\usepackage{amsmath}
\usepackage{amssymb}
\usepackage{xspace}
\usepackage{gensymb}
\usepackage{graphicx}
\usepackage{setspace}
\usepackage{hyperref}
\usepackage{multirow}

\usepackage{verbatim}

\newcommand{\detailtexcount}[1]{%
}

\newcommand{\quickwordcount}[1]{%
}

\newcommand{\quickcharcount}[1]{%
}

\usepackage{scicite}

\usepackage{times}

\newenvironment{sciabstract}{%
\begin{quote} \bf}
{\end{quote}}

\newcounter{lastnote}

\def\titlestring{Constraints on the Hubble constant from Supernova Refsdal's reappearance}

\title{\titlestring}

\author{Patrick L. Kelly$^{1}$, Steven Rodney$^{2}$, Tommaso Treu$^{3}$,\\Masamune Oguri$^{4,5,6}$, Wenlei Chen$^{1}$, Adi Zitrin$^{7}$, Simon Birrer$^{8,9}$,\\Vivien Bonvin$^{10}$, Luc Dessart$^{11}$, Jose M. Diego$^{12}$, Alexei V. Filippenko$^{13,14}$,\\Ryan J. Foley$^{15}$, Daniel Gilman$^{16,3}$, Jens Hjorth$^{17}$,\\Mathilde Jauzac$^{18,19,20,21}$, Kaisey Mandel$^{22,23}$, Martin Millon$^{10,8}$,\\Justin Pierel$^{24,2}$, Keren Sharon$^{25}$, Stephen Thorp$^{26,22}$, Liliya Williams$^{1}$,\\Tom Broadhurst$^{27}$, Alan Dressler$^{28}$, Or Graur$^{29,30}$, Saurabh Jha$^{31}$,\\Curtis McCully$^{32,33}$, Marc Postman$^{24}$, Kasper Borello Schmidt$^{34}$,\\Brad E. Tucker$^{35,36,37}$, and Anja von der Linden$^{9}$
\\
\footnotesize{$^{1}$Minnesota Institute for Astrophysics, University of Minnesota, Minneapolis, MN 55455, USA}\\
\footnotesize{$^{2}$Department of Physics and Astronomy, University of South Carolina, Columbia, SC 29208, USA}\\
\footnotesize{$^{3}$Department of Physics and Astronomy, University of California, Los Angeles, CA 90095}\\
\footnotesize{$^{4}$Center for Frontier Science, Chiba University, Chiba 263-8522, Japan}\\
\footnotesize{$^{5}$Research Center for the Early Universe, University of Tokyo, Tokyo 113-0033, Japan}\\
\footnotesize{$^{6}$Kavli Institute for the Physics and Mathematics of the Universe, University of Tokyo, Kashiwa,}\\
\footnotesize{Chiba 277-8583, Japan}\\
\footnotesize{$^{7}$Physics Department, Ben-Gurion University of the Negev, Beer-Sheva 8410501, Israel}\\
\footnotesize{$^{8}$Kavli Institute for Particle Astrophysics and Cosmology, Stanford University, Stanford, CA 94305,}\\
\footnotesize{USA}\\
\footnotesize{$^{9}$Department of Physics and Astronomy, Stony Brook University, Stony Brook, NY 11794, USA}\\
\footnotesize{$^{10}$Laboratoire d'Astrophysique, Ecole Polytechnique Federale de Lausanne, Observatoire de Sauverny,}\\
\footnotesize{CH-1290 Versoix, Switzerland}\\
\footnotesize{$^{11}$Institut d'Astrophysique de Paris,}
\footnotesize{Centre national de la recherche scientifique-Sorbonne Universit\'e,}\\
\footnotesize{F-75014 Paris, France}\\
\footnotesize{$^{12}$Instituto de F\'isica de Cantabria,}\\
\footnotesize{ Consejo Superior de Investigaciones Cient\'ificas y la Universidad de Cantabria, 39005 Santander}\\
}

\date{}

\begin{document}

\baselineskip24pt

\maketitle 

%% Replace with bibtex reference aliases when possible
%\defcitealias{insert_bibtex_name_for_paperI}{Paper I}
%\defcitealias{insert_bibtex_name_for_paperII}{Paper II}
%\defcitealias{insert_bibtex_name_for_paperIII}{Paper III}
\def\paperI{Paper I\xspace}
\def\paperII{Paper II\xspace}
\def\paperIII{Paper III\xspace}

%\makeatletter\onecolumngrid@push\makeatother
\newcommand\SR[1]{{\textcolor{red}{\bf [SR: #1]}}}
\newcommand\red[1]{{\textcolor{red}{\bf [#1]}}}

\newcommand\ionpat[2]{#1$\;${\scshape{#2}}}%                       % ion, i.e.,
%CII = \ion{C}{ii}
\def\HST{{\it HST}\xspace}
\def\hubbleconstant{71 km s$^{-1}$ Mpc$^{-1}$}
\def\veloffset{$\sim50-100$ km s$^{-1}$}
\def\absmagps{\mbox{$M_{r} \approx $ -14 mag}}
\def\foregroundext{\mbox{$A_{V} = 1.88$ mag} \citep{schlaflyfinkbeinerSFD11}}
\def\rescalefactor{0.12}
\def\instrumentalFWHM{1.85 Ang (84 km s$^{-1}$)}
\def\instrumentalsigma{$\sigma=0.79$~\AA\ (36 km s$^{-1}$)}
\def\measuredsigmaHII{$\sigma=0.90$~\AA\ (41 km s$^{-1}$)}
\def\unconvolvedsigmaHII{0.43~\AA\ (20 km s$^{-1}$)}
\def\cssname{CSS131025:044730+235856}
\def\peakRabs{$M_{R} \approx -18.4$ mag}
\def\one{111$^{\circ}$}
\def\two{119$^{\circ}$}
\def\approximatephase{??}
\def\lastlcmeas{26 May 2015}
\def\refsdalexplstwomjd{56910.6$\pm$5}
\def\refsdalexplstwo{10 September 2014}

\def\macs1149{MACS\,J1149.5+2223}

\def\maskwidthvlt{5}

\def\refsdalrmaxstwomjd{57138$\pm$20}
\def\refsdalrmaxstwo{26 April 2015}

\def\starmag{$r=15.1$ mag AB}
\def\lastphotdate{20--21 July 2015}

\def\riseafterdiscovery{$\sim$150--160 days in the observer frame (50--60 days
in the rest frame)}

\def\binwidth{100\,\AA}

\def\continuumhalpha{\mbox{$15500\,\mathrm{\AA} < \lambda <
15750\,\mathrm{\AA}$} and \mbox{$16775\,\mathrm{\AA} < \lambda <
17000\,\mathrm{\AA}$}}

\def\grismphotstartmjd{57019}
\def\grismphotstart{28 December 2014}
\def\grismphotendmjd{57022}
\def\grismphotend{31 December 2014}
\def\grismphotmjds{57019--57022}
\def\grismphot{28--31 December 2014}

\def\phasemaxgrism{-99~days}
\def\phasemaxvlt{-99~days}

\def\discoverymjd{}

\def\grismstartmjd{57014}
\def\grismstart{23 December 2014}
\def\grismmidmjd{57020.5}
\def\grismmid{29 December 2014}
\def\grismendmjd{57027} 
\def\grismend{5 January 2015}
\def\grismmjds{57014--57027}
\def\grism{23 December 2014 through 5 January 2015}

\def\mosfirephase{$\sim$\,-62$\pm$8\,d}
\def\grismphase{$\sim$\,-47$\pm$8\,d}
\def\xshooterphase{$\sim$\,+16$\pm$8\,d}

\def\mjdeightysevenaexpl{46849}
\def\mjdeightsevenasame{46892.9}
\def\mjdeightsevenadate{7 April 1987}
\def\mjdpkstwo{57128}
\def\mjddisc{56971}
% 157 days (63 days) 

\def\eightysevenaphase{+999 days after the explosion date}
\def\ninetyeightaphase{+999 days after the explosion date}

\def\ircolor{$\sim$0.15 mag AB}

\def\grismmag{25.0}

\def\magonetwentyfivemath{26.5 \pm 0.2}
\def\magonesixtymath{26.3 \pm 0.2}
\def\magonetwentyfive{26.5$\pm$0.2}
\def\magonesixty{26.3$\pm$0.2}
\def\obsdateone{30 October 2015}
\def\obsdatetwo{14 November 2015}
\def\obsdatethree{11 December 2015}

\def\einsteincrossdiscim{10 November 2014}

\def\sxcolor{0.2$\pm$0.3\,mag~AB}

\def\snrefsdal{SN\,Refsdal}

\def\coordsx{$\alpha$ = 11$^{\rm h}$49$^{\rm m}$36.022$^{\rm s}$, $\delta$ = +22$^{\circ}$23$'$48.10$''$ (J2000.0)}
\def\rasx{11$^{\rm h}$49$^{\rm m}$36.022$^{\rm s}$}
\def\decsx{22$^{\circ}$23$'$48.10$''$}

\def\chisqthreeradioactive{424.2}
\def\paramsthreeradioactive{17}
\def\aicthreeradioactive{458.2}
\def\chisqtworadioactive{421.4}
\def\paramstworadioactive{15}
\def\aictworadioactive{451.4}
\def\chisqfourradioactive{414.6}
\def\paramsfourradioactive{19}
\def\aicfourradioactive{452.6}
\def\chisq1radioactive{27202.565}
\def\params1radioactive{13}
\def\aic1radioactive{27228.565}
\def\chisqtwopoly{379.1}
\def\paramstwopoly{17}
\def\aictwopoly{413.1}
\def\doftwopoly{510}
\def\redchisqtwopoly{0.74}
\def\dofthreepoly{508}
\def\doftworadioactive{512}
\def\redchisqtworadioactive{0.82}
\def\dofthreeradioactive{510}
\def\redchisqthreeradioactive{0.83}
\def\doffourradioactive{508}
\def\redchisqfourradioactive{0.82}
\def\redchisqthreepoly{0.77}
\def\chisqthreepoly{393.2}
\def\paramsthreepoly{19}
\def\aicthreepoly{431.2}
\def\doffourpoly{506}
\def\redchisqfourpoly{0.74}
\def\chisqfourpoly{375.2}
\def\paramsfourpoly{21}
\def\aicfourpoly{417.2}

\def\delaystwosonethreetwodolphot{17.244253}
\def\mustwosonethreetwodolphot{1.089738}
\def\delaysthreesonethreetwodolphot{-40.295143}
\def\musthreesonethreetwodolphot{0.970260}
\def\delaysfoursonethreetwodolphot{8.633673}
\def\musfoursonethreetwodolphot{0.311573}
\def\delaysxsonethreetwodolphot{374.956267}
\def\musxsonethreetwodolphot{0.302446}
\def\photcoeffoneFonetwofiveWthreetwodolphot{0.412069}
\def\photcoefftwoFonetwofiveWthreetwodolphot{0.002328}
\def\photcoeffthreeFonetwofiveWthreetwodolphot{-0.000006}
\def\photcoefffourFonetwofiveWthreetwodolphot{-0.000000}
\def\nebcoeffoneFonetwofiveWthreetwodolphot{0.133161}
\def\nebcoefftwoFonetwofiveWthreetwodolphot{-0.000166}
\def\nebcoeffthreeFonetwofiveWthreetwodolphot{0.000000}
\def\photcoeffoneFonesixohWthreetwodolphot{0.466902}
\def\photcoefftwoFonesixohWthreetwodolphot{0.002843}
\def\photcoeffthreeFonesixohWthreetwodolphot{0.000000}
\def\photcoefffourFonesixohWthreetwodolphot{-0.000000}
\def\nebcoeffoneFonesixohWthreetwodolphot{0.213218}
\def\nebcoefftwoFonesixohWthreetwodolphot{-0.000222}
\def\nebcoeffthreeFonesixohWthreetwodolphot{0.000000}
\def\dofthreetwodolphot{595}
\def\redchisqthreetwodolphot{1.70}
\def\chisqthreetwodolphot{1009.7}
\def\paramsthreetwodolphot{22}
\def\aicthreetwodolphot{1053.7}
\def\delaystwosonetwoonedolphot{19.416456}
\def\mustwosonetwoonedolphot{1.114012}
\def\delaysthreesonetwoonedolphot{-40.242768}
\def\musthreesonetwoonedolphot{0.976841}
\def\delaysfoursonetwoonedolphot{14.141969}
\def\musfoursonetwoonedolphot{0.332018}
\def\delaysxsonetwoonedolphot{372.254953}
\def\musxsonetwoonedolphot{0.314290}
\def\photcoeffoneFonetwofiveWtwoonedolphot{0.410914}
\def\photcoefftwoFonetwofiveWtwoonedolphot{0.002179}
\def\photcoeffthreeFonetwofiveWtwoonedolphot{-0.000007}
\def\nebcoeffoneFonetwofiveWtwoonedolphot{0.128506}
\def\nebcoefftwoFonetwofiveWtwoonedolphot{-0.000151}
\def\photcoeffoneFonesixohWtwoonedolphot{0.481649}
\def\photcoefftwoFonesixohWtwoonedolphot{0.002653}
\def\photcoeffthreeFonesixohWtwoonedolphot{-0.000006}
\def\nebcoeffoneFonesixohWtwoonedolphot{0.197208}
\def\nebcoefftwoFonesixohWtwoonedolphot{-0.000168}
\def\doftwoonedolphot{599}
\def\redchisqtwoonedolphot{1.66}
\def\chisqtwoonedolphot{993.1}
\def\paramstwoonedolphot{18}
\def\aictwoonedolphot{1029.1}
\def\delaystwosonetwotwodolphot{19.423817}
\def\mustwosonetwotwodolphot{1.114067}
\def\delaysthreesonetwotwodolphot{-40.249498}
\def\musthreesonetwotwodolphot{0.976780}
\def\delaysfoursonetwotwodolphot{14.147393}
\def\musfoursonetwotwodolphot{0.332026}
\def\delaysxsonetwotwodolphot{372.251261}
\def\musxsonetwotwodolphot{0.314278}
\def\photcoeffoneFonetwofiveWtwotwodolphot{0.410915}
\def\photcoefftwoFonetwofiveWtwotwodolphot{0.002178}
\def\photcoeffthreeFonetwofiveWtwotwodolphot{-0.000007}
\def\nebcoeffoneFonetwofiveWtwotwodolphot{0.133821}
\def\nebcoefftwoFonetwofiveWtwotwodolphot{-0.000172}
\def\nebcoeffthreeFonetwofiveWtwotwodolphot{0.000000}
\def\photcoeffoneFonesixohWtwotwodolphot{0.481653}
\def\photcoefftwoFonesixohWtwotwodolphot{0.002653}
\def\photcoeffthreeFonesixohWtwotwodolphot{-0.000006}
\def\nebcoeffoneFonesixohWtwotwodolphot{0.207657}
\def\nebcoefftwoFonesixohWtwotwodolphot{-0.000213}
\def\nebcoeffthreeFonesixohWtwotwodolphot{0.000000}
\def\doftwotwodolphot{597}
\def\redchisqtwotwodolphot{1.66}
\def\chisqtwotwodolphot{993.1}
\def\paramstwotwodolphot{20}
\def\aictwotwodolphot{1033.1}
\def\delaystwosonethreeonedolphot{19.189314}
\def\mustwosonethreeonedolphot{1.108044}
\def\delaysthreesonethreeonedolphot{-39.861583}
\def\musthreesonethreeonedolphot{0.976608}
\def\delaysfoursonethreeonedolphot{13.980983}
\def\musfoursonethreeonedolphot{0.327841}
\def\delaysxsonethreeonedolphot{374.454781}
\def\musxsonethreeonedolphot{0.309735}
\def\photcoeffoneFonetwofiveWthreeonedolphot{0.409868}
\def\photcoefftwoFonetwofiveWthreeonedolphot{0.002266}
\def\photcoeffthreeFonetwofiveWthreeonedolphot{-0.000006}
\def\photcoefffourFonetwofiveWthreeonedolphot{-0.000000}
\def\nebcoeffoneFonetwofiveWthreeonedolphot{0.128774}
\def\nebcoefftwoFonetwofiveWthreeonedolphot{-0.000151}
\def\photcoeffoneFonesixohWthreeonedolphot{0.479086}
\def\photcoefftwoFonesixohWthreeonedolphot{0.002762}
\def\photcoeffthreeFonesixohWthreeonedolphot{-0.000004}
\def\photcoefffourFonesixohWthreeonedolphot{-0.000000}
\def\nebcoeffoneFonesixohWthreeonedolphot{0.198124}
\def\nebcoefftwoFonesixohWthreeonedolphot{-0.000169}
\def\dofthreeonedolphot{597}
\def\redchisqthreeonedolphot{1.65}
\def\chisqthreeonedolphot{984.4}
\def\paramsthreeonedolphot{20}
\def\aicthreeonedolphot{1024.4}
\def\delaystwosonefouronedolphot{20.135063}
\def\mustwosonefouronedolphot{1.114128}
\def\delaysthreesonefouronedolphot{-39.044149}
\def\musthreesonefouronedolphot{0.981040}
\def\delaysfoursonefouronedolphot{17.396990}
\def\musfoursonefouronedolphot{0.332894}
\def\delaysxsonefouronedolphot{379.730981}
\def\musxsonefouronedolphot{0.315235}
\def\photcoeffoneFonetwofiveWfouronedolphot{0.415889}
\def\photcoefftwoFonetwofiveWfouronedolphot{0.002618}
\def\photcoeffthreeFonetwofiveWfouronedolphot{-0.000008}
\def\photcoefffourFonetwofiveWfouronedolphot{-0.000000}
\def\photcoefffiveFonetwofiveWfouronedolphot{0.000000}
\def\nebcoeffoneFonetwofiveWfouronedolphot{0.128354}
\def\nebcoefftwoFonetwofiveWfouronedolphot{-0.000151}
\def\photcoeffoneFonesixohWfouronedolphot{0.482765}
\def\photcoefftwoFonesixohWfouronedolphot{0.002924}
\def\photcoeffthreeFonesixohWfouronedolphot{-0.000005}
\def\photcoefffourFonesixohWfouronedolphot{-0.000000}
\def\photcoefffiveFonesixohWfouronedolphot{0.000000}
\def\nebcoeffoneFonesixohWfouronedolphot{0.196256}
\def\nebcoefftwoFonesixohWfouronedolphot{-0.000166}
\def\doffouronedolphot{595}
\def\redchisqfouronedolphot{1.61}
\def\chisqfouronedolphot{958.0}
\def\paramsfouronedolphot{22}
\def\aicfouronedolphot{1002.0}
\def\delaystwosonefourtwodolphot{18.264046}
\def\mustwosonefourtwodolphot{1.096256}
\def\delaysthreesonefourtwodolphot{-39.935793}
\def\musthreesonefourtwodolphot{0.972036}
\def\delaysfoursonefourtwodolphot{13.217593}
\def\musfoursonefourtwodolphot{0.319154}
\def\delaysxsonefourtwodolphot{381.557306}
\def\musxsonefourtwodolphot{0.310320}
\def\photcoeffoneFonetwofiveWfourtwodolphot{0.420633}
\def\photcoefftwoFonetwofiveWfourtwodolphot{0.002929}
\def\photcoeffthreeFonetwofiveWfourtwodolphot{-0.000008}
\def\photcoefffourFonetwofiveWfourtwodolphot{-0.000000}
\def\photcoefffiveFonetwofiveWfourtwodolphot{0.000000}
\def\nebcoeffoneFonetwofiveWfourtwodolphot{0.132523}
\def\nebcoefftwoFonetwofiveWfourtwodolphot{-0.000168}
\def\nebcoeffthreeFonetwofiveWfourtwodolphot{0.000000}
\def\photcoeffoneFonesixohWfourtwodolphot{0.473150}
\def\photcoefftwoFonesixohWfourtwodolphot{0.002884}
\def\photcoeffthreeFonesixohWfourtwodolphot{-0.000002}
\def\photcoefffourFonesixohWfourtwodolphot{-0.000000}
\def\photcoefffiveFonesixohWfourtwodolphot{0.000000}
\def\nebcoeffoneFonesixohWfourtwodolphot{0.138273}
\def\nebcoefftwoFonesixohWfourtwodolphot{0.000115}
\def\nebcoeffthreeFonesixohWfourtwodolphot{-0.000000}
\def\doffourtwodolphot{593}
\def\redchisqfourtwodolphot{1.65}
\def\chisqfourtwodolphot{980.9}
\def\paramsfourtwodolphot{24}
\def\aicfourtwodolphot{1028.9}
\def\delaystwosonefiveonedolphot{20.692662}
\def\mustwosonefiveonedolphot{1.115786}
\def\delaysthreesonefiveonedolphot{-38.859320}
\def\musthreesonefiveonedolphot{0.980649}
\def\delaysfoursonefiveonedolphot{19.494317}
\def\musfoursonefiveonedolphot{0.336570}
\def\delaysxsonefiveonedolphot{375.789274}
\def\musxsonefiveonedolphot{0.316909}
\def\photcoeffoneFonetwofiveWfiveonedolphot{0.417803}
\def\photcoefftwoFonetwofiveWfiveonedolphot{0.002603}
\def\photcoeffthreeFonetwofiveWfiveonedolphot{-0.000009}
\def\photcoefffourFonetwofiveWfiveonedolphot{-0.000000}
\def\photcoefffiveFonetwofiveWfiveonedolphot{0.000000}
\def\photcoeffsixFonetwofiveWfiveonedolphot{-0.000000}
\def\nebcoeffoneFonetwofiveWfiveonedolphot{0.128658}
\def\nebcoefftwoFonetwofiveWfiveonedolphot{-0.000151}
\def\photcoeffoneFonesixohWfiveonedolphot{0.476435}
\def\photcoefftwoFonesixohWfiveonedolphot{0.002886}
\def\photcoeffthreeFonesixohWfiveonedolphot{-0.000001}
\def\photcoefffourFonesixohWfiveonedolphot{-0.000000}
\def\photcoefffiveFonesixohWfiveonedolphot{-0.000000}
\def\photcoeffsixFonesixohWfiveonedolphot{0.000000}
\def\nebcoeffoneFonesixohWfiveonedolphot{0.195701}
\def\nebcoefftwoFonesixohWfiveonedolphot{-0.000166}
\def\doffiveonedolphot{593}
\def\redchisqfiveonedolphot{1.61}
\def\chisqfiveonedolphot{952.9}
\def\paramsfiveonedolphot{24}
\def\aicfiveonedolphot{1000.9}
\def\delaystwosonefivetwodolphot{19.737085}
\def\mustwosonefivetwodolphot{1.108081}
\def\delaysthreesonefivetwodolphot{-42.384701}
\def\musthreesonefivetwodolphot{0.958264}
\def\delaysfoursonefivetwodolphot{12.316222}
\def\musfoursonefivetwodolphot{0.324742}
\def\delaysxsonefivetwodolphot{367.115460}
\def\musxsonefivetwodolphot{0.314191}
\def\photcoeffoneFonetwofiveWfivetwodolphot{0.399500}
\def\photcoefftwoFonetwofiveWfivetwodolphot{0.002659}
\def\photcoeffthreeFonetwofiveWfivetwodolphot{-0.000000}
\def\photcoefffourFonetwofiveWfivetwodolphot{-0.000000}
\def\photcoefffiveFonetwofiveWfivetwodolphot{-0.000000}
\def\photcoeffsixFonetwofiveWfivetwodolphot{0.000000}
\def\nebcoeffoneFonetwofiveWfivetwodolphot{0.191399}
\def\nebcoefftwoFonetwofiveWfivetwodolphot{-0.000403}
\def\nebcoeffthreeFonetwofiveWfivetwodolphot{0.000000}
\def\photcoeffoneFonesixohWfivetwodolphot{0.459379}
\def\photcoefftwoFonesixohWfivetwodolphot{0.002663}
\def\photcoeffthreeFonesixohWfivetwodolphot{0.000008}
\def\photcoefffourFonesixohWfivetwodolphot{0.000000}
\def\photcoefffiveFonesixohWfivetwodolphot{-0.000000}
\def\photcoeffsixFonesixohWfivetwodolphot{0.000000}
\def\nebcoeffoneFonesixohWfivetwodolphot{0.283711}
\def\nebcoefftwoFonesixohWfivetwodolphot{-0.000523}
\def\nebcoeffthreeFonesixohWfivetwodolphot{0.000000}
\def\doffivetwodolphot{591}
\def\redchisqfivetwodolphot{1.75}
\def\chisqfivetwodolphot{1036.8}
\def\paramsfivetwodolphot{26}
\def\aicfivetwodolphot{1088.8}
\def\delaystwosonesixonedolphot{14.381040}
\def\mustwosonesixonedolphot{1.074342}
\def\delaysthreesonesixonedolphot{-44.311999}
\def\musthreesonesixonedolphot{0.949849}
\def\delaysfoursonesixonedolphot{3.044176}
\def\musfoursonesixonedolphot{0.302143}
\def\delaysxsonesixonedolphot{365.728932}
\def\musxsonesixonedolphot{0.304948}
\def\photcoeffoneFonetwofiveWsixonedolphot{0.422079}
\def\photcoefftwoFonetwofiveWsixonedolphot{0.002503}
\def\photcoeffthreeFonetwofiveWsixonedolphot{-0.000014}
\def\photcoefffourFonetwofiveWsixonedolphot{0.000000}
\def\photcoefffiveFonetwofiveWsixonedolphot{0.000000}
\def\photcoeffsixFonetwofiveWsixonedolphot{-0.000000}
\def\photcoeffsevenFonetwofiveWsixonedolphot{0.000000}
\def\nebcoeffoneFonetwofiveWsixonedolphot{0.135546}
\def\nebcoefftwoFonetwofiveWsixonedolphot{-0.000162}
\def\photcoeffoneFonesixohWsixonedolphot{0.475092}
\def\photcoefftwoFonesixohWsixonedolphot{0.003091}
\def\photcoeffthreeFonesixohWsixonedolphot{0.000001}
\def\photcoefffourFonesixohWsixonedolphot{-0.000000}
\def\photcoefffiveFonesixohWsixonedolphot{-0.000000}
\def\photcoeffsixFonesixohWsixonedolphot{0.000000}
\def\photcoeffsevenFonesixohWsixonedolphot{-0.000000}
\def\nebcoeffoneFonesixohWsixonedolphot{0.187680}
\def\nebcoefftwoFonesixohWsixonedolphot{-0.000140}
\def\dofsixonedolphot{591}
\def\redchisqsixonedolphot{1.76}
\def\chisqsixonedolphot{1037.9}
\def\paramssixonedolphot{26}
\def\aicsixonedolphot{1089.9}
\def\delaystwosonesixtwodolphot{16.671806}
\def\mustwosonesixtwodolphot{1.085850}
\def\delaysthreesonesixtwodolphot{-40.156840}
\def\musthreesonesixtwodolphot{0.969007}
\def\delaysfoursonesixtwodolphot{10.006819}
\def\musfoursonesixtwodolphot{0.313659}
\def\delaysxsonesixtwodolphot{380.966961}
\def\musxsonesixtwodolphot{0.310310}
\def\photcoeffoneFonetwofiveWsixtwodolphot{0.422842}
\def\photcoefftwoFonetwofiveWsixtwodolphot{0.002937}
\def\photcoeffthreeFonetwofiveWsixtwodolphot{-0.000010}
\def\photcoefffourFonetwofiveWsixtwodolphot{-0.000000}
\def\photcoefffiveFonetwofiveWsixtwodolphot{0.000000}
\def\photcoeffsixFonetwofiveWsixtwodolphot{0.000000}
\def\photcoeffsevenFonetwofiveWsixtwodolphot{-0.000000}
\def\nebcoeffoneFonetwofiveWsixtwodolphot{0.179496}
\def\nebcoefftwoFonetwofiveWsixtwodolphot{-0.000356}
\def\nebcoeffthreeFonetwofiveWsixtwodolphot{0.000000}
\def\photcoeffoneFonesixohWsixtwodolphot{0.488157}
\def\photcoefftwoFonesixohWsixtwodolphot{0.003757}
\def\photcoeffthreeFonesixohWsixtwodolphot{-0.000002}
\def\photcoefffourFonesixohWsixtwodolphot{-0.000000}
\def\photcoefffiveFonesixohWsixtwodolphot{0.000000}
\def\photcoeffsixFonesixohWsixtwodolphot{0.000000}
\def\photcoeffsevenFonesixohWsixtwodolphot{-0.000000}
\def\nebcoeffoneFonesixohWsixtwodolphot{0.426914}
\def\nebcoefftwoFonesixohWsixtwodolphot{-0.001146}
\def\nebcoeffthreeFonesixohWsixtwodolphot{0.000001}
\def\dofsixtwodolphot{589}
\def\redchisqsixtwodolphot{1.69}
\def\chisqsixtwodolphot{993.1}
\def\paramssixtwodolphot{28}
\def\aicsixtwodolphot{1049.1}

%\bibpunct[, ]{(}{)}{;}{a}{}{,}
\newcommand{\ergA}{$\rm{erg\,cm^{-2}\,s^{-1}\,\AA^{-1}}$} 
\newcommand{\erg}{$\rm{erg\,cm^{-2}\,s^{-1}}$} 
\newcommand{\hb}{H$\beta$} 
\newcommand{\ha}{H$\alpha$} 
\newcommand{\hg}{H$\gamma$} 
\newcommand{\hii}{\ion{H}{2}} 
\newcommand{\oi}{[\ion{O}{1}]} 
\newcommand{\sii}{[\ion{S}{2}]} 
\newcommand{\oii}{[\ion{O}{2}]} 
\newcommand{\oiii}{[\ion{O}{3}]}
\newcommand{\nii}{[\ion{N}{2}]} 
\newcommand{\farc}{\hbox{$.\!\!^{\prime\prime}$}} % Fractions of arcseconds

\def\corrpmdelaysxsoneCombinedchabriermedian{$376.0_{-5.5}^{+5.6}$ days}

\def\sumpostmatrixsimsGrillogempiricalnopriornopositionsnoqualthroughtwoohtwooh{2 \times 10^{-5}}
\def\sumfracpostmatrixsimsGrillogempiricalnopriornopositionsnoqualthroughtwoohtwooh{0.041}
\def\sumpostmatrixsimsOguriaempiricalnopriornopositionsnoqualthroughtwoohtwooh{0.00019}
\def\sumfracpostmatrixsimsOguriaempiricalnopriornopositionsnoqualthroughtwoohtwooh{0.38}
\def\sumpostmatrixsimsOgurigempiricalnopriornopositionsnoqualthroughtwoohtwooh{0.00021}
\def\sumfracpostmatrixsimsOgurigempiricalnopriornopositionsnoqualthroughtwoohtwooh{0.42}
\def\sumpostmatrixsimsDiegoaempiricalnopriornopositionsnoqualthroughtwoohtwooh{1.2 \times 10^{-8}}
\def\sumfracpostmatrixsimsDiegoaempiricalnopriornopositionsnoqualthroughtwoohtwooh{2.5 \times 10^{-5}}
\def\sumpostmatrixsimsJauzaconefivedottwoempiricalnopriornopositionsnoqualthroughtwoohtwooh{8 \times 10^{-8}}
\def\sumfracpostmatrixsimsJauzaconefivedottwoempiricalnopriornopositionsnoqualthroughtwoohtwooh{0.00016}
\def\sumpostmatrixsimsSharonaempiricalnopriornopositionsnoqualthroughtwoohtwooh{2.3 \times 10^{-6}}
\def\sumfracpostmatrixsimsSharonaempiricalnopriornopositionsnoqualthroughtwoohtwooh{0.0046}
\def\sumpostmatrixsimsSharongempiricalnopriornopositionsnoqualthroughtwoohtwooh{6.9 \times 10^{-6}}
\def\sumfracpostmatrixsimsSharongempiricalnopriornopositionsnoqualthroughtwoohtwooh{0.014}
\def\sumpostmatrixsimsChentwoohtwoohempiricalnopriornopositionsnoqualthroughtwoohtwooh{1.3 \times 10^{-5}}
\def\sumfracpostmatrixsimsChentwoohtwoohempiricalnopriornopositionsnoqualthroughtwoohtwooh{0.025}
\def\sumpostmatrixsimsZitrintwoohtwoohltmempiricalnopriornopositionsnoqualthroughtwoohtwooh{1.3 \times 10^{-8}}
\def\sumfracpostmatrixsimsZitrintwoohtwoohltmempiricalnopriornopositionsnoqualthroughtwoohtwooh{2.6 \times 10^{-5}}
\def\sumpostmatrixsimsZitrintwoohtwoohpempiricalnopriornopositionsnoqualthroughtwoohtwooh{2.3 \times 10^{-5}}
\def\sumfracpostmatrixsimsZitrintwoohtwoohpempiricalnopriornopositionsnoqualthroughtwoohtwooh{0.046}
\def\sumpostmatrixsimsKeetonempiricalnopriornopositionsnoqualthroughtwoohtwooh{3.9 \times 10^{-5}}
\def\sumfracpostmatrixsimsKeetonempiricalnopriornopositionsnoqualthroughtwoohtwooh{0.078}
\def\hohmatrixsimsempiricalnopriornopositionsnoqualthroughtwoohtwooh{$63.3_{-3.2}^{+5.1}$}
\def\hohmfidmatrixsimsempiricalnopriornopositionsnoqualthroughtwoohtwooh{$+100062.3_{-3.2}^{+5.1}$}
\def\hohmatrixsimsempiricalnopriornopositionsnoqualpohtwotwoseventhroughtwoohtwooh{56.7}
\def\hohmatrixsimsempiricalnopriornopositionsnoqualponesixthroughtwoohtwooh{60.1}
\def\hohmatrixsimsempiricalnopriornopositionsnoqualpfiveohthroughtwoohtwooh{64.0}
\def\hohmatrixsimsempiricalnopriornopositionsnoqualpeightfourthroughtwoohtwooh{68.4}
\def\hohmatrixsimsempiricalnopriornopositionsnoqualpninesevenseventhreethroughtwoohtwooh{73.3}
\def\hohmatrixsimsempiricalnopriornopositionsnoqualpmaxthroughtwoohtwooh{63.3}
\def\hohmfidmatrixsimsempiricalnopriornopositionsnoqualpmaxthroughtwoohtwooh{+100062.3}
\def\sumpostmatrixsimsGrillogempiricalnopriorpositionsnoqual{2.7 \times 10^{-5}}
\def\sumfracpostmatrixsimsGrillogempiricalnopriorpositionsnoqual{0.099}
\def\sumpostmatrixsimsOguriaempiricalnopriorpositionsnoqual{0.00011}
\def\sumfracpostmatrixsimsOguriaempiricalnopriorpositionsnoqual{0.39}
\def\sumpostmatrixsimsOgurigempiricalnopriorpositionsnoqual{0.00013}
\def\sumfracpostmatrixsimsOgurigempiricalnopriorpositionsnoqual{0.46}
\def\sumpostmatrixsimsDiegoaempiricalnopriorpositionsnoqual{6.2 \times 10^{-9}}
\def\sumfracpostmatrixsimsDiegoaempiricalnopriorpositionsnoqual{2.2 \times 10^{-5}}
\def\sumpostmatrixsimsJauzaconefivedottwoempiricalnopriorpositionsnoqual{2.6 \times 10^{-8}}
\def\sumfracpostmatrixsimsJauzaconefivedottwoempiricalnopriorpositionsnoqual{9.6 \times 10^{-5}}
\def\sumpostmatrixsimsSharonaempiricalnopriorpositionsnoqual{4.1 \times 10^{-6}}
\def\sumfracpostmatrixsimsSharonaempiricalnopriorpositionsnoqual{0.015}
\def\sumpostmatrixsimsSharongempiricalnopriorpositionsnoqual{1.2 \times 10^{-5}}
\def\sumfracpostmatrixsimsSharongempiricalnopriorpositionsnoqual{0.042}
\def\sumpostmatrixsimsZitrincempiricalnopriorpositionsnoqual{4.6 \times 10^{-15}}
\def\sumfracpostmatrixsimsZitrincempiricalnopriorpositionsnoqual{1.7 \times 10^{-11}}
\def\hohmatrixsimsempiricalnopriorpositionsnoqual{$64.0_{-3.8}^{+5.0}$}
\def\hohmfidmatrixsimsempiricalnopriorpositionsnoqual{$+100063.0_{-3.8}^{+5.0}$}
\def\hohmatrixsimsempiricalnopriorpositionsnoqualpohtwotwoseven{56.3}
\def\hohmatrixsimsempiricalnopriorpositionsnoqualponesix{60.2}
\def\hohmatrixsimsempiricalnopriorpositionsnoqualpfiveoh{64.4}
\def\hohmatrixsimsempiricalnopriorpositionsnoqualpeightfour{69.0}
\def\hohmatrixsimsempiricalnopriorpositionsnoqualpninesevenseventhree{74.1}
\def\hohmatrixsimsempiricalnopriorpositionsnoqualpmax{64.0}
\def\hohmfidmatrixsimsempiricalnopriorpositionsnoqualpmax{+100063.0}
\def\sumpostmatrixsimsGrillogempiricalnopriorpositionsqualonly{2.7 \times 10^{-5}}
\def\sumfracpostmatrixsimsGrillogempiricalnopriorpositionsqualonly{0.2}
\def\sumpostmatrixsimsOguriaempiricalnopriorpositionsqualonly{0.00011}
\def\sumfracpostmatrixsimsOguriaempiricalnopriorpositionsqualonly{0.8}
\def\hohmatrixsimsempiricalnopriorpositionsqualonly{$66.3_{-3.0}^{+4.4}$}
\def\hohmfidmatrixsimsempiricalnopriorpositionsqualonly{$+100065.3_{-3.0}^{+4.4}$}
\def\hohmatrixsimsempiricalnopriorpositionsqualonlypohtwotwoseven{60.2}
\def\hohmatrixsimsempiricalnopriorpositionsqualonlyponesix{63.3}
\def\hohmatrixsimsempiricalnopriorpositionsqualonlypfiveoh{66.8}
\def\hohmatrixsimsempiricalnopriorpositionsqualonlypeightfour{70.7}
\def\hohmatrixsimsempiricalnopriorpositionsqualonlypninesevenseventhree{75.4}
\def\hohmatrixsimsempiricalnopriorpositionsqualonlypmax{66.3}
\def\hohmfidmatrixsimsempiricalnopriorpositionsqualonlypmax{+100065.3}
\def\intervalstwosonedelayRPthreerdmedianchabriersims{$-2.06^{+5.56}_{-6.62}$}
\def\rmsstwosonedelayRPthreerdmedianchabriersims{7.1}
\def\outlierfractionstwosonedelayRPthreerdmedianchabriersims{0.029}
\def\intervalstwosonedelayThorpmedianchabriersims{$0.22^{+3.76}_{-3.80}$}
\def\rmsstwosonedelayThorpmedianchabriersims{5.2}
\def\outlierfractionstwosonedelayThorpmedianchabriersims{0.031}
\def\intervalstwosonedelayKellymedianchabriersims{$-0.04^{+3.19}_{-3.19}$}
\def\rmsstwosonedelayKellymedianchabriersims{4.6}
\def\outlierfractionstwosonedelayKellymedianchabriersims{0.038}
\def\intervalstwosonedelayMillonsplmedianchabriersims{$-1.08^{+3.26}_{-3.39}$}
\def\rmsstwosonedelayMillonsplmedianchabriersims{4.9}
\def\outlierfractionstwosonedelayMillonsplmedianchabriersims{0.048}
\def\intervalsthreesonedelayRPthreerdmedianchabriersims{$-2.09^{+5.95}_{-5.71}$}
\def\rmssthreesonedelayRPthreerdmedianchabriersims{6.6}
\def\outlierfractionsthreesonedelayRPthreerdmedianchabriersims{0.008}
\def\intervalsthreesonedelayThorpmedianchabriersims{$-0.19^{+4.29}_{-3.95}$}
\def\rmssthreesonedelayThorpmedianchabriersims{5.0}
\def\outlierfractionsthreesonedelayThorpmedianchabriersims{0.015}
\def\intervalsthreesonedelayKellymedianchabriersims{$-0.08^{+3.19}_{-3.16}$}
\def\rmssthreesonedelayKellymedianchabriersims{4.3}
\def\outlierfractionsthreesonedelayKellymedianchabriersims{0.019}
\def\intervalsthreesonedelayMillonsplmedianchabriersims{$-1.12^{+3.34}_{-3.54}$}
\def\rmssthreesonedelayMillonsplmedianchabriersims{4.6}
\def\outlierfractionsthreesonedelayMillonsplmedianchabriersims{0.019}
\def\intervalsfoursonedelayRPthreerdmedianchabriersims{$-4.76^{+9.87}_{-10.51}$}
\def\rmssfoursonedelayRPthreerdmedianchabriersims{11.1}
\def\outlierfractionsfoursonedelayRPthreerdmedianchabriersims{0.019}
\def\intervalsfoursonedelayThorpmedianchabriersims{$-2.42^{+12.03}_{-8.68}$}
\def\rmssfoursonedelayThorpmedianchabriersims{10.3}
\def\outlierfractionsfoursonedelayThorpmedianchabriersims{0.026}
\def\intervalsfoursonedelayKellymedianchabriersims{$-0.38^{+4.99}_{-5.33}$}
\def\rmssfoursonedelayKellymedianchabriersims{5.8}
\def\outlierfractionsfoursonedelayKellymedianchabriersims{0.019}
\def\intervalsfoursonedelayMillonsplmedianchabriersims{$2.81^{+6.50}_{-7.50}$}
\def\rmssfoursonedelayMillonsplmedianchabriersims{8.1}
\def\outlierfractionsfoursonedelayMillonsplmedianchabriersims{0.030}
\def\intervalsxsonedelayRPthreerdmedianchabriersims{$-0.69^{+7.82}_{-8.91}$}
\def\rmssxsonedelayRPthreerdmedianchabriersims{8.6}
\def\outlierfractionsxsonedelayRPthreerdmedianchabriersims{0.009}
\def\intervalsxsonedelayThorpmedianchabriersims{$-2.29^{+6.68}_{-7.35}$}
\def\rmssxsonedelayThorpmedianchabriersims{7.7}
\def\outlierfractionsxsonedelayThorpmedianchabriersims{0.006}
\def\intervalsxsonedelayKellymedianchabriersims{$-0.18^{+5.13}_{-5.05}$}
\def\rmssxsonedelayKellymedianchabriersims{5.5}
\def\outlierfractionsxsonedelayKellymedianchabriersims{0.009}
\def\intervalsxsonedelayMillonsplmedianchabriersims{$-3.34^{+6.32}_{-6.71}$}
\def\rmssxsonedelayMillonsplmedianchabriersims{7.6}
\def\outlierfractionsxsonedelayMillonsplmedianchabriersims{0.040}
\def\weightstwosonesimsdelayRPthreerdmedianchabrier{0.005}
\def\weightstwosonesimsdelayThorpmedianchabrier{0.217}
\def\weightstwosonesimsdelayKellymedianchabrier{0.540}
\def\weightstwosonesimsdelayMillonsplmedianchabrier{0.238}
\def\uncorrsimsdelaystwosoneRPthreerd{1.47}
\def\corrsimsdelaystwosoneRPthreerdchabriermedian{4.28 (-3.10, 3.23, 8.74)}
\def\corrtwosigsimsdelaystwosoneRPthreerdchabriermedian{4.28 (-13.97, -3.10, 3.23, 8.74, 17.61)}
\def\corrtwosigsimsdelaystwosoneRPthreerdchabriermedian{4.28 (-13.97, -3.10, 3.23, 8.74, 17.61)}
\def\uncorrsimsdelaystwosoneThorp{11.61}
\def\corrsimsdelaystwosoneThorpchabriermedian{11.09 (7.95, 11.61, 15.83)}
\def\corrtwosigsimsdelaystwosoneThorpchabriermedian{11.09 (-0.80, 7.95, 11.61, 15.83, 23.48)}
\def\corrtwosigsimsdelaystwosoneThorpchabriermedian{11.09 (-0.80, 7.95, 11.61, 15.83, 23.48)}
\def\uncorrsimsdelaystwosoneKelly{8.39}
\def\corrsimsdelaystwosoneKellychabriermedian{8.49 (5.49, 8.57, 11.80)}
\def\corrtwosigsimsdelaystwosoneKellychabriermedian{8.49 (-6.67, 5.49, 8.57, 11.80, 17.75)}
\def\corrtwosigsimsdelaystwosoneKellychabriermedian{8.49 (-6.67, 5.49, 8.57, 11.80, 17.75)}
\def\uncorrsimsdelaystwosoneMillonspl{11.51}
\def\corrsimsdelaystwosoneMillonsplchabriermedian{12.89 (9.60, 12.80, 15.93)}
\def\corrtwosigsimsdelaystwosoneMillonsplchabriermedian{12.89 (-9.96, 9.60, 12.80, 15.93, 20.95)}
\def\corrtwosigsimsdelaystwosoneMillonsplchabriermedian{12.89 (-9.96, 9.60, 12.80, 15.93, 20.95)}
\def\corrsimsdelaystwosoneCombinedchabriermedian{9.69 (5.96, 9.92, 14.06)}
\def\corrtwosigsimsdelaystwosoneCombinedchabriermedian{9.69 (-8.97, 5.96, 9.92, 14.06, 20.56)}
\def\corrpmsimsdelaystwosoneCombinedchabriermedian{$9.7_{-3.7}^{+4.4}$ days}
\def\weightsthreesonesimsdelayRPthreerdmedianchabrier{0.027}
\def\weightsthreesonesimsdelayThorpmedianchabrier{0.173}
\def\weightsthreesonesimsdelayKellymedianchabrier{0.598}
\def\weightsthreesonesimsdelayMillonsplmedianchabrier{0.203}
\def\uncorrsimsdelaysthreesoneRPthreerd{11.38}
\def\corrsimsdelaysthreesoneRPthreerdchabriermedian{12.93 (7.23, 13.52, 19.58)}
\def\corrtwosigsimsdelaysthreesoneRPthreerdchabriermedian{12.93 (-0.85, 7.23, 13.52, 19.58, 28.07)}
\def\corrtwosigsimsdelaysthreesoneRPthreerdchabriermedian{12.93 (-0.85, 7.23, 13.52, 19.58, 28.07)}
\def\uncorrsimsdelaysthreesoneThorp{11.76}
\def\corrsimsdelaysthreesoneThorpchabriermedian{11.13 (8.04, 11.74, 16.25)}
\def\corrtwosigsimsdelaysthreesoneThorpchabriermedian{11.13 (2.24, 8.04, 11.74, 16.25, 23.06)}
\def\corrtwosigsimsdelaysthreesoneThorpchabriermedian{11.13 (2.24, 8.04, 11.74, 16.25, 23.06)}
\def\uncorrsimsdelaysthreesoneKelly{7.42}
\def\corrsimsdelaysthreesoneKellychabriermedian{7.52 (4.12, 7.46, 10.72)}
\def\corrtwosigsimsdelaysthreesoneKellychabriermedian{7.52 (-0.07, 4.12, 7.46, 10.72, 17.02)}
\def\corrtwosigsimsdelaysthreesoneKellychabriermedian{7.52 (-0.07, 4.12, 7.46, 10.72, 17.02)}
\def\uncorrsimsdelaysthreesoneMillonspl{10.39}
\def\corrsimsdelaysthreesoneMillonsplchabriermedian{11.73 (8.15, 11.65, 14.77)}
\def\corrtwosigsimsdelaysthreesoneMillonsplchabriermedian{11.73 (2.99, 8.15, 11.65, 14.77, 19.83)}
\def\corrtwosigsimsdelaysthreesoneMillonsplchabriermedian{11.73 (2.99, 8.15, 11.65, 14.77, 19.83)}
\def\corrsimsdelaysthreesoneCombinedchabriermedian{7.92 (5.16, 8.95, 13.56)}
\def\corrtwosigsimsdelaysthreesoneCombinedchabriermedian{7.92 (0.47, 5.16, 8.95, 13.56, 20.19)}
\def\corrpmsimsdelaysthreesoneCombinedchabriermedian{$7.9_{-2.8}^{+5.6}$ days}
\def\weightsfoursonesimsdelayRPthreerdmedianchabrier{0.023}
\def\weightsfoursonesimsdelayThorpmedianchabrier{0.015}
\def\weightsfoursonesimsdelayKellymedianchabrier{0.840}
\def\weightsfoursonesimsdelayMillonsplmedianchabrier{0.122}
\def\uncorrsimsdelaysfoursoneRPthreerd{11.32}
\def\corrsimsdelaysfoursoneRPthreerdchabriermedian{16.84 (5.24, 16.27, 27.06)}
\def\corrtwosigsimsdelaysfoursoneRPthreerdchabriermedian{16.84 (-15.09, 5.24, 16.27, 27.06, 36.89)}
\def\corrtwosigsimsdelaysfoursoneRPthreerdchabriermedian{16.84 (-15.09, 5.24, 16.27, 27.06, 36.89)}
\def\uncorrsimsdelaysfoursoneThorp{55.68}
\def\corrsimsdelaysfoursoneThorpchabriermedian{55.48 (49.51, 58.52, 70.77)}
\def\corrtwosigsimsdelaysfoursoneThorpchabriermedian{55.48 (38.12, 49.51, 58.52, 70.77, 85.43)}
\def\corrtwosigsimsdelaysfoursoneThorpchabriermedian{55.48 (38.12, 49.51, 58.52, 70.77, 85.43)}
\def\uncorrsimsdelaysfoursoneKelly{19.14}
\def\corrsimsdelaysfoursoneKellychabriermedian{19.24 (14.19, 19.29, 24.36)}
\def\corrtwosigsimsdelaysfoursoneKellychabriermedian{19.24 (4.92, 14.19, 19.29, 24.36, 31.75)}
\def\corrtwosigsimsdelaysfoursoneKellychabriermedian{19.24 (4.92, 14.19, 19.29, 24.36, 31.75)}
\def\uncorrsimsdelaysfoursoneMillonspl{34.52}
\def\corrsimsdelaysfoursoneMillonsplchabriermedian{31.25 (24.37, 31.22, 37.83)}
\def\corrtwosigsimsdelaysfoursoneMillonsplchabriermedian{31.25 (0.65, 24.37, 31.22, 37.83, 44.79)}
\def\corrtwosigsimsdelaysfoursoneMillonsplchabriermedian{31.25 (0.65, 24.37, 31.22, 37.83, 44.79)}
\def\corrsimsdelaysfoursoneCombinedchabriermedian{19.44 (14.56, 20.28, 27.45)}
\def\corrtwosigsimsdelaysfoursoneCombinedchabriermedian{19.44 (5.94, 14.56, 20.28, 27.45, 43.15)}
\def\corrpmsimsdelaysfoursoneCombinedchabriermedian{$19.4_{-4.9}^{+8.0}$ days}
\def\weightsxsonesimsdelayRPthreerdmedianchabrier{0.053}
\def\weightsxsonesimsdelayThorpmedianchabrier{0.038}
\def\weightsxsonesimsdelayKellymedianchabrier{0.885}
\def\weightsxsonesimsdelayMillonsplmedianchabrier{0.024}
\def\uncorrsimsdelaysxsoneRPthreerd{378.56}
\def\corrsimsdelaysxsoneRPthreerdchabriermedian{380.83 (370.68, 379.58, 386.68)}
\def\corrtwosigsimsdelaysxsoneRPthreerdchabriermedian{380.83 (360.53, 370.68, 379.58, 386.68, 394.58)}
\def\corrtwosigsimsdelaysxsoneRPthreerdchabriermedian{380.83 (360.53, 370.68, 379.58, 386.68, 394.58)}
\def\uncorrsimsdelaysxsoneThorp{381.23}
\def\corrsimsdelaysxsoneThorpchabriermedian{383.83 (375.96, 383.62, 390.23)}
\def\corrtwosigsimsdelaysxsoneThorpchabriermedian{383.83 (367.04, 375.96, 383.62, 390.23, 400.15)}
\def\corrtwosigsimsdelaysxsoneThorpchabriermedian{383.83 (367.04, 375.96, 383.62, 390.23, 400.15)}
\def\uncorrsimsdelaysxsoneKelly{375.32}
\def\corrsimsdelaysxsoneKellychabriermedian{376.02 (370.63, 375.74, 380.99)}
\def\corrtwosigsimsdelaysxsoneKellychabriermedian{376.02 (365.31, 370.63, 375.74, 380.99, 387.37)}
\def\corrtwosigsimsdelaysxsoneKellychabriermedian{376.02 (365.31, 370.63, 375.74, 380.99, 387.37)}
\def\uncorrsimsdelaysxsoneMillonspl{378.94}
\def\corrsimsdelaysxsoneMillonsplchabriermedian{382.43 (376.45, 382.53, 389.14)}
\def\corrtwosigsimsdelaysxsoneMillonsplchabriermedian{382.43 (367.43, 376.45, 382.53, 389.14, 450.11)}
\def\corrtwosigsimsdelaysxsoneMillonsplchabriermedian{382.43 (367.43, 376.45, 382.53, 389.14, 450.11)}
\def\corrsimsdelaysxsoneCombinedchabriermedian{376.02 (370.50, 376.03, 381.65)}
\def\corrtwosigsimsdelaysxsoneCombinedchabriermedian{376.02 (365.12, 370.50, 376.03, 381.65, 390.23)}
\def\corrpmsimsdelaysxsoneCombinedchabriermedian{$376.0_{-5.5}^{+5.6}$ days}
\def\intervalstwosonemuRPthreerdmedianchabriersims{$0.036^{+0.255}_{-0.157}$}
\def\rmsstwosonemuRPthreerdmedianchabriersims{0.30}
\def\outlierfractionstwosonemuRPthreerdmedianchabriersims{0.022}
\def\intervalstwosonemuThorpmedianchabriersims{$0.011^{+0.221}_{-0.137}$}
\def\rmsstwosonemuThorpmedianchabriersims{0.24}
\def\outlierfractionstwosonemuThorpmedianchabriersims{0.019}
\def\intervalstwosonemuKellymedianchabriersims{$0.040^{+0.257}_{-0.140}$}
\def\rmsstwosonemuKellymedianchabriersims{0.29}
\def\outlierfractionstwosonemuKellymedianchabriersims{0.028}
\def\intervalstwosonemuMillonsplmedianchabriersims{$0.028^{+0.246}_{-0.137}$}
\def\rmsstwosonemuMillonsplmedianchabriersims{0.27}
\def\outlierfractionstwosonemuMillonsplmedianchabriersims{0.023}
\def\intervalsthreesonemuRPthreerdmedianchabriersims{$0.004^{+0.312}_{-0.193}$}
\def\rmssthreesonemuRPthreerdmedianchabriersims{0.35}
\def\outlierfractionsthreesonemuRPthreerdmedianchabriersims{0.034}
\def\intervalsthreesonemuThorpmedianchabriersims{$0.007^{+0.330}_{-0.167}$}
\def\rmssthreesonemuThorpmedianchabriersims{0.32}
\def\outlierfractionsthreesonemuThorpmedianchabriersims{0.035}
\def\intervalsthreesonemuKellymedianchabriersims{$0.002^{+0.325}_{-0.164}$}
\def\rmssthreesonemuKellymedianchabriersims{0.33}
\def\outlierfractionsthreesonemuKellymedianchabriersims{0.034}
\def\intervalsthreesonemuMillonsplmedianchabriersims{$0.003^{+0.328}_{-0.159}$}
\def\rmssthreesonemuMillonsplmedianchabriersims{0.33}
\def\outlierfractionsthreesonemuMillonsplmedianchabriersims{0.035}
\def\intervalsfoursonemuRPthreerdmedianchabriersims{$0.005^{+0.061}_{-0.041}$}
\def\rmssfoursonemuRPthreerdmedianchabriersims{0.08}
\def\outlierfractionsfoursonemuRPthreerdmedianchabriersims{0.019}
\def\intervalsfoursonemuThorpmedianchabriersims{$0.082^{+0.064}_{-0.058}$}
\def\rmssfoursonemuThorpmedianchabriersims{0.11}
\def\outlierfractionsfoursonemuThorpmedianchabriersims{0.005}
\def\intervalsfoursonemuKellymedianchabriersims{$0.010^{+0.063}_{-0.036}$}
\def\rmssfoursonemuKellymedianchabriersims{0.08}
\def\outlierfractionsfoursonemuKellymedianchabriersims{0.019}
\def\intervalsfoursonemuMillonsplmedianchabriersims{$0.046^{+0.059}_{-0.036}$}
\def\rmssfoursonemuMillonsplmedianchabriersims{0.08}
\def\outlierfractionsfoursonemuMillonsplmedianchabriersims{0.010}
\def\intervalsxsonemuRPthreerdmedianchabriersims{$0.008^{+0.053}_{-0.036}$}
\def\rmssxsonemuRPthreerdmedianchabriersims{0.08}
\def\outlierfractionsxsonemuRPthreerdmedianchabriersims{0.016}
\def\intervalsxsonemuThorpmedianchabriersims{$0.032^{+0.051}_{-0.045}$}
\def\rmssxsonemuThorpmedianchabriersims{0.08}
\def\outlierfractionsxsonemuThorpmedianchabriersims{0.014}
\def\intervalsxsonemuKellymedianchabriersims{$0.005^{+0.053}_{-0.033}$}
\def\rmssxsonemuKellymedianchabriersims{0.08}
\def\outlierfractionsxsonemuKellymedianchabriersims{0.016}
\def\intervalsxsonemuMillonsplmedianchabriersims{$0.005^{+0.047}_{-0.032}$}
\def\rmssxsonemuMillonsplmedianchabriersims{0.08}
\def\outlierfractionsxsonemuMillonsplmedianchabriersims{0.016}
\def\weightstwosonesimsmuRPthreerdmedianchabrier{0.000}
\def\weightstwosonesimsmuThorpmedianchabrier{0.872}
\def\weightstwosonesimsmuKellymedianchabrier{0.128}
\def\weightstwosonesimsmuMillonsplmedianchabrier{0.000}
\def\uncorrsimsmustwosoneRPthreerd{1.096}
\def\corrsimsmustwosoneRPthreerdchabriermedian{1.021 (0.910, 1.056, 1.308)}
\def\corrtwosigsimsmustwosoneRPthreerdchabriermedian{1.021 (0.451, 0.910, 1.056, 1.308, 1.949)}
\def\corrtwosigsimsmustwosoneRPthreerdchabriermedian{1.021 (0.451, 0.910, 1.056, 1.308, 1.949)}
\def\uncorrsimsmustwosoneThorp{1.093}
\def\corrsimsmustwosoneThorpchabriermedian{1.057 (0.944, 1.081, 1.296)}
\def\corrtwosigsimsmustwosoneThorpchabriermedian{1.057 (0.767, 0.944, 1.081, 1.296, 1.794)}
\def\corrtwosigsimsmustwosoneThorpchabriermedian{1.057 (0.767, 0.944, 1.081, 1.296, 1.794)}
\def\uncorrsimsmustwosoneKelly{1.090}
\def\corrsimsmustwosoneKellychabriermedian{1.021 (0.915, 1.048, 1.287)}
\def\corrtwosigsimsmustwosoneKellychabriermedian{1.021 (0.618, 0.915, 1.048, 1.287, 1.921)}
\def\corrtwosigsimsmustwosoneKellychabriermedian{1.021 (0.618, 0.915, 1.048, 1.287, 1.921)}
\def\uncorrsimsmustwosoneMillonspl{1.109}
\def\corrsimsmustwosoneMillonsplchabriermedian{1.051 (0.952, 1.079, 1.315)}
\def\corrtwosigsimsmustwosoneMillonsplchabriermedian{1.051 (0.667, 0.952, 1.079, 1.315, 1.885)}
\def\corrtwosigsimsmustwosoneMillonsplchabriermedian{1.051 (0.667, 0.952, 1.079, 1.315, 1.885)}
\def\corrsimsmustwosoneCombinedchabriermedian{1.057 (0.941, 1.079, 1.304)}
\def\corrtwosigsimsmustwosoneCombinedchabriermedian{1.057 (0.630, 0.941, 1.079, 1.304, 1.775)}
\def\corrpmsimsmustwosoneCombinedchabriermedian{$1.06_{-0.12}^{+0.25}$}
\def\weightsthreesonesimsmuRPthreerdmedianchabrier{0.092}
\def\weightsthreesonesimsmuThorpmedianchabrier{0.535}
\def\weightsthreesonesimsmuKellymedianchabrier{0.373}
\def\weightsthreesonesimsmuMillonsplmedianchabrier{0.000}
\def\uncorrsimsmusthreesoneRPthreerd{0.943}
\def\corrsimsmusthreesoneRPthreerdchabriermedian{0.889 (0.721, 0.933, 1.206)}
\def\corrtwosigsimsmusthreesoneRPthreerdchabriermedian{0.889 (0.281, 0.721, 0.933, 1.206, 2.008)}
\def\corrtwosigsimsmusthreesoneRPthreerdchabriermedian{0.889 (0.281, 0.721, 0.933, 1.206, 2.008)}
\def\uncorrsimsmusthreesoneThorp{0.983}
\def\corrsimsmusthreesoneThorpchabriermedian{0.919 (0.803, 0.977, 1.285)}
\def\corrtwosigsimsmusthreesoneThorpchabriermedian{0.919 (0.506, 0.803, 0.977, 1.285, 2.025)}
\def\corrtwosigsimsmusthreesoneThorpchabriermedian{0.919 (0.506, 0.803, 0.977, 1.285, 2.025)}
\def\uncorrsimsmusthreesoneKelly{0.970}
\def\corrsimsmusthreesoneKellychabriermedian{0.907 (0.799, 0.969, 1.250)}
\def\corrtwosigsimsmusthreesoneKellychabriermedian{0.907 (0.357, 0.799, 0.969, 1.250, 2.021)}
\def\corrtwosigsimsmusthreesoneKellychabriermedian{0.907 (0.357, 0.799, 0.969, 1.250, 2.021)}
\def\uncorrsimsmusthreesoneMillonspl{0.993}
\def\corrsimsmusthreesoneMillonsplchabriermedian{0.931 (0.827, 0.996, 1.279)}
\def\corrtwosigsimsmusthreesoneMillonsplchabriermedian{0.931 (0.396, 0.827, 0.996, 1.279, 2.044)}
\def\corrtwosigsimsmusthreesoneMillonsplchabriermedian{0.931 (0.396, 0.827, 0.996, 1.279, 2.044)}
\def\corrsimsmusthreesoneCombinedchabriermedian{0.907 (0.802, 0.969, 1.298)}
\def\corrtwosigsimsmusthreesoneCombinedchabriermedian{0.907 (0.498, 0.802, 0.969, 1.298, 2.068)}
\def\corrpmsimsmusthreesoneCombinedchabriermedian{$0.91_{-0.11}^{+0.39}$}
\def\weightsfoursonesimsmuRPthreerdmedianchabrier{0.308}
\def\weightsfoursonesimsmuThorpmedianchabrier{0.054}
\def\weightsfoursonesimsmuKellymedianchabrier{0.583}
\def\weightsfoursonesimsmuMillonsplmedianchabrier{0.055}
\def\uncorrsimsmusfoursoneRPthreerd{0.290}
\def\corrsimsmusfoursoneRPthreerdchabriermedian{0.278 (0.247, 0.285, 0.345)}
\def\corrtwosigsimsmusfoursoneRPthreerdchabriermedian{0.278 (0.176, 0.247, 0.285, 0.345, 0.495)}
\def\corrtwosigsimsmusfoursoneRPthreerdchabriermedian{0.278 (0.176, 0.247, 0.285, 0.345, 0.495)}
\def\uncorrsimsmusfoursoneThorp{0.438}
\def\corrsimsmusfoursoneThorpchabriermedian{0.357 (0.298, 0.356, 0.420)}
\def\corrtwosigsimsmusfoursoneThorpchabriermedian{0.357 (0.239, 0.298, 0.356, 0.420, 0.525)}
\def\corrtwosigsimsmusfoursoneThorpchabriermedian{0.357 (0.239, 0.298, 0.356, 0.420, 0.525)}
\def\uncorrsimsmusfoursoneKelly{0.311}
\def\corrsimsmusfoursoneKellychabriermedian{0.296 (0.268, 0.303, 0.363)}
\def\corrtwosigsimsmusfoursoneKellychabriermedian{0.296 (0.199, 0.268, 0.303, 0.363, 0.529)}
\def\corrtwosigsimsmusfoursoneKellychabriermedian{0.296 (0.199, 0.268, 0.303, 0.363, 0.529)}
\def\uncorrsimsmusfoursoneMillonspl{0.389}
\def\corrsimsmusfoursoneMillonsplchabriermedian{0.333 (0.310, 0.344, 0.397)}
\def\corrtwosigsimsmusfoursoneMillonsplchabriermedian{0.333 (0.248, 0.310, 0.344, 0.397, 0.538)}
\def\corrtwosigsimsmusfoursoneMillonsplchabriermedian{0.333 (0.248, 0.310, 0.344, 0.397, 0.538)}
\def\corrsimsmusfoursoneCombinedchabriermedian{0.284 (0.259, 0.300, 0.368)}
\def\corrtwosigsimsmusfoursoneCombinedchabriermedian{0.284 (0.176, 0.259, 0.300, 0.368, 0.522)}
\def\corrpmsimsmusfoursoneCombinedchabriermedian{$0.28_{-0.03}^{+0.08}$}
\def\weightsxsonesimsmuRPthreerdmedianchabrier{0.000}
\def\weightsxsonesimsmuThorpmedianchabrier{0.014}
\def\weightsxsonesimsmuKellymedianchabrier{0.252}
\def\weightsxsonesimsmuMillonsplmedianchabrier{0.734}
\def\uncorrsimsmusxsoneRPthreerd{0.297}
\def\corrsimsmusxsoneRPthreerdchabriermedian{0.275 (0.253, 0.289, 0.343)}
\def\corrtwosigsimsmusxsoneRPthreerdchabriermedian{0.275 (0.198, 0.253, 0.289, 0.343, 0.447)}
\def\corrtwosigsimsmusxsoneRPthreerdchabriermedian{0.275 (0.198, 0.253, 0.289, 0.343, 0.447)}
\def\uncorrsimsmusxsoneThorp{0.342}
\def\corrsimsmusxsoneThorpchabriermedian{0.311 (0.264, 0.311, 0.362)}
\def\corrtwosigsimsmusxsoneThorpchabriermedian{0.311 (0.203, 0.264, 0.311, 0.362, 0.455)}
\def\corrtwosigsimsmusxsoneThorpchabriermedian{0.311 (0.203, 0.264, 0.311, 0.362, 0.455)}
\def\uncorrsimsmusxsoneKelly{0.302}
\def\corrsimsmusxsoneKellychabriermedian{0.287 (0.264, 0.297, 0.350)}
\def\corrtwosigsimsmusxsoneKellychabriermedian{0.287 (0.211, 0.264, 0.297, 0.350, 0.459)}
\def\corrtwosigsimsmusxsoneKellychabriermedian{0.287 (0.211, 0.264, 0.297, 0.350, 0.459)}
\def\uncorrsimsmusxsoneMillonspl{0.313}
\def\corrsimsmusxsoneMillonsplchabriermedian{0.305 (0.274, 0.309, 0.355)}
\def\corrtwosigsimsmusxsoneMillonsplchabriermedian{0.305 (0.194, 0.274, 0.309, 0.355, 0.437)}
\def\corrtwosigsimsmusxsoneMillonsplchabriermedian{0.305 (0.194, 0.274, 0.309, 0.355, 0.437)}
\def\corrsimsmusxsoneCombinedchabriermedian{0.299 (0.272, 0.305, 0.354)}
\def\corrtwosigsimsmusxsoneCombinedchabriermedian{0.299 (0.216, 0.272, 0.305, 0.354, 0.447)}
\def\corrpmsimsmusxsoneCombinedchabriermedian{$0.30_{-0.03}^{+0.05}$}
\def\sumpostmatrixsimsGrillogempiricalpriornopositionsnoqualthroughtwoohtwooh{1.9 \times 10^{-6}}
\def\sumfracpostmatrixsimsGrillogempiricalpriornopositionsnoqualthroughtwoohtwooh{0.058}
\def\sumpostmatrixsimsOguriaempiricalpriornopositionsnoqualthroughtwoohtwooh{4.4 \times 10^{-6}}
\def\sumfracpostmatrixsimsOguriaempiricalpriornopositionsnoqualthroughtwoohtwooh{0.7}
\def\sumpostmatrixsimsOgurigempiricalpriornopositionsnoqualthroughtwoohtwooh{4.6 \times 10^{-7}}
\def\sumfracpostmatrixsimsOgurigempiricalpriornopositionsnoqualthroughtwoohtwooh{0.074}
\def\sumpostmatrixsimsDiegoaempiricalpriornopositionsnoqualthroughtwoohtwooh{2 \times 10^{-10}}
\def\sumfracpostmatrixsimsDiegoaempiricalpriornopositionsnoqualthroughtwoohtwooh{3.3 \times 10^{-5}}
\def\sumpostmatrixsimsJauzaconefivedottwoempiricalpriornopositionsnoqualthroughtwoohtwooh{3.5 \times 10^{-9}}
\def\sumfracpostmatrixsimsJauzaconefivedottwoempiricalpriornopositionsnoqualthroughtwoohtwooh{0.00057}
\def\sumpostmatrixsimsSharonaempiricalpriornopositionsnoqualthroughtwoohtwooh{1.2 \times 10^{-8}}
\def\sumfracpostmatrixsimsSharonaempiricalpriornopositionsnoqualthroughtwoohtwooh{0.002}
\def\sumpostmatrixsimsSharongempiricalpriornopositionsnoqualthroughtwoohtwooh{\times 10^{-7}}
\def\sumfracpostmatrixsimsSharongempiricalpriornopositionsnoqualthroughtwoohtwooh{0.017}
\def\sumpostmatrixsimsChentwoohtwoohempiricalpriornopositionsnoqualthroughtwoohtwooh{1.9 \times 10^{-11}}
\def\sumfracpostmatrixsimsChentwoohtwoohempiricalpriornopositionsnoqualthroughtwoohtwooh{3.1 \times 10^{-6}}
\def\sumpostmatrixsimsZitrintwoohtwoohltmempiricalpriornopositionsnoqualthroughtwoohtwooh{1.4 \times 10^{-14}}
\def\sumfracpostmatrixsimsZitrintwoohtwoohltmempiricalpriornopositionsnoqualthroughtwoohtwooh{2.3 \times 10^{-9}}
\def\sumpostmatrixsimsZitrintwoohtwoohpempiricalpriornopositionsnoqualthroughtwoohtwooh{1.6 \times 10^{-8}}
\def\sumfracpostmatrixsimsZitrintwoohtwoohpempiricalpriornopositionsnoqualthroughtwoohtwooh{0.0025}
\def\sumpostmatrixsimsKeetonempiricalpriornopositionsnoqualthroughtwoohtwooh{2.9 \times 10^{-7}}
\def\sumfracpostmatrixsimsKeetonempiricalpriornopositionsnoqualthroughtwoohtwooh{0.047}
\def\hohmatrixsimsempiricalpriornopositionsnoqualthroughtwoohtwooh{$72.6_{-1.0}^{+1.0}$}
\def\hohmfidmatrixsimsempiricalpriornopositionsnoqualthroughtwoohtwooh{$+100071.6_{-1.0}^{+1.0}$}
\def\hohmatrixsimsempiricalpriornopositionsnoqualpohtwotwoseventhroughtwoohtwooh{70.6}
\def\hohmatrixsimsempiricalpriornopositionsnoqualponesixthroughtwoohtwooh{71.6}
\def\hohmatrixsimsempiricalpriornopositionsnoqualpfiveohthroughtwoohtwooh{72.6}
\def\hohmatrixsimsempiricalpriornopositionsnoqualpeightfourthroughtwoohtwooh{73.6}
\def\hohmatrixsimsempiricalpriornopositionsnoqualpninesevenseventhreethroughtwoohtwooh{74.7}
\def\hohmatrixsimsempiricalpriornopositionsnoqualpmaxthroughtwoohtwooh{72.6}
\def\hohmfidmatrixsimsempiricalpriornopositionsnoqualpmaxthroughtwoohtwooh{+100071.6}
\def\sumpostmatrixsimsGrillogempiricalpriorpositionsnoqual{1.3 \times 10^{-6}}
\def\sumfracpostmatrixsimsGrillogempiricalpriorpositionsnoqual{0.31}
\def\sumpostmatrixsimsOguriaempiricalpriorpositionsnoqual{2.5 \times 10^{-6}}
\def\sumfracpostmatrixsimsOguriaempiricalpriorpositionsnoqual{0.58}
\def\sumpostmatrixsimsOgurigempiricalpriorpositionsnoqual{2.8 \times 10^{-7}}
\def\sumfracpostmatrixsimsOgurigempiricalpriorpositionsnoqual{0.066}
\def\sumpostmatrixsimsDiegoaempiricalpriorpositionsnoqual{\times 10^{-10}}
\def\sumfracpostmatrixsimsDiegoaempiricalpriorpositionsnoqual{2.4 \times 10^{-5}}
\def\sumpostmatrixsimsJauzaconefivedottwoempiricalpriorpositionsnoqual{1.2 \times 10^{-9}}
\def\sumfracpostmatrixsimsJauzaconefivedottwoempiricalpriorpositionsnoqual{0.00028}
\def\sumpostmatrixsimsSharonaempiricalpriorpositionsnoqual{2.1 \times 10^{-8}}
\def\sumfracpostmatrixsimsSharonaempiricalpriorpositionsnoqual{0.0051}
\def\sumpostmatrixsimsSharongempiricalpriorpositionsnoqual{1.8 \times 10^{-7}}
\def\sumfracpostmatrixsimsSharongempiricalpriorpositionsnoqual{0.041}
\def\sumpostmatrixsimsZitrincempiricalpriorpositionsnoqual{1.3 \times 10^{-16}}
\def\sumfracpostmatrixsimsZitrincempiricalpriorpositionsnoqual{3.2 \times 10^{-11}}
\def\hohmatrixsimsempiricalpriorpositionsnoqual{$72.6_{-1.0}^{+1.1}$}
\def\hohmfidmatrixsimsempiricalpriorpositionsnoqual{$+100071.6_{-1.0}^{+1.1}$}
\def\hohmatrixsimsempiricalpriorpositionsnoqualpohtwotwoseven{70.6}
\def\hohmatrixsimsempiricalpriorpositionsnoqualponesix{71.6}
\def\hohmatrixsimsempiricalpriorpositionsnoqualpfiveoh{72.6}
\def\hohmatrixsimsempiricalpriorpositionsnoqualpeightfour{73.7}
\def\hohmatrixsimsempiricalpriorpositionsnoqualpninesevenseventhree{74.7}
\def\hohmatrixsimsempiricalpriorpositionsnoqualpmax{72.6}
\def\hohmfidmatrixsimsempiricalpriorpositionsnoqualpmax{+100071.6}
\def\sumpostmatrixsimsGrillogempiricalpriorpositionsqualonly{1.3 \times 10^{-6}}
\def\sumfracpostmatrixsimsGrillogempiricalpriorpositionsqualonly{0.34}
\def\sumpostmatrixsimsOguriaempiricalpriorpositionsqualonly{2.5 \times 10^{-6}}
\def\sumfracpostmatrixsimsOguriaempiricalpriorpositionsqualonly{0.66}
\def\hohmatrixsimsempiricalpriorpositionsqualonly{$72.7_{-1.1}^{+1.0}$}
\def\hohmfidmatrixsimsempiricalpriorpositionsqualonly{$+100071.7_{-1.1}^{+1.0}$}
\def\hohmatrixsimsempiricalpriorpositionsqualonlypohtwotwoseven{70.6}
\def\hohmatrixsimsempiricalpriorpositionsqualonlyponesix{71.6}
\def\hohmatrixsimsempiricalpriorpositionsqualonlypfiveoh{72.7}
\def\hohmatrixsimsempiricalpriorpositionsqualonlypeightfour{73.7}
\def\hohmatrixsimsempiricalpriorpositionsqualonlypninesevenseventhree{74.7}
\def\hohmatrixsimsempiricalpriorpositionsqualonlypmax{72.7}
\def\hohmfidmatrixsimsempiricalpriorpositionsqualonlypmax{+100071.7}
\def\sumpostmatrixsimsGrillogempiricalPlancknopositionsnoqualthroughtwoohtwooh{1.9 \times 10^{-6}}
\def\sumfracpostmatrixsimsGrillogempiricalPlancknopositionsnoqualthroughtwoohtwooh{0.058}
\def\sumpostmatrixsimsOguriaempiricalPlancknopositionsnoqualthroughtwoohtwooh{2 \times 10^{-5}}
\def\sumfracpostmatrixsimsOguriaempiricalPlancknopositionsnoqualthroughtwoohtwooh{0.61}
\def\sumpostmatrixsimsOgurigempiricalPlancknopositionsnoqualthroughtwoohtwooh{7.4 \times 10^{-6}}
\def\sumfracpostmatrixsimsOgurigempiricalPlancknopositionsnoqualthroughtwoohtwooh{0.23}
\def\sumpostmatrixsimsDiegoaempiricalPlancknopositionsnoqualthroughtwoohtwooh{3.4 \times 10^{-10}}
\def\sumfracpostmatrixsimsDiegoaempiricalPlancknopositionsnoqualthroughtwoohtwooh{1.1 \times 10^{-5}}
\def\sumpostmatrixsimsJauzaconefivedottwoempiricalPlancknopositionsnoqualthroughtwoohtwooh{4 \times 10^{-9}}
\def\sumfracpostmatrixsimsJauzaconefivedottwoempiricalPlancknopositionsnoqualthroughtwoohtwooh{0.00012}
\def\sumpostmatrixsimsSharonaempiricalPlancknopositionsnoqualthroughtwoohtwooh{4.4 \times 10^{-8}}
\def\sumfracpostmatrixsimsSharonaempiricalPlancknopositionsnoqualthroughtwoohtwooh{0.0014}
\def\sumpostmatrixsimsSharongempiricalPlancknopositionsnoqualthroughtwoohtwooh{3.2 \times 10^{-7}}
\def\sumfracpostmatrixsimsSharongempiricalPlancknopositionsnoqualthroughtwoohtwooh{0.0099}
\def\sumpostmatrixsimsChentwoohtwoohempiricalPlancknopositionsnoqualthroughtwoohtwooh{6.8 \times 10^{-8}}
\def\sumfracpostmatrixsimsChentwoohtwoohempiricalPlancknopositionsnoqualthroughtwoohtwooh{0.0021}
\def\sumpostmatrixsimsZitrintwoohtwoohltmempiricalPlancknopositionsnoqualthroughtwoohtwooh{9.7 \times 10^{-13}}
\def\sumfracpostmatrixsimsZitrintwoohtwoohltmempiricalPlancknopositionsnoqualthroughtwoohtwooh{3 \times 10^{-8}}
\def\sumpostmatrixsimsZitrintwoohtwoohpempiricalPlancknopositionsnoqualthroughtwoohtwooh{5.8 \times 10^{-7}}
\def\sumfracpostmatrixsimsZitrintwoohtwoohpempiricalPlancknopositionsnoqualthroughtwoohtwooh{0.018}
\def\sumpostmatrixsimsKeetonempiricalPlancknopositionsnoqualthroughtwoohtwooh{2.3 \times 10^{-6}}
\def\sumfracpostmatrixsimsKeetonempiricalPlancknopositionsnoqualthroughtwoohtwooh{0.07}
\def\hohmatrixsimsempiricalPlancknopositionsnoqualthroughtwoohtwooh{$67.3_{-0.6}^{+0.6}$}
\def\hohmfidmatrixsimsempiricalPlancknopositionsnoqualthroughtwoohtwooh{$+100066.3_{-0.6}^{+0.6}$}
\def\hohmatrixsimsempiricalPlancknopositionsnoqualpohtwotwoseventhroughtwoohtwooh{66.1}
\def\hohmatrixsimsempiricalPlancknopositionsnoqualponesixthroughtwoohtwooh{66.7}
\def\hohmatrixsimsempiricalPlancknopositionsnoqualpfiveohthroughtwoohtwooh{67.3}
\def\hohmatrixsimsempiricalPlancknopositionsnoqualpeightfourthroughtwoohtwooh{67.9}
\def\hohmatrixsimsempiricalPlancknopositionsnoqualpninesevenseventhreethroughtwoohtwooh{68.5}
\def\hohmatrixsimsempiricalPlancknopositionsnoqualpmaxthroughtwoohtwooh{67.3}
\def\hohmfidmatrixsimsempiricalPlancknopositionsnoqualpmaxthroughtwoohtwooh{+100066.3}
\def\sumpostmatrixsimsGrillogempiricalPlanckpositionsnoqual{2.5 \times 10^{-6}}
\def\sumfracpostmatrixsimsGrillogempiricalPlanckpositionsnoqual{0.13}
\def\sumpostmatrixsimsOguriaempiricalPlanckpositionsnoqual{1.1 \times 10^{-5}}
\def\sumfracpostmatrixsimsOguriaempiricalPlanckpositionsnoqual{0.6}
\def\sumpostmatrixsimsOgurigempiricalPlanckpositionsnoqual{4.5 \times 10^{-6}}
\def\sumfracpostmatrixsimsOgurigempiricalPlanckpositionsnoqual{0.24}
\def\sumpostmatrixsimsDiegoaempiricalPlanckpositionsnoqual{1.7 \times 10^{-10}}
\def\sumfracpostmatrixsimsDiegoaempiricalPlanckpositionsnoqual{9.2 \times 10^{-6}}
\def\sumpostmatrixsimsJauzaconefivedottwoempiricalPlanckpositionsnoqual{1.3 \times 10^{-9}}
\def\sumfracpostmatrixsimsJauzaconefivedottwoempiricalPlanckpositionsnoqual{7 \times 10^{-5}}
\def\sumpostmatrixsimsSharonaempiricalPlanckpositionsnoqual{7.8 \times 10^{-8}}
\def\sumfracpostmatrixsimsSharonaempiricalPlanckpositionsnoqual{0.0041}
\def\sumpostmatrixsimsSharongempiricalPlanckpositionsnoqual{5.4 \times 10^{-7}}
\def\sumfracpostmatrixsimsSharongempiricalPlanckpositionsnoqual{0.028}
\def\sumpostmatrixsimsZitrincempiricalPlanckpositionsnoqual{9.8 \times 10^{-17}}
\def\sumfracpostmatrixsimsZitrincempiricalPlanckpositionsnoqual{5.2 \times 10^{-12}}
\def\hohmatrixsimsempiricalPlanckpositionsnoqual{$67.3_{-0.5}^{+0.6}$}
\def\hohmfidmatrixsimsempiricalPlanckpositionsnoqual{$+100066.3_{-0.5}^{+0.6}$}
\def\hohmatrixsimsempiricalPlanckpositionsnoqualpohtwotwoseven{66.2}
\def\hohmatrixsimsempiricalPlanckpositionsnoqualponesix{66.8}
\def\hohmatrixsimsempiricalPlanckpositionsnoqualpfiveoh{67.3}
\def\hohmatrixsimsempiricalPlanckpositionsnoqualpeightfour{67.9}
\def\hohmatrixsimsempiricalPlanckpositionsnoqualpninesevenseventhree{68.5}
\def\hohmatrixsimsempiricalPlanckpositionsnoqualpmax{67.3}
\def\hohmfidmatrixsimsempiricalPlanckpositionsnoqualpmax{+100066.3}
\def\sumpostmatrixsimsGrillogempiricalPlanckpositionsqualonly{2.5 \times 10^{-6}}
\def\sumfracpostmatrixsimsGrillogempiricalPlanckpositionsqualonly{0.18}
\def\sumpostmatrixsimsOguriaempiricalPlanckpositionsqualonly{1.1 \times 10^{-5}}
\def\sumfracpostmatrixsimsOguriaempiricalPlanckpositionsqualonly{0.82}
\def\hohmatrixsimsempiricalPlanckpositionsqualonly{$67.4_{-0.6}^{+0.6}$}
\def\hohmfidmatrixsimsempiricalPlanckpositionsqualonly{$+100066.4_{-0.6}^{+0.6}$}
\def\hohmatrixsimsempiricalPlanckpositionsqualonlypohtwotwoseven{66.2}
\def\hohmatrixsimsempiricalPlanckpositionsqualonlyponesix{66.8}
\def\hohmatrixsimsempiricalPlanckpositionsqualonlypfiveoh{67.4}
\def\hohmatrixsimsempiricalPlanckpositionsqualonlypeightfour{68.0}
\def\hohmatrixsimsempiricalPlanckpositionsqualonlypninesevenseventhree{68.6}
\def\hohmatrixsimsempiricalPlanckpositionsqualonlypmax{67.4}
\def\hohmfidmatrixsimsempiricalPlanckpositionsqualonlypmax{+100066.4}
\def\sumpostmatrixsimsGrillogempiricalSHohESnopositionsnoqualthroughtwoohtwooh{9.6 \times 10^{-7}}
\def\sumfracpostmatrixsimsGrillogempiricalSHohESnopositionsnoqualthroughtwoohtwooh{0.16}
\def\sumpostmatrixsimsOguriaempiricalSHohESnopositionsnoqualthroughtwoohtwooh{4.4 \times 10^{-6}}
\def\sumfracpostmatrixsimsOguriaempiricalSHohESnopositionsnoqualthroughtwoohtwooh{0.7}
\def\sumpostmatrixsimsOgurigempiricalSHohESnopositionsnoqualthroughtwoohtwooh{4.6 \times 10^{-7}}
\def\sumfracpostmatrixsimsOgurigempiricalSHohESnopositionsnoqualthroughtwoohtwooh{0.074}
\def\sumpostmatrixsimsDiegoaempiricalSHohESnopositionsnoqualthroughtwoohtwooh{2 \times 10^{-10}}
\def\sumfracpostmatrixsimsDiegoaempiricalSHohESnopositionsnoqualthroughtwoohtwooh{3.3 \times 10^{-5}}
\def\sumpostmatrixsimsJauzaconefivedottwoempiricalSHohESnopositionsnoqualthroughtwoohtwooh{3.5 \times 10^{-9}}
\def\sumfracpostmatrixsimsJauzaconefivedottwoempiricalSHohESnopositionsnoqualthroughtwoohtwooh{0.00057}
\def\sumpostmatrixsimsSharonaempiricalSHohESnopositionsnoqualthroughtwoohtwooh{1.2 \times 10^{-8}}
\def\sumfracpostmatrixsimsSharonaempiricalSHohESnopositionsnoqualthroughtwoohtwooh{0.002}
\def\sumpostmatrixsimsSharongempiricalSHohESnopositionsnoqualthroughtwoohtwooh{\times 10^{-7}}
\def\sumfracpostmatrixsimsSharongempiricalSHohESnopositionsnoqualthroughtwoohtwooh{0.017}
\def\sumpostmatrixsimsChentwoohtwoohempiricalSHohESnopositionsnoqualthroughtwoohtwooh{1.9 \times 10^{-11}}
\def\sumfracpostmatrixsimsChentwoohtwoohempiricalSHohESnopositionsnoqualthroughtwoohtwooh{3.1 \times 10^{-6}}
\def\sumpostmatrixsimsZitrintwoohtwoohltmempiricalSHohESnopositionsnoqualthroughtwoohtwooh{1.4 \times 10^{-14}}
\def\sumfracpostmatrixsimsZitrintwoohtwoohltmempiricalSHohESnopositionsnoqualthroughtwoohtwooh{2.3 \times 10^{-9}}
\def\sumpostmatrixsimsZitrintwoohtwoohpempiricalSHohESnopositionsnoqualthroughtwoohtwooh{1.6 \times 10^{-8}}
\def\sumfracpostmatrixsimsZitrintwoohtwoohpempiricalSHohESnopositionsnoqualthroughtwoohtwooh{0.0025}
\def\sumpostmatrixsimsKeetonempiricalSHohESnopositionsnoqualthroughtwoohtwooh{2.9 \times 10^{-7}}
\def\sumfracpostmatrixsimsKeetonempiricalSHohESnopositionsnoqualthroughtwoohtwooh{0.047}
\def\hohmatrixsimsempiricalSHohESnopositionsnoqualthroughtwoohtwooh{$72.6_{-1.0}^{+1.0}$}
\def\hohmfidmatrixsimsempiricalSHohESnopositionsnoqualthroughtwoohtwooh{$+100071.6_{-1.0}^{+1.0}$}
\def\hohmatrixsimsempiricalSHohESnopositionsnoqualpohtwotwoseventhroughtwoohtwooh{70.6}
\def\hohmatrixsimsempiricalSHohESnopositionsnoqualponesixthroughtwoohtwooh{71.6}
\def\hohmatrixsimsempiricalSHohESnopositionsnoqualpfiveohthroughtwoohtwooh{72.6}
\def\hohmatrixsimsempiricalSHohESnopositionsnoqualpeightfourthroughtwoohtwooh{73.6}
\def\hohmatrixsimsempiricalSHohESnopositionsnoqualpninesevenseventhreethroughtwoohtwooh{74.7}
\def\hohmatrixsimsempiricalSHohESnopositionsnoqualpmaxthroughtwoohtwooh{72.6}
\def\hohmfidmatrixsimsempiricalSHohESnopositionsnoqualpmaxthroughtwoohtwooh{+100071.6}
\def\sumpostmatrixsimsGrillogempiricalSHohESpositionsnoqual{1.3 \times 10^{-6}}
\def\sumfracpostmatrixsimsGrillogempiricalSHohESpositionsnoqual{0.31}
\def\sumpostmatrixsimsOguriaempiricalSHohESpositionsnoqual{2.5 \times 10^{-6}}
\def\sumfracpostmatrixsimsOguriaempiricalSHohESpositionsnoqual{0.58}
\def\sumpostmatrixsimsOgurigempiricalSHohESpositionsnoqual{2.8 \times 10^{-7}}
\def\sumfracpostmatrixsimsOgurigempiricalSHohESpositionsnoqual{0.066}
\def\sumpostmatrixsimsDiegoaempiricalSHohESpositionsnoqual{\times 10^{-10}}
\def\sumfracpostmatrixsimsDiegoaempiricalSHohESpositionsnoqual{2.4 \times 10^{-5}}
\def\sumpostmatrixsimsJauzaconefivedottwoempiricalSHohESpositionsnoqual{1.2 \times 10^{-9}}
\def\sumfracpostmatrixsimsJauzaconefivedottwoempiricalSHohESpositionsnoqual{0.00028}
\def\sumpostmatrixsimsSharonaempiricalSHohESpositionsnoqual{2.1 \times 10^{-8}}
\def\sumfracpostmatrixsimsSharonaempiricalSHohESpositionsnoqual{0.0051}
\def\sumpostmatrixsimsSharongempiricalSHohESpositionsnoqual{1.8 \times 10^{-7}}
\def\sumfracpostmatrixsimsSharongempiricalSHohESpositionsnoqual{0.041}
\def\sumpostmatrixsimsZitrincempiricalSHohESpositionsnoqual{1.3 \times 10^{-16}}
\def\sumfracpostmatrixsimsZitrincempiricalSHohESpositionsnoqual{3.2 \times 10^{-11}}
\def\hohmatrixsimsempiricalSHohESpositionsnoqual{$72.6_{-1.0}^{+1.1}$}
\def\hohmfidmatrixsimsempiricalSHohESpositionsnoqual{$+100071.6_{-1.0}^{+1.1}$}
\def\hohmatrixsimsempiricalSHohESpositionsnoqualpohtwotwoseven{70.6}
\def\hohmatrixsimsempiricalSHohESpositionsnoqualponesix{71.6}
\def\hohmatrixsimsempiricalSHohESpositionsnoqualpfiveoh{72.6}
\def\hohmatrixsimsempiricalSHohESpositionsnoqualpeightfour{73.7}
\def\hohmatrixsimsempiricalSHohESpositionsnoqualpninesevenseventhree{74.7}
\def\hohmatrixsimsempiricalSHohESpositionsnoqualpmax{72.6}
\def\hohmfidmatrixsimsempiricalSHohESpositionsnoqualpmax{+100071.6}
\def\sumpostmatrixsimsGrillogempiricalSHohESpositionsqualonly{1.3 \times 10^{-6}}
\def\sumfracpostmatrixsimsGrillogempiricalSHohESpositionsqualonly{0.34}
\def\sumpostmatrixsimsOguriaempiricalSHohESpositionsqualonly{2.5 \times 10^{-6}}
\def\sumfracpostmatrixsimsOguriaempiricalSHohESpositionsqualonly{0.66}
\def\hohmatrixsimsempiricalSHohESpositionsqualonly{$72.7_{-1.1}^{+1.0}$}
\def\hohmfidmatrixsimsempiricalSHohESpositionsqualonly{$+100071.7_{-1.1}^{+1.0}$}
\def\hohmatrixsimsempiricalSHohESpositionsqualonlypohtwotwoseven{70.6}
\def\hohmatrixsimsempiricalSHohESpositionsqualonlyponesix{71.6}
\def\hohmatrixsimsempiricalSHohESpositionsqualonlypfiveoh{72.7}
\def\hohmatrixsimsempiricalSHohESpositionsqualonlypeightfour{73.7}
\def\hohmatrixsimsempiricalSHohESpositionsqualonlypninesevenseventhree{74.7}
\def\hohmatrixsimsempiricalSHohESpositionsqualonlypmax{72.7}
\def\hohmfidmatrixsimsempiricalSHohESpositionsqualonlypmax{+100071.7}
\def\sumposthybridsimsGrillogempiricalnopriornopositionsnoqualthroughtwoohtwooh{3.2 \times 10^{-5}}
\def\sumfracposthybridsimsGrillogempiricalnopriornopositionsnoqualthroughtwoohtwooh{0.018}
\def\sumposthybridsimsOguriaempiricalnopriornopositionsnoqualthroughtwoohtwooh{0.00076}
\def\sumfracposthybridsimsOguriaempiricalnopriornopositionsnoqualthroughtwoohtwooh{0.42}
\def\sumposthybridsimsOgurigempiricalnopriornopositionsnoqualthroughtwoohtwooh{0.00083}
\def\sumfracposthybridsimsOgurigempiricalnopriornopositionsnoqualthroughtwoohtwooh{0.47}
\def\sumposthybridsimsDiegoaempiricalnopriornopositionsnoqualthroughtwoohtwooh{1.4 \times 10^{-8}}
\def\sumfracposthybridsimsDiegoaempiricalnopriornopositionsnoqualthroughtwoohtwooh{7.8 \times 10^{-6}}
\def\sumposthybridsimsJauzaconefivedottwoempiricalnopriornopositionsnoqualthroughtwoohtwooh{3 \times 10^{-8}}
\def\sumfracposthybridsimsJauzaconefivedottwoempiricalnopriornopositionsnoqualthroughtwoohtwooh{1.7 \times 10^{-5}}
\def\sumposthybridsimsSharonaempiricalnopriornopositionsnoqualthroughtwoohtwooh{4 \times 10^{-7}}
\def\sumfracposthybridsimsSharonaempiricalnopriornopositionsnoqualthroughtwoohtwooh{0.00022}
\def\sumposthybridsimsSharongempiricalnopriornopositionsnoqualthroughtwoohtwooh{3.6 \times 10^{-6}}
\def\sumfracposthybridsimsSharongempiricalnopriornopositionsnoqualthroughtwoohtwooh{0.002}
\def\sumposthybridsimsChentwoohtwoohempiricalnopriornopositionsnoqualthroughtwoohtwooh{9.1 \times 10^{-6}}
\def\sumfracposthybridsimsChentwoohtwoohempiricalnopriornopositionsnoqualthroughtwoohtwooh{0.0051}
\def\sumposthybridsimsZitrintwoohtwoohltmempiricalnopriornopositionsnoqualthroughtwoohtwooh{9 \times 10^{-7}}
\def\sumfracposthybridsimsZitrintwoohtwoohltmempiricalnopriornopositionsnoqualthroughtwoohtwooh{0.0005}
\def\sumposthybridsimsZitrintwoohtwoohpempiricalnopriornopositionsnoqualthroughtwoohtwooh{1.6 \times 10^{-5}}
\def\sumfracposthybridsimsZitrintwoohtwoohpempiricalnopriornopositionsnoqualthroughtwoohtwooh{0.0088}
\def\sumposthybridsimsKeetonempiricalnopriornopositionsnoqualthroughtwoohtwooh{0.00014}
\def\sumfracposthybridsimsKeetonempiricalnopriornopositionsnoqualthroughtwoohtwooh{0.077}
\def\hohhybridsimsempiricalnopriornopositionsnoqualthroughtwoohtwooh{$64.6_{-4.2}^{+4.4}$}
\def\hohmfidhybridsimsempiricalnopriornopositionsnoqualthroughtwoohtwooh{$-1.3_{-4.2}^{+4.4}$}
\def\hohhybridsimsempiricalnopriornopositionsnoqualpohtwotwoseventhroughtwoohtwooh{56.4}
\def\hohhybridsimsempiricalnopriornopositionsnoqualponesixthroughtwoohtwooh{60.4}
\def\hohhybridsimsempiricalnopriornopositionsnoqualpfiveohthroughtwoohtwooh{64.6}
\def\hohhybridsimsempiricalnopriornopositionsnoqualpeightfourthroughtwoohtwooh{69.0}
\def\hohhybridsimsempiricalnopriornopositionsnoqualpninesevenseventhreethroughtwoohtwooh{73.4}
\def\hohhybridsimsempiricalnopriornopositionsnoqualpmaxthroughtwoohtwooh{64.6}
\def\hohmfidhybridsimsempiricalnopriornopositionsnoqualpmaxthroughtwoohtwooh{-1.3}
\def\sumposthybridsimsGrillogempiricalnopriorpositionsnoqualthroughtwoohtwooh{2.1 \times 10^{-5}}
\def\sumfracposthybridsimsGrillogempiricalnopriorpositionsnoqualthroughtwoohtwooh{0.065}
\def\sumposthybridsimsOguriaempiricalnopriorpositionsnoqualthroughtwoohtwooh{0.00011}
\def\sumfracposthybridsimsOguriaempiricalnopriorpositionsnoqualthroughtwoohtwooh{0.36}
\def\sumposthybridsimsOgurigempiricalnopriorpositionsnoqualthroughtwoohtwooh{0.00013}
\def\sumfracposthybridsimsOgurigempiricalnopriorpositionsnoqualthroughtwoohtwooh{1.0}
\def\sumposthybridsimsDiegoaempiricalnopriorpositionsnoqualthroughtwoohtwooh{6.7 \times 10^{-9}}
\def\sumfracposthybridsimsDiegoaempiricalnopriorpositionsnoqualthroughtwoohtwooh{2.1 \times 10^{-5}}
\def\sumposthybridsimsJauzaconefivedottwoempiricalnopriorpositionsnoqualthroughtwoohtwooh{3.5 \times 10^{-8}}
\def\sumfracposthybridsimsJauzaconefivedottwoempiricalnopriorpositionsnoqualthroughtwoohtwooh{0.00011}
\def\sumposthybridsimsSharonaempiricalnopriorpositionsnoqualthroughtwoohtwooh{5 \times 10^{-6}}
\def\sumfracposthybridsimsSharonaempiricalnopriorpositionsnoqualthroughtwoohtwooh{0.016}
\def\sumposthybridsimsSharongempiricalnopriorpositionsnoqualthroughtwoohtwooh{\times 10^{-5}}
\def\sumfracposthybridsimsSharongempiricalnopriorpositionsnoqualthroughtwoohtwooh{0.031}
\def\sumposthybridsimsKeetonempiricalnopriorpositionsnoqualthroughtwoohtwooh{3.9 \times 10^{-5}}
\def\sumfracposthybridsimsKeetonempiricalnopriorpositionsnoqualthroughtwoohtwooh{0.12}
\def\hohhybridsimsempiricalnopriorpositionsnoqualthroughtwoohtwooh{$62.1_{-3.4}^{+4.1}$}
\def\hohmfidhybridsimsempiricalnopriorpositionsnoqualthroughtwoohtwooh{$-3.8_{-3.4}^{+4.1}$}
\def\hohhybridsimsempiricalnopriorpositionsnoqualpohtwotwoseventhroughtwoohtwooh{55.3}
\def\hohhybridsimsempiricalnopriorpositionsnoqualponesixthroughtwoohtwooh{58.7}
\def\hohhybridsimsempiricalnopriorpositionsnoqualpfiveohthroughtwoohtwooh{62.3}
\def\hohhybridsimsempiricalnopriorpositionsnoqualpeightfourthroughtwoohtwooh{66.2}
\def\hohhybridsimsempiricalnopriorpositionsnoqualpninesevenseventhreethroughtwoohtwooh{70.6}
\def\hohhybridsimsempiricalnopriorpositionsnoqualpmaxthroughtwoohtwooh{62.1}
\def\hohmfidhybridsimsempiricalnopriorpositionsnoqualpmaxthroughtwoohtwooh{-3.8}
\def\sumposthybridsimsGrillogempiricalnopriorpositionsnoqual{4.3 \times 10^{-5}}
\def\sumfracposthybridsimsGrillogempiricalnopriorpositionsnoqual{0.044}
\def\sumposthybridsimsOguriaempiricalnopriorpositionsnoqual{0.00043}
\def\sumfracposthybridsimsOguriaempiricalnopriorpositionsnoqual{0.44}
\def\sumposthybridsimsOgurigempiricalnopriorpositionsnoqual{0.0005}
\def\sumfracposthybridsimsOgurigempiricalnopriorpositionsnoqual{0.51}
\def\sumposthybridsimsDiegoaempiricalnopriorpositionsnoqual{7 \times 10^{-9}}
\def\sumfracposthybridsimsDiegoaempiricalnopriorpositionsnoqual{7.1 \times 10^{-6}}
\def\sumposthybridsimsJauzaconefivedottwoempiricalnopriorpositionsnoqual{9.9 \times 10^{-9}}
\def\sumfracposthybridsimsJauzaconefivedottwoempiricalnopriorpositionsnoqual{1.0 \times 10^{-5}}
\def\sumposthybridsimsSharonaempiricalnopriorpositionsnoqual{7.1 \times 10^{-7}}
\def\sumfracposthybridsimsSharonaempiricalnopriorpositionsnoqual{0.00072}
\def\sumposthybridsimsSharongempiricalnopriorpositionsnoqual{6.1 \times 10^{-6}}
\def\sumfracposthybridsimsSharongempiricalnopriorpositionsnoqual{0.0062}
\def\sumposthybridsimsZitrincempiricalnopriorpositionsnoqual{3 \times 10^{-12}}
\def\sumfracposthybridsimsZitrincempiricalnopriorpositionsnoqual{3 \times 10^{-9}}
\def\hohhybridsimsempiricalnopriorpositionsnoqual{$64.8_{-4.3}^{+4.4}$}
\def\hohmfidhybridsimsempiricalnopriorpositionsnoqual{$-1.1_{-4.3}^{+4.4}$}
\def\hohhybridsimsempiricalnopriorpositionsnoqualpohtwotwoseven{56.5}
\def\hohhybridsimsempiricalnopriorpositionsnoqualponesix{60.5}
\def\hohhybridsimsempiricalnopriorpositionsnoqualpfiveoh{64.8}
\def\hohhybridsimsempiricalnopriorpositionsnoqualpeightfour{69.2}
\def\hohhybridsimsempiricalnopriorpositionsnoqualpninesevenseventhree{73.6}
\def\hohhybridsimsempiricalnopriorpositionsnoqualpmax{64.8}
\def\hohmfidhybridsimsempiricalnopriorpositionsnoqualpmax{-1.1}
\def\sumposthybridsimsGrillogempiricalnopriorpositionsqualonly{4.3 \times 10^{-5}}
\def\sumfracposthybridsimsGrillogempiricalnopriorpositionsqualonly{0.091}
\def\sumposthybridsimsOguriaempiricalnopriorpositionsqualonly{0.00043}
\def\sumfracposthybridsimsOguriaempiricalnopriorpositionsqualonly{0.91}
\def\hohhybridsimsempiricalnopriorpositionsqualonly{$66.6_{-3.3}^{+4.1}$}
\def\hohmfidhybridsimsempiricalnopriorpositionsqualonly{$+0.7_{-3.3}^{+4.1}$}
\def\hohhybridsimsempiricalnopriorpositionsqualonlypohtwotwoseven{59.8}
\def\hohhybridsimsempiricalnopriorpositionsqualonlyponesix{63.3}
\def\hohhybridsimsempiricalnopriorpositionsqualonlypfiveoh{66.9}
\def\hohhybridsimsempiricalnopriorpositionsqualonlypeightfour{70.7}
\def\hohhybridsimsempiricalnopriorpositionsqualonlypninesevenseventhree{74.8}
\def\hohhybridsimsempiricalnopriorpositionsqualonlypmax{66.6}
\def\hohmfidhybridsimsempiricalnopriorpositionsqualonlypmax{+0.7}
\def\sumposthybridsimsGrillogempiricalPlancknopositionsnoqualthroughtwoohtwooh{3 \times 10^{-6}}
\def\sumfracposthybridsimsGrillogempiricalPlancknopositionsnoqualthroughtwoohtwooh{0.023}
\def\sumposthybridsimsOguriaempiricalPlancknopositionsnoqualthroughtwoohtwooh{7.9 \times 10^{-5}}
\def\sumfracposthybridsimsOguriaempiricalPlancknopositionsnoqualthroughtwoohtwooh{0.6}
\def\sumposthybridsimsOgurigempiricalPlancknopositionsnoqualthroughtwoohtwooh{4 \times 10^{-5}}
\def\sumfracposthybridsimsOgurigempiricalPlancknopositionsnoqualthroughtwoohtwooh{0.3}
\def\sumposthybridsimsDiegoaempiricalPlancknopositionsnoqualthroughtwoohtwooh{3.4 \times 10^{-10}}
\def\sumfracposthybridsimsDiegoaempiricalPlancknopositionsnoqualthroughtwoohtwooh{2.6 \times 10^{-6}}
\def\sumposthybridsimsJauzaconefivedottwoempiricalPlancknopositionsnoqualthroughtwoohtwooh{1.4 \times 10^{-9}}
\def\sumfracposthybridsimsJauzaconefivedottwoempiricalPlancknopositionsnoqualthroughtwoohtwooh{1.1 \times 10^{-5}}
\def\sumposthybridsimsSharonaempiricalPlancknopositionsnoqualthroughtwoohtwooh{1.7 \times 10^{-8}}
\def\sumfracposthybridsimsSharonaempiricalPlancknopositionsnoqualthroughtwoohtwooh{0.00013}
\def\sumposthybridsimsSharongempiricalPlancknopositionsnoqualthroughtwoohtwooh{2.1 \times 10^{-7}}
\def\sumfracposthybridsimsSharongempiricalPlancknopositionsnoqualthroughtwoohtwooh{0.0016}
\def\sumposthybridsimsChentwoohtwoohempiricalPlancknopositionsnoqualthroughtwoohtwooh{5.4 \times 10^{-8}}
\def\sumfracposthybridsimsChentwoohtwoohempiricalPlancknopositionsnoqualthroughtwoohtwooh{0.00041}
\def\sumposthybridsimsZitrintwoohtwoohltmempiricalPlancknopositionsnoqualthroughtwoohtwooh{2.2 \times 10^{-9}}
\def\sumfracposthybridsimsZitrintwoohtwoohltmempiricalPlancknopositionsnoqualthroughtwoohtwooh{1.7 \times 10^{-5}}
\def\sumposthybridsimsZitrintwoohtwoohpempiricalPlancknopositionsnoqualthroughtwoohtwooh{6.7 \times 10^{-7}}
\def\sumfracposthybridsimsZitrintwoohtwoohpempiricalPlancknopositionsnoqualthroughtwoohtwooh{0.0051}
\def\sumposthybridsimsKeetonempiricalPlancknopositionsnoqualthroughtwoohtwooh{8.4 \times 10^{-6}}
\def\sumfracposthybridsimsKeetonempiricalPlancknopositionsnoqualthroughtwoohtwooh{0.064}
\def\hohhybridsimsempiricalPlancknopositionsnoqualthroughtwoohtwooh{$67.3_{-0.5}^{+0.6}$}
\def\hohmfidhybridsimsempiricalPlancknopositionsnoqualthroughtwoohtwooh{$+1.4_{-0.5}^{+0.6}$}
\def\hohhybridsimsempiricalPlancknopositionsnoqualpohtwotwoseventhroughtwoohtwooh{66.2}
\def\hohhybridsimsempiricalPlancknopositionsnoqualponesixthroughtwoohtwooh{66.8}
\def\hohhybridsimsempiricalPlancknopositionsnoqualpfiveohthroughtwoohtwooh{67.3}
\def\hohhybridsimsempiricalPlancknopositionsnoqualpeightfourthroughtwoohtwooh{67.9}
\def\hohhybridsimsempiricalPlancknopositionsnoqualpninesevenseventhreethroughtwoohtwooh{68.5}
\def\hohhybridsimsempiricalPlancknopositionsnoqualpmaxthroughtwoohtwooh{67.3}
\def\hohmfidhybridsimsempiricalPlancknopositionsnoqualpmaxthroughtwoohtwooh{+1.4}
\def\sumposthybridsimsGrillogempiricalPlanckpositionsnoqualthroughtwoohtwooh{1.9 \times 10^{-6}}
\def\sumfracposthybridsimsGrillogempiricalPlanckpositionsnoqualthroughtwoohtwooh{0.086}
\def\sumposthybridsimsOguriaempiricalPlanckpositionsnoqualthroughtwoohtwooh{1.2 \times 10^{-5}}
\def\sumfracposthybridsimsOguriaempiricalPlanckpositionsnoqualthroughtwoohtwooh{0.54}
\def\sumposthybridsimsOgurigempiricalPlanckpositionsnoqualthroughtwoohtwooh{5.3 \times 10^{-6}}
\def\sumfracposthybridsimsOgurigempiricalPlanckpositionsnoqualthroughtwoohtwooh{0.24}
\def\sumposthybridsimsDiegoaempiricalPlanckpositionsnoqualthroughtwoohtwooh{1.6 \times 10^{-10}}
\def\sumfracposthybridsimsDiegoaempiricalPlanckpositionsnoqualthroughtwoohtwooh{7.1 \times 10^{-6}}
\def\sumposthybridsimsJauzaconefivedottwoempiricalPlanckpositionsnoqualthroughtwoohtwooh{1.8 \times 10^{-9}}
\def\sumfracposthybridsimsJauzaconefivedottwoempiricalPlanckpositionsnoqualthroughtwoohtwooh{7.9 \times 10^{-5}}
\def\sumposthybridsimsSharonaempiricalPlanckpositionsnoqualthroughtwoohtwooh{2 \times 10^{-7}}
\def\sumfracposthybridsimsSharonaempiricalPlanckpositionsnoqualthroughtwoohtwooh{0.0089}
\def\sumposthybridsimsSharongempiricalPlanckpositionsnoqualthroughtwoohtwooh{5.7 \times 10^{-7}}
\def\sumfracposthybridsimsSharongempiricalPlanckpositionsnoqualthroughtwoohtwooh{0.026}
\def\sumposthybridsimsKeetonempiricalPlanckpositionsnoqualthroughtwoohtwooh{2.3 \times 10^{-6}}
\def\sumfracposthybridsimsKeetonempiricalPlanckpositionsnoqualthroughtwoohtwooh{0.1}
\def\hohhybridsimsempiricalPlanckpositionsnoqualthroughtwoohtwooh{$67.3_{-0.5}^{+0.6}$}
\def\hohmfidhybridsimsempiricalPlanckpositionsnoqualthroughtwoohtwooh{$+1.4_{-0.5}^{+0.6}$}
\def\hohhybridsimsempiricalPlanckpositionsnoqualpohtwotwoseventhroughtwoohtwooh{66.2}
\def\hohhybridsimsempiricalPlanckpositionsnoqualponesixthroughtwoohtwooh{66.8}
\def\hohhybridsimsempiricalPlanckpositionsnoqualpfiveohthroughtwoohtwooh{67.3}
\def\hohhybridsimsempiricalPlanckpositionsnoqualpeightfourthroughtwoohtwooh{67.9}
\def\hohhybridsimsempiricalPlanckpositionsnoqualpninesevenseventhreethroughtwoohtwooh{68.5}
\def\hohhybridsimsempiricalPlanckpositionsnoqualpmaxthroughtwoohtwooh{67.3}
\def\hohmfidhybridsimsempiricalPlanckpositionsnoqualpmaxthroughtwoohtwooh{+1.4}
\def\sumposthybridsimsGrillogempiricalPlanckpositionsnoqual{4.1 \times 10^{-6}}
\def\sumfracposthybridsimsGrillogempiricalPlanckpositionsnoqual{0.056}
\def\sumposthybridsimsOguriaempiricalPlanckpositionsnoqual{4.5 \times 10^{-5}}
\def\sumfracposthybridsimsOguriaempiricalPlanckpositionsnoqual{0.61}
\def\sumposthybridsimsOgurigempiricalPlanckpositionsnoqual{2.4 \times 10^{-5}}
\def\sumfracposthybridsimsOgurigempiricalPlanckpositionsnoqual{0.33}
\def\sumposthybridsimsDiegoaempiricalPlanckpositionsnoqual{1.7 \times 10^{-10}}
\def\sumfracposthybridsimsDiegoaempiricalPlanckpositionsnoqual{2.3 \times 10^{-6}}
\def\sumposthybridsimsJauzaconefivedottwoempiricalPlanckpositionsnoqual{4.8 \times 10^{-10}}
\def\sumfracposthybridsimsJauzaconefivedottwoempiricalPlanckpositionsnoqual{6.5 \times 10^{-6}}
\def\sumposthybridsimsSharonaempiricalPlanckpositionsnoqual{3.1 \times 10^{-8}}
\def\sumfracposthybridsimsSharonaempiricalPlanckpositionsnoqual{0.00042}
\def\sumposthybridsimsSharongempiricalPlanckpositionsnoqual{3.6 \times 10^{-7}}
\def\sumfracposthybridsimsSharongempiricalPlanckpositionsnoqual{0.0049}
\def\sumposthybridsimsZitrincempiricalPlanckpositionsnoqual{3.6 \times 10^{-19}}
\def\sumfracposthybridsimsZitrincempiricalPlanckpositionsnoqual{4.9 \times 10^{-15}}
\def\hohhybridsimsempiricalPlanckpositionsnoqual{$67.4_{-0.6}^{+0.5}$}
\def\hohmfidhybridsimsempiricalPlanckpositionsnoqual{$+1.5_{-0.6}^{+0.5}$}
\def\hohhybridsimsempiricalPlanckpositionsnoqualpohtwotwoseven{66.2}
\def\hohhybridsimsempiricalPlanckpositionsnoqualponesix{66.8}
\def\hohhybridsimsempiricalPlanckpositionsnoqualpfiveoh{67.4}
\def\hohhybridsimsempiricalPlanckpositionsnoqualpeightfour{67.9}
\def\hohhybridsimsempiricalPlanckpositionsnoqualpninesevenseventhree{68.5}
\def\hohhybridsimsempiricalPlanckpositionsnoqualpmax{67.4}
\def\hohmfidhybridsimsempiricalPlanckpositionsnoqualpmax{+1.5}
\def\sumposthybridsimsGrillogempiricalPlanckpositionsqualonly{4.1 \times 10^{-6}}
\def\sumfracposthybridsimsGrillogempiricalPlanckpositionsqualonly{0.084}
\def\sumposthybridsimsOguriaempiricalPlanckpositionsqualonly{4.5 \times 10^{-5}}
\def\sumfracposthybridsimsOguriaempiricalPlanckpositionsqualonly{0.92}
\def\hohhybridsimsempiricalPlanckpositionsqualonly{$67.4_{-0.6}^{+0.6}$}
\def\hohmfidhybridsimsempiricalPlanckpositionsqualonly{$+1.5_{-0.6}^{+0.6}$}
\def\hohhybridsimsempiricalPlanckpositionsqualonlypohtwotwoseven{66.2}
\def\hohhybridsimsempiricalPlanckpositionsqualonlyponesix{66.8}
\def\hohhybridsimsempiricalPlanckpositionsqualonlypfiveoh{67.4}
\def\hohhybridsimsempiricalPlanckpositionsqualonlypeightfour{68.0}
\def\hohhybridsimsempiricalPlanckpositionsqualonlypninesevenseventhree{68.6}
\def\hohhybridsimsempiricalPlanckpositionsqualonlypmax{67.4}
\def\hohmfidhybridsimsempiricalPlanckpositionsqualonlypmax{+1.5}
\def\sumposthybridsimsGrillogempiricalSHohESnopositionsnoqualthroughtwoohtwooh{1.1 \times 10^{-6}}
\def\sumfracposthybridsimsGrillogempiricalSHohESnopositionsnoqualthroughtwoohtwooh{0.056}
\def\sumposthybridsimsOguriaempiricalSHohESnopositionsnoqualthroughtwoohtwooh{1.5 \times 10^{-5}}
\def\sumfracposthybridsimsOguriaempiricalSHohESnopositionsnoqualthroughtwoohtwooh{0.78}
\def\sumposthybridsimsOgurigempiricalSHohESnopositionsnoqualthroughtwoohtwooh{2.4 \times 10^{-6}}
\def\sumfracposthybridsimsOgurigempiricalSHohESnopositionsnoqualthroughtwoohtwooh{0.12}
\def\sumposthybridsimsDiegoaempiricalSHohESnopositionsnoqualthroughtwoohtwooh{1.8 \times 10^{-10}}
\def\sumfracposthybridsimsDiegoaempiricalSHohESnopositionsnoqualthroughtwoohtwooh{9.5 \times 10^{-6}}
\def\sumposthybridsimsJauzaconefivedottwoempiricalSHohESnopositionsnoqualthroughtwoohtwooh{1.3 \times 10^{-9}}
\def\sumfracposthybridsimsJauzaconefivedottwoempiricalSHohESnopositionsnoqualthroughtwoohtwooh{6.5 \times 10^{-5}}
\def\sumposthybridsimsSharonaempiricalSHohESnopositionsnoqualthroughtwoohtwooh{7.1 \times 10^{-9}}
\def\sumfracposthybridsimsSharonaempiricalSHohESnopositionsnoqualthroughtwoohtwooh{0.00036}
\def\sumposthybridsimsSharongempiricalSHohESnopositionsnoqualthroughtwoohtwooh{5.8 \times 10^{-8}}
\def\sumfracposthybridsimsSharongempiricalSHohESnopositionsnoqualthroughtwoohtwooh{0.003}
\def\sumposthybridsimsChentwoohtwoohempiricalSHohESnopositionsnoqualthroughtwoohtwooh{2.1 \times 10^{-11}}
\def\sumfracposthybridsimsChentwoohtwoohempiricalSHohESnopositionsnoqualthroughtwoohtwooh{1.1 \times 10^{-6}}
\def\sumposthybridsimsZitrintwoohtwoohltmempiricalSHohESnopositionsnoqualthroughtwoohtwooh{2.1 \times 10^{-10}}
\def\sumfracposthybridsimsZitrintwoohtwoohltmempiricalSHohESnopositionsnoqualthroughtwoohtwooh{1.1 \times 10^{-5}}
\def\sumposthybridsimsZitrintwoohtwoohpempiricalSHohESnopositionsnoqualthroughtwoohtwooh{2.7 \times 10^{-8}}
\def\sumfracposthybridsimsZitrintwoohtwoohpempiricalSHohESnopositionsnoqualthroughtwoohtwooh{0.0014}
\def\sumposthybridsimsKeetonempiricalSHohESnopositionsnoqualthroughtwoohtwooh{6.8 \times 10^{-7}}
\def\sumfracposthybridsimsKeetonempiricalSHohESnopositionsnoqualthroughtwoohtwooh{0.035}
\def\hohhybridsimsempiricalSHohESnopositionsnoqualthroughtwoohtwooh{$73.1_{-1.3}^{+1.4}$}
\def\hohmfidhybridsimsempiricalSHohESnopositionsnoqualthroughtwoohtwooh{$+7.2_{-1.3}^{+1.4}$}
\def\hohhybridsimsempiricalSHohESnopositionsnoqualpohtwotwoseventhroughtwoohtwooh{70.4}
\def\hohhybridsimsempiricalSHohESnopositionsnoqualponesixthroughtwoohtwooh{71.8}
\def\hohhybridsimsempiricalSHohESnopositionsnoqualpfiveohthroughtwoohtwooh{73.1}
\def\hohhybridsimsempiricalSHohESnopositionsnoqualpeightfourthroughtwoohtwooh{74.5}
\def\hohhybridsimsempiricalSHohESnopositionsnoqualpninesevenseventhreethroughtwoohtwooh{75.9}
\def\hohhybridsimsempiricalSHohESnopositionsnoqualpmaxthroughtwoohtwooh{73.1}
\def\hohmfidhybridsimsempiricalSHohESnopositionsnoqualpmaxthroughtwoohtwooh{+7.2}
\def\sumposthybridsimsGrillogempiricalSHohESpositionsnoqualthroughtwoohtwooh{7 \times 10^{-7}}
\def\sumfracposthybridsimsGrillogempiricalSHohESpositionsnoqualthroughtwoohtwooh{0.2}
\def\sumposthybridsimsOguriaempiricalSHohESpositionsnoqualthroughtwoohtwooh{2 \times 10^{-6}}
\def\sumfracposthybridsimsOguriaempiricalSHohESpositionsnoqualthroughtwoohtwooh{0.59}
\def\sumposthybridsimsOgurigempiricalSHohESpositionsnoqualthroughtwoohtwooh{3.6 \times 10^{-7}}
\def\sumfracposthybridsimsOgurigempiricalSHohESpositionsnoqualthroughtwoohtwooh{0.1}
\def\sumposthybridsimsDiegoaempiricalSHohESpositionsnoqualthroughtwoohtwooh{8.5 \times 10^{-11}}
\def\sumfracposthybridsimsDiegoaempiricalSHohESpositionsnoqualthroughtwoohtwooh{2.4 \times 10^{-5}}
\def\sumposthybridsimsJauzaconefivedottwoempiricalSHohESpositionsnoqualthroughtwoohtwooh{1.4 \times 10^{-9}}
\def\sumfracposthybridsimsJauzaconefivedottwoempiricalSHohESpositionsnoqualthroughtwoohtwooh{0.00041}
\def\sumposthybridsimsSharonaempiricalSHohESpositionsnoqualthroughtwoohtwooh{7.5 \times 10^{-8}}
\def\sumfracposthybridsimsSharonaempiricalSHohESpositionsnoqualthroughtwoohtwooh{0.022}
\def\sumposthybridsimsSharongempiricalSHohESpositionsnoqualthroughtwoohtwooh{1.3 \times 10^{-7}}
\def\sumfracposthybridsimsSharongempiricalSHohESpositionsnoqualthroughtwoohtwooh{0.036}
\def\sumposthybridsimsKeetonempiricalSHohESpositionsnoqualthroughtwoohtwooh{1.8 \times 10^{-7}}
\def\sumfracposthybridsimsKeetonempiricalSHohESpositionsnoqualthroughtwoohtwooh{0.051}
\def\hohhybridsimsempiricalSHohESpositionsnoqualthroughtwoohtwooh{$73.2_{-1.3}^{+1.4}$}
\def\hohmfidhybridsimsempiricalSHohESpositionsnoqualthroughtwoohtwooh{$+7.3_{-1.3}^{+1.4}$}
\def\hohhybridsimsempiricalSHohESpositionsnoqualpohtwotwoseventhroughtwoohtwooh{70.5}
\def\hohhybridsimsempiricalSHohESpositionsnoqualponesixthroughtwoohtwooh{71.9}
\def\hohhybridsimsempiricalSHohESpositionsnoqualpfiveohthroughtwoohtwooh{73.2}
\def\hohhybridsimsempiricalSHohESpositionsnoqualpeightfourthroughtwoohtwooh{74.6}
\def\hohhybridsimsempiricalSHohESpositionsnoqualpninesevenseventhreethroughtwoohtwooh{76.0}
\def\hohhybridsimsempiricalSHohESpositionsnoqualpmaxthroughtwoohtwooh{73.2}
\def\hohmfidhybridsimsempiricalSHohESpositionsnoqualpmaxthroughtwoohtwooh{+7.3}
\def\sumposthybridsimsGrillogempiricalSHohESpositionsnoqual{1.5 \times 10^{-6}}
\def\sumfracposthybridsimsGrillogempiricalSHohESpositionsnoqual{0.13}
\def\sumposthybridsimsOguriaempiricalSHohESpositionsnoqual{8.6 \times 10^{-6}}
\def\sumfracposthybridsimsOguriaempiricalSHohESpositionsnoqual{0.74}
\def\sumposthybridsimsOgurigempiricalSHohESpositionsnoqual{1.5 \times 10^{-6}}
\def\sumfracposthybridsimsOgurigempiricalSHohESpositionsnoqual{0.13}
\def\sumposthybridsimsDiegoaempiricalSHohESpositionsnoqual{9.3 \times 10^{-11}}
\def\sumfracposthybridsimsDiegoaempiricalSHohESpositionsnoqual{8 \times 10^{-6}}
\def\sumposthybridsimsJauzaconefivedottwoempiricalSHohESpositionsnoqual{4.2 \times 10^{-10}}
\def\sumfracposthybridsimsJauzaconefivedottwoempiricalSHohESpositionsnoqual{3.6 \times 10^{-5}}
\def\sumposthybridsimsSharonaempiricalSHohESpositionsnoqual{1.2 \times 10^{-8}}
\def\sumfracposthybridsimsSharonaempiricalSHohESpositionsnoqual{0.0011}
\def\sumposthybridsimsSharongempiricalSHohESpositionsnoqual{9.7 \times 10^{-8}}
\def\sumfracposthybridsimsSharongempiricalSHohESpositionsnoqual{0.0084}
\def\sumposthybridsimsZitrincempiricalSHohESpositionsnoqual{1.3 \times 10^{-24}}
\def\sumfracposthybridsimsZitrincempiricalSHohESpositionsnoqual{1.1 \times 10^{-19}}
\def\hohhybridsimsempiricalSHohESpositionsnoqual{$73.1_{-1.3}^{+1.4}$}
\def\hohmfidhybridsimsempiricalSHohESpositionsnoqual{$+7.2_{-1.3}^{+1.4}$}
\def\hohhybridsimsempiricalSHohESpositionsnoqualpohtwotwoseven{70.5}
\def\hohhybridsimsempiricalSHohESpositionsnoqualponesix{71.8}
\def\hohhybridsimsempiricalSHohESpositionsnoqualpfiveoh{73.1}
\def\hohhybridsimsempiricalSHohESpositionsnoqualpeightfour{74.5}
\def\hohhybridsimsempiricalSHohESpositionsnoqualpninesevenseventhree{75.9}
\def\hohhybridsimsempiricalSHohESpositionsnoqualpmax{73.1}
\def\hohmfidhybridsimsempiricalSHohESpositionsnoqualpmax{+7.2}
\def\sumposthybridsimsGrillogempiricalSHohESpositionsqualonly{1.5 \times 10^{-6}}
\def\sumfracposthybridsimsGrillogempiricalSHohESpositionsqualonly{0.14}
\def\sumposthybridsimsOguriaempiricalSHohESpositionsqualonly{8.6 \times 10^{-6}}
\def\sumfracposthybridsimsOguriaempiricalSHohESpositionsqualonly{0.86}
\def\hohhybridsimsempiricalSHohESpositionsqualonly{$73.2_{-1.4}^{+1.3}$}
\def\hohmfidhybridsimsempiricalSHohESpositionsqualonly{$+7.3_{-1.4}^{+1.3}$}
\def\hohhybridsimsempiricalSHohESpositionsqualonlypohtwotwoseven{70.5}
\def\hohhybridsimsempiricalSHohESpositionsqualonlyponesix{71.8}
\def\hohhybridsimsempiricalSHohESpositionsqualonlypfiveoh{73.2}
\def\hohhybridsimsempiricalSHohESpositionsqualonlypeightfour{74.5}
\def\hohhybridsimsempiricalSHohESpositionsqualonlypninesevenseventhree{75.9}
\def\hohhybridsimsempiricalSHohESpositionsqualonlypmax{73.2}
\def\hohmfidhybridsimsempiricalSHohESpositionsqualonlypmax{+7.3}

\begin{centering}
\linespread{1.37}
\footnotesize{$^{13}$Department of Astronomy, University of California, Berkeley, CA 94720-3411, USA}\\
\footnotesize{$^{14}$Senior Miller Fellow, Miller Institute for Basic Research in Science, University of California,}\\
\footnotesize{Berkeley, CA 94720, USA}\\
\footnotesize{$^{15}$Department of Astronomy and Astrophysics, University of California Observatories/Lick Observatory,}\\
\footnotesize{University of California, Santa Cruz, CA 95064, USA}\\
\footnotesize{$^{16}$Department of Astronomy and Astrophysics, University of Toronto, Toronto, ON, M5S 3H4, Canada}\\
\footnotesize{$^{17}$Dark Cosmology Centre, Niels Bohr Institute, University of Copenhagen, DK-2100 Copenhagen, Denmark }\\
\footnotesize{$^{18}$Centre for Extragalactic Astronomy, Department of Physics, Durham University, Durham DH1 3LE, U.K.}\\
\footnotesize{$^{19}$Institute for Computational Cosmology, Durham University, Durham DH1 3LE, U.K.}\\
\footnotesize{$^{20}$Astrophysics Research Center, University of KwaZulu-Natal, Durban 4041, South Africa}\\
\footnotesize{$^{21}$School of Mathematics, Statistics, and Computer Science, University of KwaZulu-Natal, Durban 4041,}\\
\footnotesize{South Africa}\\
\footnotesize{$^{22}$Institute of Astronomy and Kavli Institute for Cosmology, Cambridge, CB3 0HA, UK}\\
\footnotesize{$^{23}$Statistical Laboratory, Department of Pure Mathematics and Mathematical Statistics,}\\
\footnotesize{University of Cambridge, Cambridge, CB3 0WB, UK}\\
\footnotesize{$^{24}$Space Telescope Science Institute, Baltimore, MD 21218, USA}\\
\footnotesize{$^{25}$University of Michigan, Department of Astronomy, Ann Arbor, MI 48109-1107, USA}\\
\footnotesize{$^{26}$The Oskar Klein Centre, Department of Physics, Stockholm University, AlbaNova, 10691, Stockholm,}\\
\footnotesize{Sweden }\\
\footnotesize{$^{27}$Ikerbasque Foundation, University of the Basque Country, Donosita Interational Physics Center,}\\
\footnotesize{Donostia, Spain}\\
\footnotesize{$^{28}$The Observatories of the Carnegie Institution for Science, Pasadena, CA 91101, USA}\\
\footnotesize{$^{29}$Institute of Cosmology and Gravitation, University of Portsmouth, Portsmouth, PO1 3FX, UK}\\
\footnotesize{$^{30}$Department of Astrophysics, American Museum of Natural History, New York, NY 10024, USA}\\
\footnotesize{$^{31}$Department of Physics and Astronomy, Rutgers, The State University of New Jersey, Piscataway,}\\
\footnotesize{NJ 08854, USA}\\
\footnotesize{$^{32}$Las Cumbres Observatory, Goleta, CA 93117, USA}\\
\footnotesize{$^{33}$Department of Physics, University of California, Santa Barbara, CA 93106-9530, USA}\\
\footnotesize{$^{34}$Leibniz-Institut fur Astrophysik Potsdam, 14482 Potsdam, Germany}\\
\footnotesize{$^{35}$The Research School of Astronomy and Astrophysics, Mount Stromlo Observatory,}\\
\footnotesize{Australian National University, Canberra, ACT 2611, Australia}\\
\footnotesize{$^{36}$National Centre for the Public Awareness of Science, Australian National University, Canberra,}\\
\footnotesize{ACT 2611, Australia}\\
\footnotesize{$^{37}$The Australian Research Council Centre of Excellence for All-Sky Astrophysics in 3 Dimensions}\\
\footnotesize{ACT 2611, Australia}\\
\normalsize{$^\ast$To whom correspondence should be addressed; E-mail:  plkelly@umn.edu.}
\linespread{1}
\end{centering}

\begin{sciabstract}
The gravitationally lensed Supernova Refsdal has successively appeared  in multiple images formed by a massive foreground galaxy-cluster lens. After the supernova (SN) appeared in 2014, lens models of the galaxy cluster   predicted an additional image of the SN would appear in 2015, which was subsequently observed. We use the time delays between the appearances to perform a blinded measurement of the expansion rate of the Universe, known as the Hubble constant ($H_0$). 
Using eight cluster lens models, we infer $H_0$\,=\,{\rm \hohhybridsimsempiricalnopriorpositionsnoqual}\ km\,s$^{-1}$\,Mpc$^{-1}$, where Mpc is the megaparsec. Using the  two models most consistent with the observations, we find $H_0$\,=\,{\rm\hohhybridsimsempiricalnopriorpositionsqualonly}\ km\,s$^{-1}$\,Mpc$^{-1}$. Models that assign dark-matter halos to individual galaxies and the overall cluster best reproduce the observations.
\end{sciabstract}

Strong gravitational lensing refers to the action of a foreground mass such as a galaxy cluster to produce multiple images of a well-aligned background source.
In principle, the time delays between the images of a strongly lensed supernova (SN) provide a one-step geometric distance, enabling a measurement of Hubble constant, $H_0$ \cite{refsdal64}. Although originally proposed for supernovae (SNe), this technique, known as time-delay cosmography \cite{treumarshall16}, has only been applied to quasars strongly lensed by foreground, single-galaxy lenses \cite{tewescourbin13, suyutreuhilbert14,birrershajib20}.  
The strongly lensed Supernova Refsdal appeared in late 2014 in four resolved images, designated S1--S4 (coordinates listed in Table~\ref{tab:positions}), arranged in a cross-like configuration, known as an Einstein cross, around an early-type member of the galaxy cluster MACS\,J1149.5+2223 (11$^{\rm h}$49$^{\rm m}$35.8$^{\rm s}$ 22$^{\circ}$23$'$55$''$ (J2000); hereafter MACS\,J1149) (Fig.~\ref{fig:mosaic}A) \cite{kellyrodneytreu15}. %In an accompanying paper \cite{paperone}, %\cite[][hereafter, Paper I]{citepaperI}, 
Models of the gravitational lens predicted that an additional image would appear, designated SX (Table~\ref{tab:positions}), which was observed in 2015 (Fig.~\ref{fig:mosaic}B) \cite{kellyrodneytreu16}. An accompanying paper \cite{paperone} measures the relative time delay between S1--S4 and SX as \corrpmdelaysxsoneCombinedchabriermedian, a precision of 1.5\%, from Hubble Space Telescope (HST) observations through the near-infrared F125W ({\it J}) and F160W ({\it H}) wide-band filters.  If the matter distribution in the foreground MACS\,J1149 cluster lens were known exactly, time delay cosmography could provide a measurement of $H_0$ with equivalent 1.5\% precision.

The value of $H_0$ is currently debated, due to a tension between early-time and late-time probes of the expansion rate of the Universe. Assuming a standard cosmological model with flat geometry, a cosmological constant $\Lambda$, and cold dark matter (CDM), the cosmic microwave background (CMB) measurements using the {\it Planck} satellite imply $H_0 = 67.4 \pm 0.6$\,km\,s$^{-1}$\,Mpc$^{-1}$, where Mpc is the megaparsec \cite{planckhubble18}. In contrast, the alternative local distance ladder method employed by the Supernova H0 for the Equation of State (SH0ES) team yields $H_0=73.04 \pm 1.04$\,km\,s$^{-1}$\,Mpc$^{-1}$ \cite{riessyuanmacri22}.
 This tension between the SH0ES and {\it Planck} $H_0$ measurements has $>$5$\sigma$ significance, indicating a potential problem with standard cosmology \cite{verdetreuriess19}.

Measurements of $H_0$ using independent techniques are needed to confirm or refute the apparent tension. A local distance ladder measurement using the tip of the red giant branch (TRGB) method yields $H_0=69.8 \pm 0.8 \rm{(stat)} \pm 1.7 \rm{(sys)}$ \,km\,s$^{-1}$\,Mpc$^{-1}$ \cite{freedmanmadorehatt19}.
Time-delay cosmography using quasar systems multiply imaged by foreground galaxy-scale lenses find $H_0 = 73.7^{+1.4}_{-1.5}$\,km\,s$^{-1}$\,Mpc$^{-1}$ \cite{millongalancourbin20}, or $H_0 = 74.5^{+5.6}_{-6.1}$\,km\,s$^{-1}$\,Mpc$^{-1}$ (with broader assumptions) for time-delay lenses, and $H_0 = 68.4^{+4.1}_{-3.2}$\,km\,s$^{-1}$\,Mpc$^{-1}$ when combining time-delay lenses with non-time-delay lenses at lower redshift, assuming both are drawn from the same parent population \cite{birrertreurusu19}.
 %The relative time delay of SN Refsdal provides an opportunity to place a new constraint on the value of $H_0$
The appearance of the final image of SN Refsdal provides an opportunity to make an independent measurement of $H_0$, using models of MAC\,J1149 \cite{vegaferrerodiegomiranda18,grillorosatisuyu18}.  The systematic uncertainties of cluster-scale models \cite{grillorosatisuyu20} differ from the galaxy-scale models used for quasar time-delay measurements.

Lensed SNe \cite{oguri19} are predicted to provide comparatively more precise time-delay measurements than quasar observations, and require shorter observations spanning months or years \cite{ogurikawano03,doblerkeeton06,goobaramanullahkulkarni17}.  
Time delays of months or years are expected to be unusual among SNe that are strongly lensed by clusters with extensive pre-existing observations \cite{lihjorthrichard12}. 
The massive galaxy cluster MACS\,J1149 [total mass ($1.4 \pm 0.3) \times 10^{15}$\,M$_{\odot}$  \cite{vdlallen14,kellyvonderlinden14,applegatevdl14}] was observed numerous times before the appearance of SN Refsdal, placing strong constraints on its mass distribution \cite{ebelingedgehenry01,zitrinbroadhurst09,kawamataoguriishigaki16}. Numerous teams used lens models to predict the reappearance of SN Refsdal, prior to the observation of SX \cite{treubrammerdiego16,jauzacrichardlimousin16}.
These models used several different assumptions, so the appearance of SX can also be used to test those assumptions.

The reappearance of SN Refsdal, in image SX, was detected in 
HST observations taken on 2015 December 11 UT
\cite{kellyrodneytreu15}, and was reported to the community the following day \cite{kellyrodneybrammer15}. This observation was used to estimate the relative time delay between SX and S1 as 320--380\,days \cite{rodneystrolgerkelly16}.

Each lens model yielded predictions $\Delta t_{\rm X,1}^{\rm pred, 70}$ for the relative time delay between images SX and S1 (corresponding to subscript 1 and X of $\Delta t_{\rm X,1}^{\rm pred, 70}$, respectively), as well as between other image pairs, assuming $H_0 = 70$\,km\,s$^{-1}$\,Mpc$^{-1}$ (denoted by superscript 70 of $\Delta t_{\rm X,1}^{\rm pred, 70}$). 
For fixed values of the cosmological matter density parameter $\Omega_M$ and the dark energy density $\Omega_{\Lambda}$, the predicted time delay $\Delta t_{\rm X,1}^{\rm pred}(H_0)$ for a given value of $H_0$ depends inversely on the value of $H_0$:
\begin{equation}
\label{eq:rescale}
\Delta t_{\rm X,1}^{\rm pred}(H_0) = \Delta t_{\rm X,1}^{\rm pred, 70} \times \frac{70\,{\rm km}\,{\rm s}^{-1}\,{\rm Mpc}^{-1}}{H_0},
\end{equation}
allowing us to compute the predicted time delay for a given value of $H_0$ in the case for each model. % for $H_0 = 70$\,km\,s$^{-1}$\,Mpc$^{-1}$, 
The uncertainty associated with this rescaling of the predictions is smaller than $\sim 0.7$\,km\,s$^{-1}$\,Mpc$^{-1}$, given the weak dependence of $H_0$ on $\Omega_M$ and $\Omega_{\Lambda}$.
Using a simulation of a galaxy cluster \cite{meneghettinatarajancoe17}, we calculate that the single model that receives the greatest weight in our analysis can recover $H_0$ with at least 5\% accuracy \cite{supplementarymaterial}. We also determine that our error budget is consistent with the model's astrometric errors \cite{birrertreu19}. 

  %blind and post-blind.

We use two sets of lens models to compute parallel estimates of the value of $H_0$, using our observations of SN Refsdal. We first obtain an estimate of the value of $H_0$ using a set of predictions from eight models - which we refer to as the Diego-a free-form model \cite{treubrammerdiego16}, the Zitrin-c* light-traces-mass (LTM) model \cite{treubrammerdiego16}, and the Grillo-g \cite{treubrammerdiego16}, Oguri-a* and Oguri-g* \cite{treubrammerdiego16}, Jauzac-15.2* \cite{jauzacrichardlimousin16}, and Sharon-a* and Sharon-g* \cite{treubrammerdiego16} simply parameterized models.
Models that are simply parameterized use easy-to-compute parametric forms to describe the distribution of dark matter, using priors inferred from the luminous tracers and cosmological simulations. They include one or more 
cluster-scale halos and assign a dark-matter halo
to each luminous galaxy-cluster member, after scaling the 
halo mass according to the galaxies' properties such as stellar mass or
velocity dispersion. 
The second, parallel estimate includes only the Grillo-g and Oguri-a* models, which we selected before unblinding the time delay, as explained below. %The Grillo-g prediction has not been updated following the reappearance, and the Oguri-a* model used to calculate predictions before unblinding has only been updated to include the reappearance's position as a constraint.
In both cases, each model's contribution to the $H_0$ inference is weighted according to its ability to reproduce $H_0$-independent observables.

\paragraph*{Blinded analysis:}
\label{sec:blind}

To avoid human bias, we carried out our analysis without knowledge of the time-delay measurements and the implied value of $H_0$.
For the light curve of each of the images S1--SX, we selected a random number, whose value was stored but kept hidden, which was added to the dates associated with the flux measurements, shifting the light curve in time by an unknown amount \cite{paperone}.
Before unblinding the time delay, we were only aware that the relative delay of SX and S1 was in the range 320—-380\,days ($68$\% confidence level) from a previously published analysis of two epochs of imaging \cite{kellyrodneydiego16}. 
This range corresponds to $\sim 17$\% in $H_0$, before unblinding.

\paragraph*{Observables used for likelihood functions:}
\label{sec:observables}
We use constraints on the time delay between images A and B from the observed light curves (LC) expressed as a probability $P(\Delta t_{\rm A,B}| {\rm LC})$. The prediction for the time delay, $\Delta t_{\rm A,B}^{\rm pred, 70}$, from each lens model $M_l$ is then used with
Bayes' theorem to constrain $H_0$:
\begin{equation}
P(H_0; M_l \mid {\rm LC}) \propto P(H_0)  P(M_l) \int P(\Delta t_{\rm A,B} \mid M_l; H_0) P( \Delta t_{\rm A,B} \mid {\rm LC} )\,d\Delta t_{\rm A,B},
\end{equation}
where $P(H_0)$ is a prior probability for $H_0$, $P(M_l)$ is a prior probability for model $l$, and $P(\Delta t_{\rm A,B} \mid M_l; H_0)$ is the likelihood of $\Delta t_{\rm A,B}$, given the model prediction and a value of $H_0$. 

We include in our likelihood function 
the relative time delays between four image pairs S2--S1, S3--S1, S4--S1, and SX--S1, and magnification ratios of three image pairs S2/S1, S3/S1, and SX/S1  (the magnification ratio S4/S1 is not included, which we explain below), and the position of the reappearance. For the set of observables, $\mathcal{O}$, 
\begin{equation}
P(H_0 \mid LC ) \propto P(H_0) \sum_l P(M_l) \int  P( \mathcal{O} \mid M_l; H_0) P( \mathcal{O} \mid {\rm LC} )\,d\mathcal{O}_1 ... d\mathcal{O}_n.
\label{eq:probabilty}
\end{equation}
Including these observables, in addition to the SX--S1 time delay, allows us to weight each model's contribution to the $H_0$ inference, according to its ability to reproduce the observables \cite{vegaferrerodiegomiranda18}.
We only include observables in the likelihood function if their values were not available to lens modelers before they made their pre-reappearance predictions.

\begin{figure}
\centering
\includegraphics[width=6.25in]{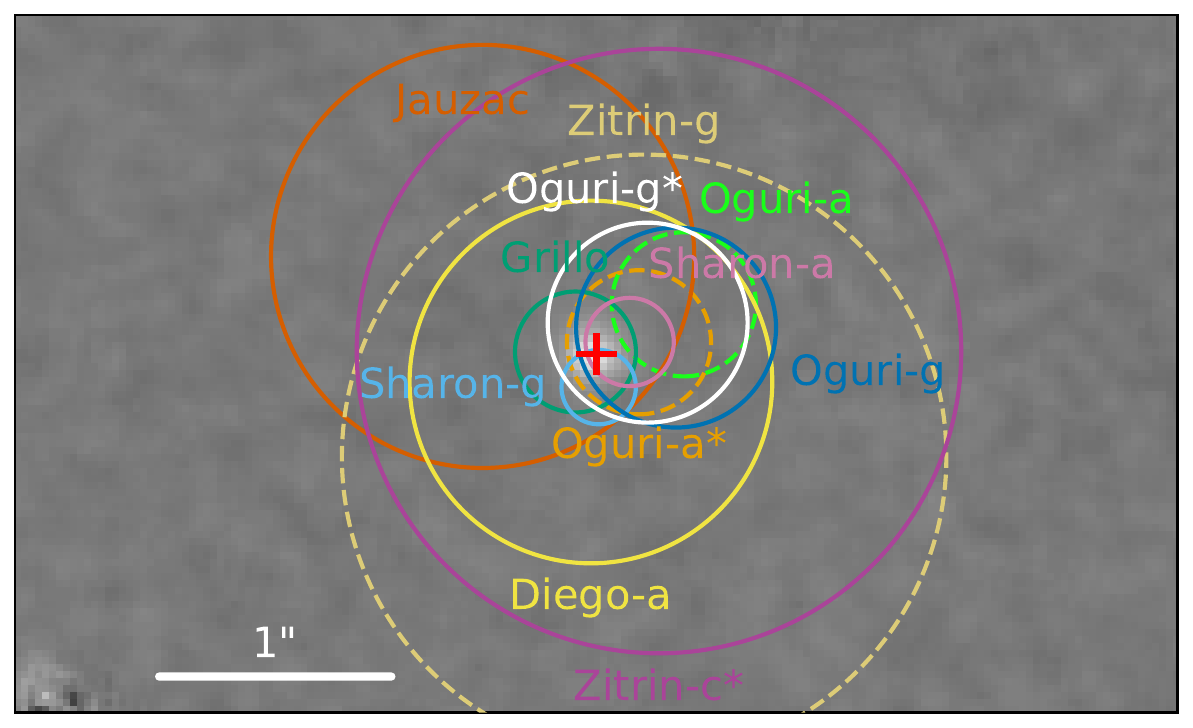}
\caption{{\bf Comparison between the predicted and actual positions of the reappearance of SN Refsdal image SX in late 2015.} 
Panel shows the difference between a coaddition of HST F125W images taken after the appearance of SX (December 2015 through May 2016) and a coaddition of F125W imaging acquired prior to the reappearance of SX (before July 2015). Table~\ref{tab:positions} lists the coordinates of SX.
The radius of the circle for each model indicates the root-mean-square (RMS) of the offsets between the predicted and observed positions of all multiply imaged galaxies used as constraints on the lens model \cite{treubrammerdiego16,jauzacrichardlimousin16}. 
The pre-reappearance Oguri-a prediction is dashed bright green, and the revised position of the Oguri-a* model, after including SX's position as a constraint, is  dashed orange.
The location of the circle shown for Zitrin-c* in solid purple was not computed before the reappearance.
}
\label{fig:ReappearanceLocation}
\end{figure}

Fig.~\ref{fig:ReappearanceLocation} shows the predicted position of the final image (SX) from each lens model, overlaid on a difference image showing the detected position of SX. 
We use the SX position in the likelihood calculation, because it was not known when the pre-reappearance models were constructed.

The HST F125W and F160W light curves of image S4 have very strong evidence ($>5\sigma$) that the SN overlaps a region of high magnification due to one or more stars (or stellar remnants) in the foreground cluster lens \cite{paperone}.  The apparent F125W--F160W color of S4 is bluer than the other four images of the SN by $0.28 \pm 0.05$\,mag through 150 days past its peak brightness (in the observer frame) and $1.8 \pm 0.4$\,mag after 150 days [\cite{paperone}, their figure 12]. This bluer color implies that i) the SN does not have a uniform color and ii) that it overlaps a region where the magnification due to stars changes abruptly called a caustic, which causes differential magnification \cite{paperone}. 
Simulations of lensing of an azimuthally symmetric simulated SN with properties similar to Supernova Refsdal \cite{dessarthillier19} by objects in the foreground cluster lens do not predict microlensing events with color differences as large as that observed for S4 [\cite{paperone}, their figure 6]. Therefore, we exclude the magnification ratio of S4 to S1, because we cannot model accurately the extreme microlensing event, cannot correct its magnification or assign an uncertainty.

The observables $\mathcal{O}$ we use to compute likelihoods are a set inferred from the light curve,
\begin{equation}
\mathcal{O}_{1,j}=
\begin{bmatrix}
\Delta t_{\rm 2,1}, &
\Delta t_{\rm 3,1}, &
\Delta t_{\rm 4,1}, & 
\Delta t_{\rm X,1}, &
\mu_{2}/\mu_{1}, &
\mu_{3}/\mu_{1}, & 
\mu_{X}/\mu_{1}, 
\end{bmatrix},
\label{eq:observables}
\end{equation}
\noindent where $\Delta t_{j,1}$ is the relative time delay between the $j$th image and S1,
$\mu_{j}/\mu_{1}$ is the ratio of the magnifications of image $j$th image and S1. Two additional observables describe SX's position,
\begin{equation}
\mathcal{O_{\rm pos}}= 
\begin{bmatrix}
\alpha_{\rm X}, \delta_{\rm X}
\end{bmatrix}
\end{equation}
\noindent where $\alpha_{\rm X}$ and $\delta_{\rm X}$ are (respectively) the Right Ascension and Declination coordinates of SX.
The ratios of the relative time delays of the images, magnification ratios, and relative positions of the images are independent of $H_0$. %(The unlensed source position and flux are not observable, so the 

\paragraph*{Lens models:}
\label{sec:lens_models}
Table~\ref{tab:hfflensmodels} lists the lens models of MACS\,J1149 that were published before the appearance of SX: there are eight models from six research groups \cite{treubrammerdiego16, jauzacrichardlimousin16}. We use these models, with updates described below, to constrain $H_0$. The Bradac and Merten models were not used to make predictions and were excluded from the time-delay calculations \cite{supplementarymaterial}.

Several of the published pre-reappearance calculations have required revisions. The first Zitrin calculation was amended to address an issue affecting the time-delay surface \cite{treubrammerdiego16}. While their underlying mass models have remained the same, the Sharon-a, Sharon-g, and Jauzac15.1 time-delay predictions required revisions by $\sim 20$\% because of technical issues affecting their time-delay calculations \cite{jauzacrichardlimousin16,supplementarymaterial}.  The observed location of SX differed by $>1\sigma$ from that predicted by the Oguri-a and Oguri-g models (Fig.~\ref{fig:ReappearanceLocation}), so we also updated the Oguri-a and Oguri-g models by adding SX's position as a constraint. 
To preserve blinding, the direction of the $\sim 1\sigma$ shift in the SX--S1 time delay was not disclosed until after we unblinded the time-delay measurement. We use an asterisk to denote models whose predictions were updated or first made after the reappearance. When computing the likelihood of each model, we use pre-reappearance Oguri predictions for SX's position.

For the first of our parallel estimates of $H_0$, we employ the full set of eight pre-reappearance models, after the corrections above. The Diego-a and Zitrin-c* models were not able to reproduce the Einstein cross.
The second parallel estimate for the value of $H_0$ uses only the Grillo-g and Oguri-a* models, because their time-delay calculations did not require large  corrections ($\sim20$\%), and they reproduced the positions of the four images S1-S4 with $<0.1''$ precision.  %To avoid introducing bias into our constraints on $H_0$, 
These two models were selected before unblinding the time-delay measurement.
With the exception of the Jauzac model, all other model predictions were published as part of an organized effort \cite{treubrammerdiego16}. These models were given the choice to utilize only the images of strongly lensed galaxies with spectroscopic redshifts referred to as the gold sample or all the images of strongly lensed galaxies. We label models using the gold sample with a ``g'' suffix, and those that used all images with an ``a'' suffix.
The Grillo team only produced a model using the gold set of images, while Oguri models were created with both sets. However, M. Oguri preferred the Oguri-a* model over the Oguri-g* model  \cite{kawamataoguriishigaki16}. We therefore chose to adopt the Oguri-a* model. % before unblinding.

Table~\ref{tab:hfflensmodels} also lists several
models that were produced after the reappearance so  could not make pre-reappearance predictions for SX. For completeness, we also calculated a value of $H_0$ that includes these models without using SX's position, but do not regard it as a blinded result.

\paragraph*{Likelihood calculation:}
\label{sec:likelihood_hybridsims}

We used light-curve simulations to compute the likelihood of the data, given a value of $H_0$.  
1000 sets of simulated light curves of S1--SX were produced \cite{paper one} for random relative magnifications and time delays.
These light curves have noise characteristics and measurement cadence that closely approximate the SN Refsdal observations, and include the lensing effects of the expected population of dark-matter subhalos and stars in the foreground cluster \cite{paperone}.

We use the simulated light curves to measure and correct the bias associated with the four light-curve fitting algorithms \cite{paperone}, and construct a Gaussian mixture model for the likelihood of the observations that accounts for the covariance among the measurements \cite{supplementarymaterial}.

  %\tablehead{  \multicolumn{1}{c}{Name} & \multicolumn{1}{c}{$\kappa$}  & \multicolumn{1}{c}{$\gamma$} & \multicolumn{1}{c}{$\mu$} & \multicolumn{1}{c}{Parity}}

\paragraph*{$H_0$ constraints:}
\label{sec:hnaught}

For the two parallel estimates of $H_0$, Table~\ref{tab:measurements} lists the values of our unblinded measurements of the relative time delays and magnification ratios, including our constraint on the SX--S1 time delay of \corrpmdelaysxsoneCombinedchabriermedian\ \cite{paperone}.

Gravitational lensing magnification has a sign (positive or negative) associated with it, but only the absolute value of the magnification can be measured directly through observations of images' brightnesses. %, which depends upon the determinant of the amplification matrix.
When we unblinded the revised Oguri-a* and Oguri-g*
model predictions by analyzing Markov Chain Monte Carlo (MCMC) chains of lens model parameters, we did not anticipate that the predicted magnification values (which we had not inspected) included the magnification's sign. After unblinding, the negative magnifications affected the weights of Oguri-a* and Oguri-g* models, but the issue was immediately apparent. 
We report it for transparency; we addressed it by applying an absolute value to the magnification used in the weighting process. %was made to the analysis.
The only other post-blinding changes to our analysis were chosing to present an additional estimate of $H_0$ that includes all pre-reappearance models (not just the preferred models), and to use SX's position instead of the SX--S1 angular separation to weigh models, which changed the results by 0.1\,km\,s$^{-1}$\,Mpc$^{-1}$.

\begin{figure}
\centering
\includegraphics[width=5.5in]{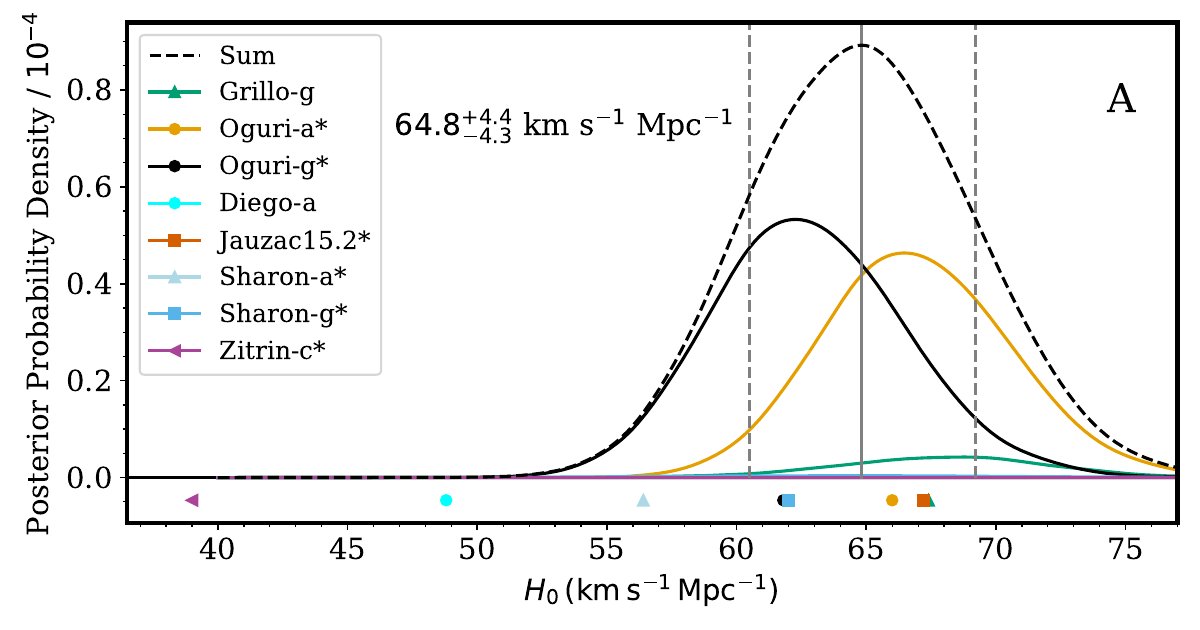}
\includegraphics[width=5.5in]{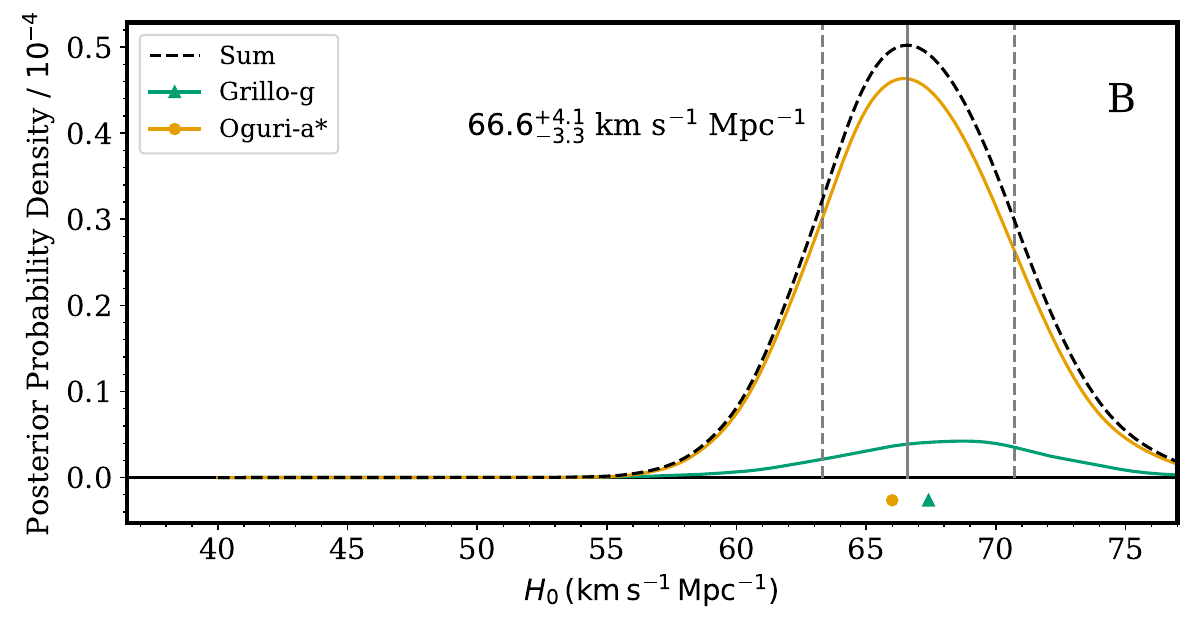}
\caption{{\bf Constraints on $H_0$ from SN Refsdal.} (A) Constraints using the full set of eight models constructed before the reappearance, some with subsequent updates. (B) Parallel constraints using the subset of two pre-appearance models: Grillo-g and the Oguri-a* model, with the latter only updated using the reappearance's position. Both panels show the posterior probability densities for each model as colored lines (see legend), and their sum (dashed black line), assuming a uniform prior. The colored markers beneath each plot show the most probable value of $H_0$ for each of the models, calculated from their predicted SX--S1 time delays. The vertical gray lines show the most likely value from the summed distribution, with the pair of dashed vertical lines marking the 16th and 84th percentile confidence intervals. Models were weighted by their ability to describe the data (Table ~\ref{tab:modweights}), so for most models in the panel A the probability densities are very small, with the lines overlapping zero. The purple arrow for Zitrin-c* in (B) points towards the model's most probable value of $H_0$, which is smaller than 39\,km\,s$^{-1}$\,Mpc$^{-1}$. }
\label{fig:empirical_prior_positions_qualonly}
\end{figure}

\begin{figure}
\centering
\includegraphics[width=6.25in]{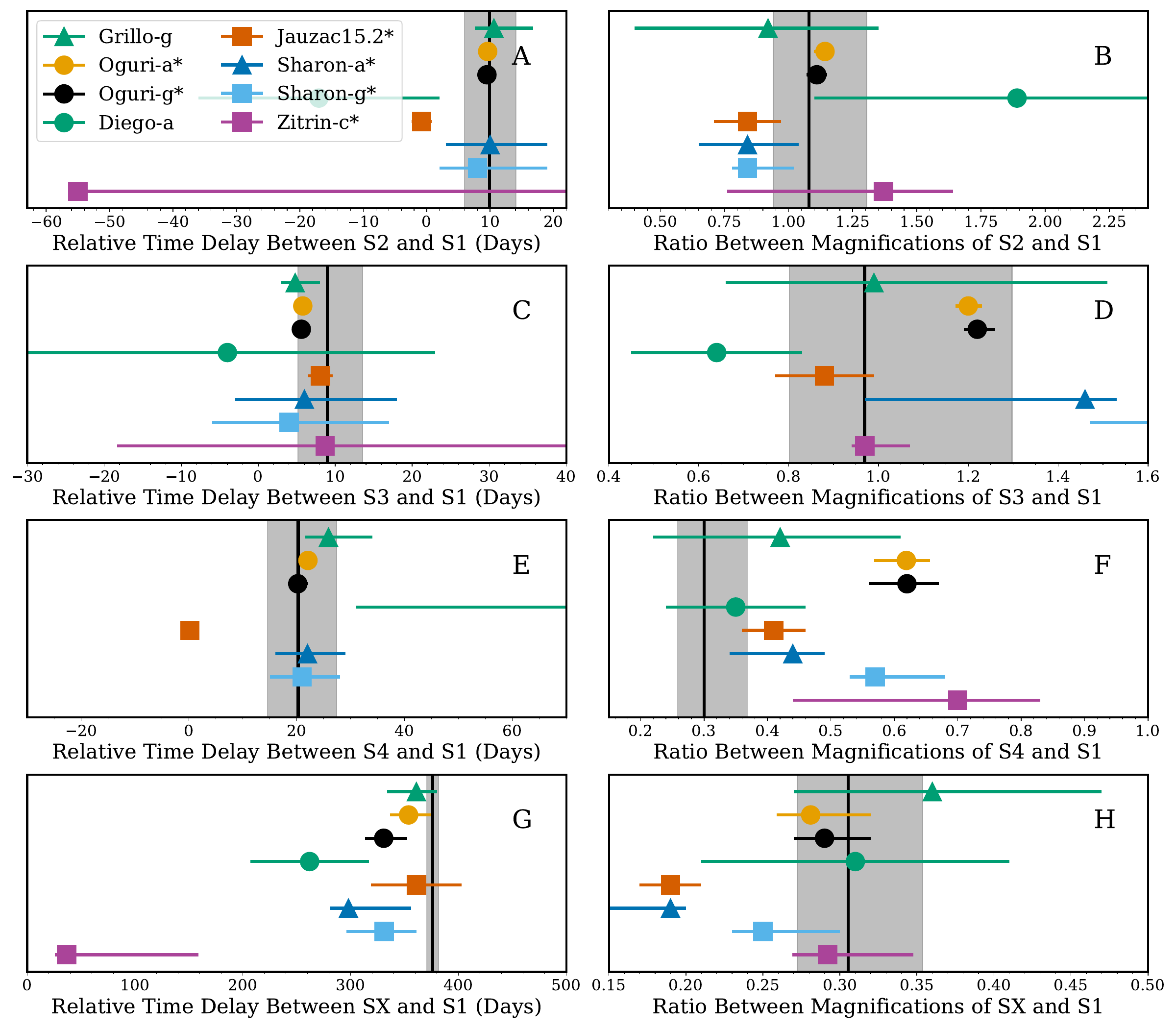}
\caption{{\bf Model predictions and measured relative time delays and magnification ratios between images S2--SX and S1.} (A, C, E, and G) Relative time delays between images S2--SX and S1. Coloured bars indicate the model predictions (see legend), with dots being the median value and bars extending to the 16th and 84th percentiles \cite{paperone}. (B, D, F, H) Same but for the ratios of the magnification of images S2--SX and S1.
Panels A and B compare S1 to S2, panels C and D compare S1 to S3, panels E and F compare S1 to S4, and panels G and H compare S1 to SX.
Models with an asterisk were revised after SX was observed. The predicted time delays shown were computed for $H_0 = 70$\,km\,s$^{-1}$\,Mpc$^{-1}$.} 
\label{fig:delay_combined}
\end{figure}

When we consider all eight pre-reappearance models,
we find $H_0 =$ \hohhybridsimsempiricalnopriorpositionsnoqual\,km\,s$^{-1}$\,Mpc$^{-1}$ (Fig.~\ref{fig:empirical_prior_positions_qualonly}).
The weights for all eight models, listed in Table~\ref{tab:modweights}, show that the Oguri-a* and Oguri-g* models receive \sumfracposthybridsimsOguriaempiricalnopriorpositionsnoqual\ and \sumfracposthybridsimsOgurigempiricalnopriorpositionsnoqual, respectively, while the Grillo-g model receives \sumfracposthybridsimsGrillogempiricalnopriorpositionsnoqual\ (from a total of 1.0).
Fig.~\ref{fig:delay_combined}, A to H, shows a comparison between the constraints on the observables and
the model predictions. The Oguri models are the best match to the observations, which is why they received higher weights. 
When we consider only the Grillo-g and Oguri-a* models,
we find $H_0 =$ \hohhybridsimsempiricalnopriorpositionsqualonly\,km\,s$^{-1}$\,Mpc$^{-1}$.
In this sample, Grillo-g receives a weight of \sumfracposthybridsimsGrillogempiricalnopriorpositionsqualonly\, and Oguri-a* a weight of \sumfracposthybridsimsOguriaempiricalnopriorpositionsqualonly.
All models listed in Table~\ref{tab:hfflensmodels} that can be used to make time-delay calculations even those produced after the appearance of SX, we find $H_0 = $ \hohhybridsimsempiricalnopriornopositionsnoqualthroughtwoohtwooh\,km\,s$^{-1}$\,Mpc$^{-1}$ (Fig.~\ref{fig:empirical_prior_positions_through_2020}). 
The latter calculation does not use the position of SX to weight models, because it was already known when some of those models were produced.

 The measured value of the delay is $376.0^{+5.6}_{-5.5}$\,days \cite{paperone}, which yields $H_0 = $~\hohhybridsimsempiricalnopriorpositionsnoqual\,km\,s$^{-1}$\,Mpc$^{-1}$ considering the eight pre-reappearance models. We next calculate the time delay expected if $H_0$ is equal to the value found by the SH0ES local distance ladder. % of $H_0=73.04 \pm 1.04$\,km\,s$^{-1}$\,Mpc$^{-1}$.  
For $H_0=73.04$\,km\,s$^{-1}$\,Mpc$^{-1}$, the weighted combination of eight pre-reappearance models  yield a delay between SX and S1 of $333.8$\,days. 

\begin{figure}
\centering
\includegraphics[width=4.4in]{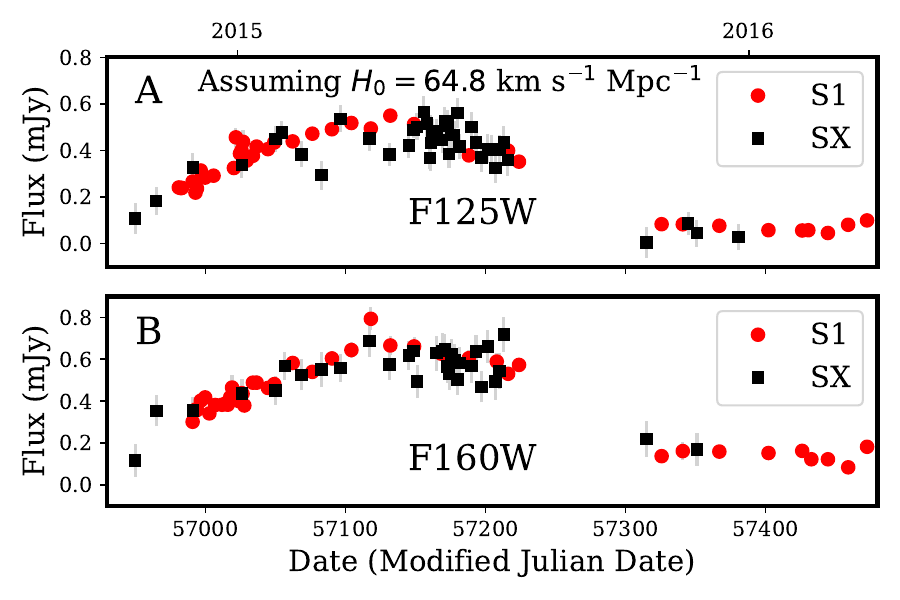}
\includegraphics[width=4.4in]{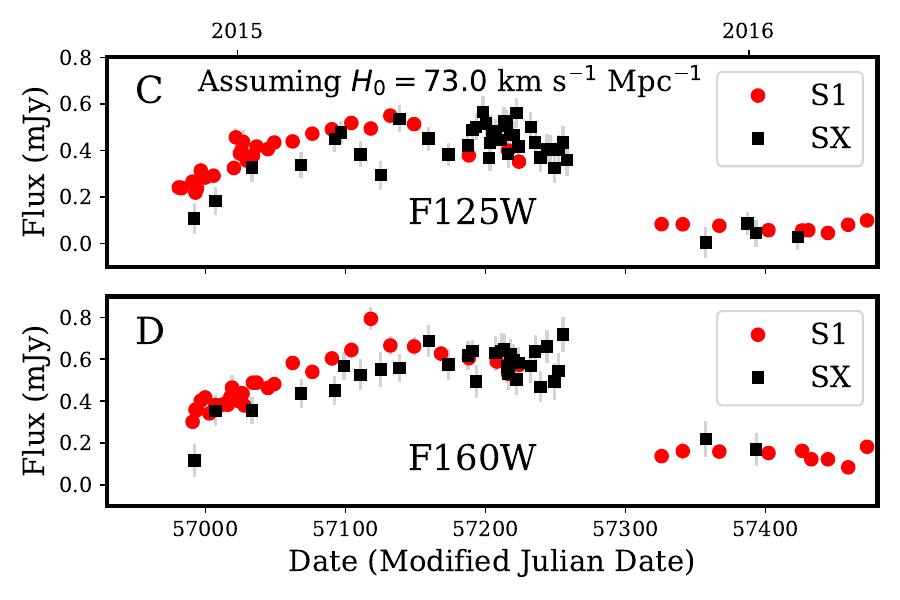}
\caption{{\bf Shifted light curves of images S1 and SX.}
Fluxes of the images S1 (red circles) and SX (black squares) in (A) the F125W filter and (B) the F160W filter, shown as a function of time after removing the measured 376.0\,day delay \cite{paperone}, which corresponds to a constraint on the value of $H_0$ of \hohhybridsimsempiricalnopriorpositionsnoqual\,km\,s$^{-1}$\,Mpc$^{-1}$. (C and D) the same light curves, but using the delay of 333.8\,days expected for $H_0 = 73.04$\,km\,s$^{-1}$\,Mpc$^{-1}$, the value inferred from the supernova distance ladder \cite{riessyuanmacri22} given the eight pre-reappearance models weighted according to their ability to reproduce the $H_0$-independent observables.
While we fixed the SX--S1 time delay, we allow the magnification ratio of SX and S1 to vary in our fit. The  ratios of the magnifications of SX and S1 that minimize the $\chi^2$ value between a piecewise polynomial model of the light curve \cite{paperone} and the measured photometry are 0.31 and 0.32, respectively. Error bars show 1$\sigma$ uncertainties. The $\chi^2$ difference between the calculations for the two time delays is 58.1 with 178 degrees of freedom, favouring panels A \& B over C \& D. %We note that ime-delay measurements we report are derived from a fit to the light curves of all five images. % 996.99 vs. 1048.44 
}
\label{fig:h0_vals}
\end{figure}

 While holding the time delay between SX and S1 fixed but allowing the ratio of their magnifications to vary, we fit the F125W and F160W light curves of SX and S1 using a piecewise polynomial model \cite{paperone}. This model describes the light curve in each filter as two polynomials. The first, a third order polynomial, applies before 150 days after peak brightess, and the second, a second-order polynomial, applies at later epochs. We first fix the SX--S1 time delay first to 333.8 days ($H_0=73$\,km\,s$^{-1}$\,Mpc$^{-1}$) and then to 376.0 days ($H_0 = $~\hohhybridsimsempiricalnopriorpositionsnoqual\,km\,s$^{-1}$\,Mpc$^{-1}$). 
 %The SX--S1 time delay is fixed, but the  magnification ratio of SX and S1 is allowed to vary. 
 The fixed 333.8 day SX--S1 delay yields a worse $\chi^2$ greater by 58.1 for 178 degrees of freedom. %for the light curves of all observed images. 
 We conclude that a shorter time delay of 333.8 days for $H_0=73$\,km\,s$^{-1}$\,Mpc$^{-1}$ and a flexible model provide a significantly worse model for  the observations of Supernova Refsdal. Fig.~\ref{fig:h0_vals} shows the light curves of images S1 and SX shifted by 376.0 days (panels A and B), and 333.8 days (panels C and D).
 The ratios of the magnifications of SX and S1 that minimize the $\chi^2$ value between a piecewise polynomial model of the light curve \cite{paperone} and the measured photometry are 0.31 and 0.32, respectively. 
 
 %which yields $H_0 = $~\hohhybridsimsempiricalnopriorpositionsnoqual\,km\,s$^{-1}$\,Mpc$^{-1}$. The SH0ES local distance ladder finds $H_0=73.04 \pm 1.04$\,km\,s$^{-1}$\,Mpc$^{-1}$. Given the weighted combination of the Grillo-g and Oguri-a* models, $H_0=73.04$\,km\,s$^{-1}$\,Mpc$^{-1}$ corresponds to a most probable predicted delay between SX and S1 of $339.4$\,days.

 %We also carry out the same fit after fixing the delay to 376.0 days. A shorter time delay of 339.4 days and a flexible model provide a significantly worse model for the observations of S1 and SX. 

\begin{figure}
\centering
\includegraphics[height=5.3in]{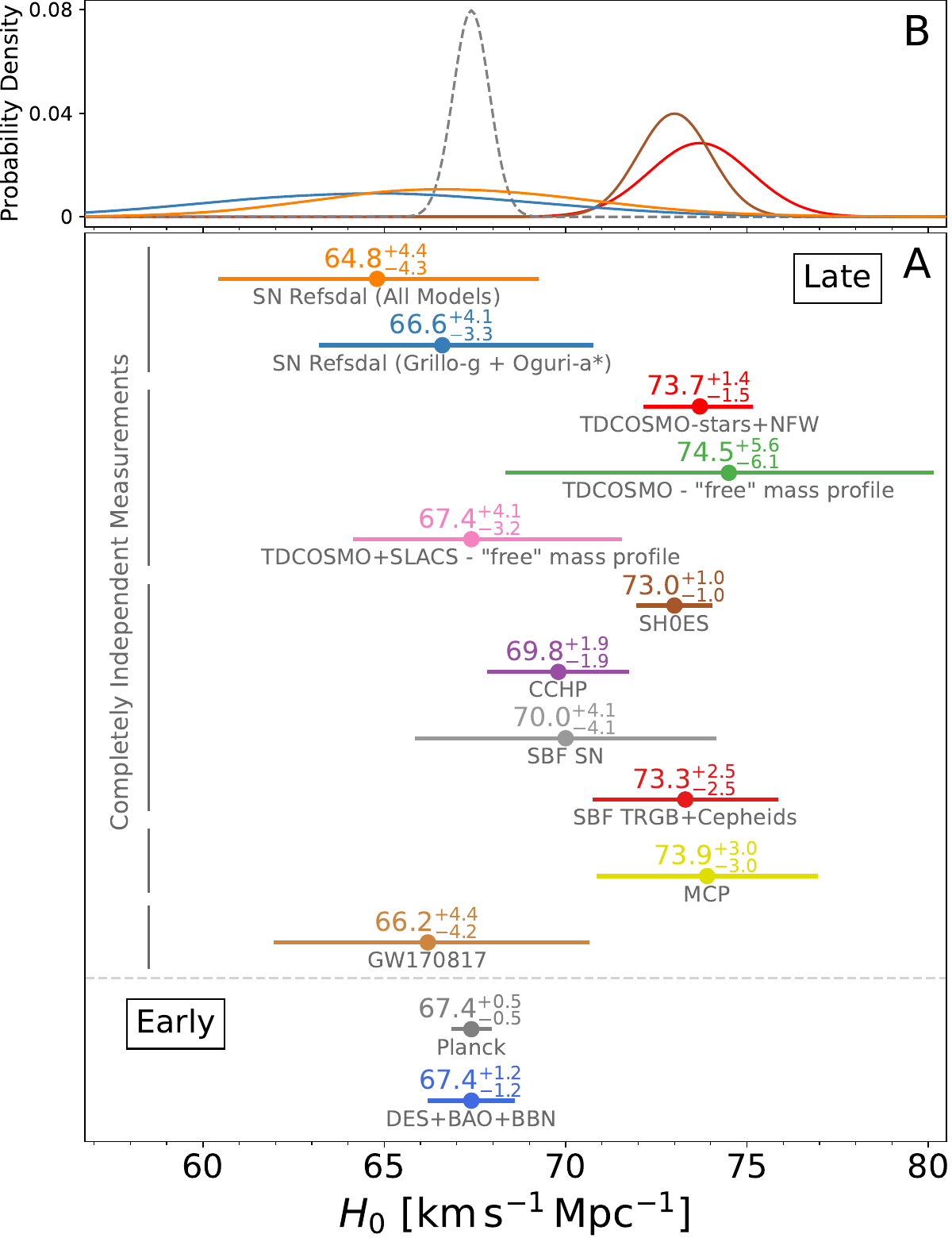}
\caption{{\bf Comparison of the $H_0$ measurement using SN Refsdal to previous measurements.} 
(A) Constraints from our measurements of SN Refsdal, for the full set of eight models (orange) and our preferred subset of the two best models (blue). These are compared to previous constraints from SH0ES + Gaia \cite{riessyuanmacri22}, the Carnegie-Chicago Hubble Program (CCHP) \cite{freedmanmadorehatt19}, H0 Lenses in COSMOGRAIL's Wellspring (H0LICOW) \cite{millongalancourbin20}, STRong lensing Insight into DES (STRIDES) \cite{shajibbirrertreu20}, %Mira variables \cite{huangriessyuan20}, 
surface brightness fluctuations (SBF) SN \cite{khetanizzobranchesi21}, 
SBF Tip of the Red Giant Branch (TRGB)+Cepheids \cite{blakesleejensenma21},
Megamaser Cosmology Project (MCP) \cite{pescebraatzreid20},
gravitational wave (GW) event 170817 \cite{dietrichcoughlinpang20},
Planck \cite{planckhubble18}, and Dark Energy Survey (DES) + Baryon Acoustic Oscillation (BAO) + Big Bang nucleosynthesis (BBN) \cite{deshubble18}.
(B) posterior probability densities for SN Refsdal (orange and blue), Planck \cite[dashed gray]{planckhubble18}, SH0ES \cite{riessyuanmacri22}, and H0LICOW \cite{millongalancourbin20}. Error bars show the 16th, 50th, and 84th percentile confidence levels. Dashed horizontal line separates measurements from observations of the universe early in its evolution from those late in its evolution. $H_0$ measurements bracketed by different vertical gray bars are entirely independent of each other.
Figure generated using a previous comparison \cite{vivien_bonvin_2020_3635517}.
}
\label{fig:h0_tension}
\end{figure}

Fig.~\ref{fig:h0_tension}A compares our measurements of $H_0$ from SN Refsdal to current results from both the late-time and early-time Universe. The posterior probability distributions from our $H_0$ measurement are shown in Fig.~\ref{fig:h0_tension}B, along with and several previous measurements of the expansion rate. Our two estimates of $H_0$ using SN Refsdal favor values for $H_0$ that are smaller than the SH0ES estimate \cite{riessyuanmacri22} by 8.2\,km\,s$^{-1}$\,Mpc$^{-1}$ and 6.4\,km\,s$^{-1}$\,Mpc$^{-1}$, which correspond to   1.8$\sigma$ and 1.5$\sigma$, respectively. Our most probable values of $H_0$ are smaller by 2.6\,km\,s$^{-1}$\,Mpc$^{-1}$ and 0.8\,km\,s$^{-1}$\,Mpc$^{-1}$, corresponding to differences of 0.6$\sigma$ and 0.2$\sigma$, respectively, from the value of $H_0$ inferred from early-Universe observations \cite{planckhubble18}.

\paragraph*{Systematic uncertainties and error budget:}
\label{sec:linearerr}
We considered two sets of models for our parallel estimates of $H_0$, but all models separately prefer smaller values of $H_0$ than 68\,km\,s$^{-1}$\,Mpc$^{-1}$ given the SX-S1 time delay (Fig.~\ref{fig:empirical_prior_positions_qualonly}).

We check whether our measurement of $H_0$ is robust to the assumptions of the simply-parameterized models using a model produced after the appearance of SX.
The Chen2020 model \cite{chenkellywilliams20} makes only minimal assumptions about the distribution of matter at the cluster scale; it calculates an SX--S1 time delay of $332.35 \pm 9.32$\,days assuming $H_0=70$\,km\,s$^{-1}$\,Mpc$^{-1}$ \cite{chenkellywilliams20}, consistent with those of the simply parameterized models (Table~\ref{tab:mod_time_delays}). The Chen2020 model also reproduces the Einstein cross.

The closely related  Oguri-a*  and Oguri-g* models account for 95\% of the weight for our primary estimate  of $H_0$, while the  Oguri-a*  model accounts for 91\% of the weight of our parallel estimate (Table~\ref{tab:modweights}).  Given the models' weight, we use the MCMC chain for the Oguri-a* model to construct an error budget for the model.
We split the Oguri-a*  model parameters \cite{kawamataoguriishigaki16} into six sets of related parameters, and construct a second-order model of the predicted SX--S1 time delay in terms of these variables, after subtracting the mean value of each parameter from the values of the parameters \cite{supplementarymaterial}.  
We next compute the reduction in the variance of the predicted SX--S1 time delay when we add each group of model parameters in succession.

There is covariance among the groups of model parameters, so we compute the mean of the reduction of the variance after repeating the calculation for all permutations of the groups of parameters. The resulting error budget (Fig.~\ref{fig:errorbudget}) shows that the cluster's principal dark-matter halo which dominates its mass has the largest contribution to the uncertainty in $H_0$. 

We used the Hera galaxy cluster simulation \cite{meneghettinatarajancoe17} to assess the ability of the Oguri models to recover the value of $H_0$, by applying the same modelling code [{\tt GLAFIC} \cite{kawamataoguriishigaki16,oguri10}] to the simulation.
In Fig.~\ref{fig:hera_glafic_comparison}, we compare the time delays and uncertainties estimated from the Hera simulation \cite{meneghettinatarajancoe17} with those  estimated using the {\tt GLAFIC} code.  This comparison assumes the same cosmological parameters that were used to construct the Hera simulations ($H_0 = 72$\,km\,s$^{-1}$\,Mpc$^{-1}$ and $\Omega_M = 0.24$). We find the recovered and actual time delays in Fig.~\ref{fig:hera_glafic_comparison} are correlated, with a best-fitting slope of $1.046 \pm 0.021$. We infer that {\tt GLAFIC}'s can be used to infer $H_0$ within $\sim 5$\%, consistent with the error budget calculated for SN Refsdal.

Both the Grillo and Oguri teams performed their time-delay calculations using the positions of images S1--SX as predicted by their models. We examined whether the models' SX--S1 time delays would shift if the images' observed positions instead are used \cite{supplementarymaterial}; we find differences $<$2 days (Table~\ref{tab:mod_time_delays}), less than the precision of the measured time delay.

An independent investigation by the Grillo team into the accuracy of $H_0$ estimated 
using SN Refsdal found that a 3\% constraint on the
relative S1--SX time delay corresponds to 6\% uncertainty in
$H_0$ \cite{grillorosatisuyu20}, which is consistent with our uncertainty calculations.

\paragraph*{Analytical expectation for $H_0$ uncertainty:}
\label{sec:linearerr}
The change in $H_0$ due to an overprediction or underprediction of the separation of two images of a strongly lensed object can be described by a simple differential equation \cite{birrertreu19}. 
We use this relationship, and the predicted image positions for S1 and SX in the MCMC chain for the Oguri-a* model, to compute the expected uncertainty in $H_0$. We find that the precision with which the  Oguri-a*  model reproduces the separation of SX and S1 corresponds to a 5.5\% uncertainty in $H_0$. We also find that $\sim 92$\% of the variance in the relative time delay of SX and S1 can be accounted for by the predicted separation of the two images. Therefore, we conclude that our $\sim 5$\% error budget for the  Oguri-a*  model is consistent with the analytic expectation.

\paragraph*{SN cosmography with a cluster lens:}
\label{sec:conclusion}
We have used observations of SN Refsdal to perform  cosmography and measure the cosmic expansion rate. 
Using two sets of models, we derived constraints on $H_0$ of
\hohhybridsimsempiricalnopriorpositionsnoqual\
 km\,s$^{-1}$\,Mpc$^{-1}$   
and
\hohhybridsimsempiricalnopriorpositionsqualonly\
 km\,s$^{-1}$\,Mpc$^{-1}$. 

We expect our results have different systematic uncertainties to other methods for measuring $H_0$, including time-delay cosmography using quasars which instead employ galaxy-scale lenses.
The dominant source of uncertainty in our $H_0$ measurement is the cluster lens model.
The 1.5\% uncertainty in the measurement of the SX--S1 time delay \cite{paperone} is sufficiently small that it would provide an equally precise constraint on the value of $H_0$, if the cluster model were perfect.

Our analysis uses lens models that were blind to the time-delay measurement.  
The models that account for almost all of the weight in our estimate of $H_0$ are consistent with the observations (Fig.~\ref{fig:delay_combined}), with the exception of the relative magnification of image S4 which appears to have experienced a microlensing event.  
The simply parameterized models best reproduce the observables, even when the CMB and SN constraints on $H_0$ are employed as priors (see Supplementary Text).
We analyzed our error budget using the Oguri-a* model, finding that our $\sim 5.5$\% uncertainty on $H_0$ is consistent with analytic expectations \cite{birrertreu19} given the model's astrometric accuracy. Further tests using a simulated galaxy cluster also indicated an expected uncertainty of $\sim 5$\%.

\bibliography{ms}

\bibliographystyle{Science}

\noindent {\bf Acknowledgements:} 
We thank Program Coordinators Beth Periello and Tricia Royle, as well as Contact Scientists Norbert Pirzkal, Ivelina Momcheva, and Kailash Sahu of the Space Telescope Science Institute (STScI) for their help carrying out the HST observations. We also appreciate useful discussions with P. Rosati, C. Grillo, S. Suyu, and M. Meneghetti. We used the HFF Lens Model website, hosted by the Mikulski Archive for Space Telescopes
(MAST) at STScI. 
{\bf Funding:} NASA/STScI grants GO-14041, 14199, 14208, 14528, 14872,  14922, 15050, 15791, 16134 provided financial support for P.K., A.V.F., T.T., J.P., and S.R..
P.K. is supported by NSF grant AST-1908823. % and MRI-1908823. %, and NASA/Keck JPL RSA 1644110.
M.O. acknowledges support by the WPI Initiative (MEXT, Japan) and by JSPS KAKENHI grants JP18K03693, JP20H00181, JP20H05856, and JP22H01260.
T.T. acknowledges the support of NSF grant AST-1906976.
S.T. was supported by the Cambridge Centre for Doctoral Training in Data-Intensive Science funded by the UK Science and Technology Facilities Council (STFC).
J.D.R.P. was supported by the NASA FINESST award 80NSSC19K1414.
J.M.D. acknowledges the support of project PGC2018-101814-B-100 (MCIU/AEI/MINECO/FEDER, UE) Ministerio de Ciencia, Investigaci\'on y Universidades.  J.M.D. was funded by the Agencia Estatal de Investigaci\'on, Unidad de Excelencia Mar\'ia de Maeztu, ref. MDM-2017-0765. 
M.J. is supported by the United Kingdom Research and Innovation (UKRI) Future Leaders Fellowship (grant MR/S017216/1).
M.M. and V.B. acknowledge support from the European Research Council (ERC) under the Horizon 2020 research and innovation program (COSMICLENS: grant agreement 787886).
A.V.F. is grateful for funding from the Christopher R. Redlich Fund, the TABASGO Foundation, the U.C. Berkeley Miller Institute for Basic Research in Science (as a Miller Senior Fellow), and financial contributions many additional individual donors.
R.J.F. is supported in part by NASA grant NNG17PX03C, NSF grants AST-1518052 and AST-1815935, the Gordon \& Betty Moore Foundation, the Heising-Simons Foundation, and the David and Lucile Packard Foundation. 
J.H. acknowledges support from VILLUM FONDEN Investigator Grant 16599.
S.W.J. is supported by NSF CAREER award AST-0847157, as well as by NASA/Keck JPL RSA 1508337 and 1520634. K.S.M. acknowledges funding from the European Union’s Horizon 2020 research and innovation programme under ERC Grant Agreement No. 101002652 and Marie Skłodowska-Curie Grant Agreement No. 873089.
A.Z. acknowledges support by Grant No. 2020750 from the United States-Israel Binational Science Foundation (BSF) and Grant No. 2109066 from the United States National Science Foundation (NSF), and by the Ministry of Science \& Technology, Israel. {\bf Author contributions:} P.K. measured the SN photometry, developed the statistical framework and analysis code, and drafted the manuscript. P.K., S.R., T.T., A.V.F., R.J.F., J.H., T.B., O.G., S.J., C.M., M.P., K.B.S, B.E.T., and A.v.d.L. obtained follow-up HST imaging. P.K., S.R., T.T., S.B., V.B., L.D., D.G., K.M., M.M., J.P., K.S., and S.T. contributed to the time-delay measurement. P.K., S.R., T.T., V.B., A.V.F., K.M., and S.T. helped prepare the manuscript. S.R., T.T., L.W., T.B., and A.D. aided the interpretation. M.O., W.C., A.Z., J.M.D., and M.J.    modeled the galaxy cluster. {\bf Competing interests:} We declare no competing interests.

\section*{Supplementary Materials}
Materials and Methods\\
Supplementary Text\\
Tables S1--S15\\
Figs. S1--S6\\
References (47--67)

\clearpage

\setlength\tabcolsep{3pt}

\clearpage

\begin{table}
\footnotesize
\caption{\normalsize \label{tab:modweights} {\bf $H_0$ constraints and weights determined by posterior probabilities for each set of lens-model predictions.} First column lists the most probable value of $H_0$ and our 16th, 84th, and 99.7th percentile confidence levels.
Other columns list the weights (out of 1.0) we calculated using the constraints from Supernova Refsdal. The first and second rows are to our parallel estimates of the value of $H_0$ using all eight models and our two preferred models, respectively. %The third and fourth rows list model weights when applying different priors from previous SN and CMB measurements of $H_0$.
The relative weights are proportional to the models' posterior probabilities. } % $^{\dagger}$We note that the Zitrin-c* problem presented here was affected by a calculation problem and was subsequently corrected in Zitrin2020-ltm (see Table~\ref{tab:posteriorprobsrecent}).}
\begin{tabular}{cc|cccccccc}
\multicolumn{2}{c|}{$H_0$}  & \multicolumn{8}{c}{Weight of Each Model} \\
  \multicolumn{2}{c|}{(km s$^{-1}$ Mpc$^{-1}$)} &   \multicolumn{8}{c}{ } \\
  \hline
  \multicolumn{2}{c|}{Inference} &  Diego-a & Grillo-g & Jauzac15.2* & Oguri-a* & Oguri-g* & Sharon-a* & Sharon-g* & Zitrin-c* \\  
  16-MP-84 & 99.7 &  &  &   &  &  &  &  & \\
\hline
 \hohhybridsimsempiricalnopriorpositionsnoqual & \hohhybridsimsempiricalnopriorpositionsnoqualpninesevenseventhree &  $\sumfracposthybridsimsDiegoaempiricalnopriorpositionsnoqual$ & $\sumfracposthybridsimsGrillogempiricalnopriorpositionsnoqual$ &
$\sumfracposthybridsimsJauzaconefivedottwoempiricalnopriorpositionsnoqual$ &
$\sumfracposthybridsimsOguriaempiricalnopriorpositionsnoqual$ & $\sumfracposthybridsimsOgurigempiricalnopriorpositionsnoqual$ &
$\sumfracposthybridsimsSharonaempiricalnopriorpositionsnoqual$ &
$\sumfracposthybridsimsSharongempiricalnopriorpositionsnoqual$ &
 $\sumfracposthybridsimsZitrincempiricalnopriorpositionsnoqual$ \\
\hline
 \hohhybridsimsempiricalnopriorpositionsqualonly
 & \hohhybridsimsempiricalnopriorpositionsqualonlypninesevenseventhree
 & ... &
$\sumfracposthybridsimsGrillogempiricalnopriorpositionsqualonly$ &
... &
$\sumfracposthybridsimsOguriaempiricalnopriorpositionsqualonly$ & %$\sumfracposthybridsimsOgurigempiricalnopriorpositionsqualonly$ & 
... &
... &
... &
... \\
 \hline
\end{tabular}

\end{table}

\begin{table}
\footnotesize
\centering
\caption{\label{tab:posteriorprobs} {\bf Posterior probability for each set of lens-model predictions.} 
} % $^{\dagger}$We note that the Zitrin-c* model presented here contains an uncorrected problem, subsequently revised in Zitrin2020-ltm (see Table~\ref{tab:posteriorprobsrecent}).}
\begin{tabular}{cccccccc}
\hline
Diego-a & Grillo-g & Jauzac15.  2* & Oguri-g* & Oguri-a* & Sharon-a* & Sharon-g* & Zitrin-c* \\  
\hline
$\sumposthybridsimsDiegoaempiricalnopriorpositionsnoqual$ & $\sumposthybridsimsGrillogempiricalnopriorpositionsnoqual$ &
$\sumposthybridsimsJauzaconefivedottwoempiricalnopriorpositionsnoqual$ &
$\sumposthybridsimsOguriaempiricalnopriorpositionsnoqual$ & $\sumposthybridsimsOgurigempiricalnopriorpositionsnoqual$ &
$\sumposthybridsimsSharonaempiricalnopriorpositionsnoqual$ &
$\sumposthybridsimsSharongempiricalnopriorpositionsnoqual$ &
 $\sumposthybridsimsZitrincempiricalnopriorpositionsnoqual$ \\
\hline
\end{tabular}
\end{table}

\renewcommand{\thetable}{S\arabic{table}}
\renewcommand{\thefigure}{S\arabic{figure}}
\renewcommand{\theequation}{S\arabic{equation}}
\setcounter{figure}{0} 
\setcounter{table}{0} 
\setcounter{equation}{0}

\clearpage
  \singlespacing 
  \setcounter{page}{1}

\section*{\Large \center Supplementary Materials for}

\begin{center}
{\large \titlestring}
\end{center}

\begin{center}
Patrick L. Kelly*, Steven Rodney, Tommaso Treu, Masamune Oguri, Wenlei Chen, Adi Zitrin, Simon Birrer, Vivien Bonvin, Luc Dessart, Jose M. Diego, Alexei V. Filippenko, Ryan~J.~Foley, Daniel Gilman, Jens Hjorth, Mathilde Jauzac, Kaisey Mandel, Martin Millon, Justin Pierel, Keren Sharon, Stephen Thorp, Liliya Williams, Tom Broadhurst, Alan Dressler, Or Graur, Saurabh W. Jha, Curtis McCully, Marc Postman, Kasper Borello Schmidt, Brad~E.~Tucker, Anja von der Linden
\end{center}

\begin{center}
*Corresponding author. Email: plkelly@umn.edu
\end{center}

\paragraph{This PDF file includes the following:} % \mbox{} \\

\newenvironment{myitemize}
{ \begin{itemize}
    \setlength{\itemsep}{0pt}
    \setlength{\parskip}{0pt}
    \setlength{\parsep}{0pt}     }
{ \end{itemize}             }

\begin{myitemize}
\itemsep0em 
\setlength{\itemindent}{4.5pt}
\item[] Materials and Methods
\item[] Supplementary Text
\item[] Figs. S1 to S6
\item[] Tables S1 to S15
\end{myitemize}

\paragraph*{\large Materials and Methods}

\paragraph*{Observables used for likelihood function:}

The galaxy-cluster lens models have mass distributions that are smooth on subkiloparsec scales, and, consequently, their time-delay and magnification predictions do not include the effects of microlensing by stars and stellar remnants (neutron stars and black holes), and millilensing by dark-matter subhalos. 
Constraints on the values of these observables in the companion paper \cite{paperone} use simulations that include microlensing and millilensing to make inferences.  Consequently, the confidence intervals for the observables can be compared directly to the cluster models' predictions.

We use the image position for SX measured from HST imaging \cite{kellyrodneydiego16} listed in Table~\ref{tab:positions}, and the root-mean-square (RMS) uncertainties are $\lesssim 0.03''$. The RMS residuals of the lens models' predicted positions from the observed positions are taken from [\cite{treubrammerdiego16}, their table 2;  \cite{jauzacrichardlimousin16}], 
as the uncertainty associated with each image's position.
When computing the uncertainties for the models' predicted values of $\alpha_{\rm X}$ and $\delta_{\rm X}$, we assume equal variance along the RA and Dec axes.

\paragraph*{Cluster lens models:}
Current galaxy-cluster lens models adopt different assumptions about 
how the distribution of dark matter connects with that of luminous matter.
Broadly speaking, there are three classes of models. 
The main strength of the simply parameterized approach is that the number of free parameters can be kept comparable to the amount of information, since the potentials and other lensing quantities can be computed with arbitrary precision. 

The second class of models are known as free-form models that aim to make minimal assumptions about how dark matter is distributed by representing the mass or gravitational potential using highly flexible basis sets \cite{diegobroadhurstchen16,williamsliesenborgs19,bradactreuapplegate09}. 
This approach generally results in more free parameters than
constraints, so different strategies have been proposed
and implemented to address this issue. The main strength of this approach is that it is flexible and allows one to adopt fewer assumptions about the distribution of dark matter. The weakness is that there could be too much freedom, resulting in models that are unphysical.  

A third approach was developed to obtain fast lens models, under the assumption that light approximately traces mass, and that the dark-matter component is some smoothed version of the light.
This light-traces-matter (LTM) technique \cite{zitrinbroadhurstumetsu09,zitrinmeneghettiumetsu13} is very fast and has predictive power for finding multiple images. However, the method has a finite resolution, and as such, was not originally designed for high-precision applications such as computing time delays. That functionality was only added following the appearance of SN Refsdal.

Implementation of the modeling approach also matters in determining the outcome. Each investigator is faced with a number of choices, including which data are used to constrain the models. For cluster-scale lenses, most modelers use multiple image positions as constraints, rather than the full surface brightness distribution of each source as is done for galaxy-scale lenses. %PK: not sure footnotes are allowed in Science's format \footnote{The reason is mainly computational, i.e. lensed features are dispersed over a much larger number of pixels in clusters, compared to galaxies. Effort is underway to bridge to construct cluster-scale pixel based models but it is not a trivial one (Birrer et al. 2021, in prep.)}.

Even within this approach, however, the selection of images is critical. At the depth of the Hubble Frontier Fields program \cite{lotzkoekemoercoe17}, some teams have identified and used $\sim100$ multiple images \cite{jauzacrichardlimousin16}, most of which are too faint for spectroscopic confirmation. Other teams have focused on the smaller number of spectroscopically confirmed images \cite{grillokarmansuyu16} or adopted stringent quality criteria \cite{wanghoaghuang15} that also reduce substantially the number of images.
Those two choices are illustrated in a comparison exercise \cite{treubrammerdiego16}, in which modelers were given the choice to utilize only the spectroscopic confirmed images (gold sample) or all the images (all). As described in the main text, we label models using the gold sample with a g suffix, and the all sample with an a suffix.

Technical choices, such as whether the predicted or observed image positions are used when computing time delays, can also affect the results. 

\paragraph*{Cluster models used for $H_0$ inference:}
\label{sec:models}

To compute a posterior probability distribution for $H_0$ using Bayes' theorem, we require a prior probability to weight each of the eight models used to make predictions before the reappearance. 
We follow two approaches: inclusive and selective. In the first, we include all the pre-reappearance models that could be used to make predictions, as listed in Table~\ref{tab:hfflensmodels}. In the second approach, we select two models, Grillo-g and Oguri-a*, which were produced using codes developed for time-delay cosmography. The parallel constraints on the value of $H_0$ we obtain are consistent across the two choices, and that  $H_0$-independent observables favor the Grillo-g, Oguri-a*, and Oguri-g* models. We also show, in Fig.~\ref{fig:empirical_prior_positions_through_2020}, that including all published models, even those produced after the appearance of SX, yields consistent constraints on $H_0$.

We apply a strong prior and select only a subset of two models for the exclusive estimate, for the reasons described below. Because we made the decision to apply this prior for the exclusive estimate before unblinding the time delay, we could not anticipate how our choice of models would affect the agreement between our estimate of $H_0$ and existing SN~Ia \cite{riessyuanmacri22} or CMB \cite{planckhubble18} constraints. %Since the 1.5\% time delay was not known to us, we could select models without introducing bias into the inference on $H_0$, because we could not anticipate how model choices would affect agreement between our estimate of $H_0$ and existing constraints.

Fig.~\ref{fig:timeline} shows a timeline of the decisions we made and of the unblinding of the relative time-delay constraints. After the reappearance, relative time delays and magnification ratios for images S1--S4 were published, but none of the models we employ used these as constraints \cite{treubrammerdiego16,rodneystrolgerkelly16}.

For a source at a specific location and a model of a foreground lens, the time delay can be calculated across the lens plane, which is called a time-delay surface. Images of the source form at the extrema of this surface. However, it was not possible to identify the position of S1 from the time-delay surfaces produced by the Diego-a model. The time-delay surface computed for the Zitrin-c* model contained a numerical artifact and extrema were not present at the locations of the Einstein cross. 
Moreover, an ability to reproduce the time-delay surface is necessary given the dependence of the predicted time delay on the source position \cite{birrertreu19}.
After the appearance of SX, the Zitrin-c* model was reimplemented and now reproduces the Einstein cross \cite{zitrinrevised21}, so we include this updated model in Fig.~\ref{fig:empirical_prior_positions_through_2020} when we compute constraints using all published models.

Because time-delay cosmography had previously only been carried out using galaxy-scale lenses, many cluster-modeling groups not previously computed time delays from their lens models. %The short time period before the reappearance meant that several of the teams without prior experience produced initial predictions that either were not optimal or else contained clerical errors.
In reproducing the time-delay predictions of 
the Sharon-a and Sharon-g lens models \cite{treubrammerdiego16}, we found that the time delays required revision by $\sim 27$\%, due to a clerical error entering the redshifts used to compute angular-diameter distances that affected the original time-delay distance calculation \cite{sharonjohnson15,treubrammerdiego16}. %The predictions reported in our work are the ones we compute directly from the original ``Sharon-a'' and ``Sharon-g'' models, which do not required any revision, except for rescaling. 
In the case of the Sharon-g model, the SX--S1 time delay, after rescaling, is $298_{-17}^{+58}$\,days, while the Sharon-a model prediction, after rescaling, is $352_{-57}^{+9}$\,days.
However, we found that the realization of the Sharon-a models that had the minimum chi-squared value was misidentified.
The value for the best-fitting realization, which we adopt in this paper, is $331_{-35}^{+30}$\,days.

The Jauzac team made two predictions for the reappearance. For the first, they used the deflection field of their galaxy-cluster model and the observed positions of S1--S4 to compute a separate source-plane position for each of the images. They next computed time delays using the distinct source-plane positions for S1--S4  \cite{jauzacrichardlimousin16}.
This calculation yielded an SX--S1 predicted time delay of $449 \pm 45$\,days which we refer to as Jauzac15.1 \cite{jauzacrichardlimousin16v3}.

The Jauzac team's predicted time delay was late updated by adopting a single, shared source-plane position for the SN, which was the average of the source-plane images' positions weighting according to the images' respective magnifications. The authors considered this method to be more accurate, and it yielded an SX--S1 predicted time delay, dubbed Jauzac15.2, of $361 \pm 42$\,days. % which differs by 24\% from the prediction in ``Jauzac15.1.'' 
However, this value was not published until after the appearance of SX \cite{jauzacrichardlimousin16v3}.

Given the magnitude of the adjustments to the Jauzac and Sharon time-delay calculations, we elected to place constraints on the value of $H_0$ using only model calculations from two teams, Grillo and Oguri. 
The Grillo team published only the Grillo-g model, while there were two  model predictions published by the Oguri team,  Oguri-a  and Oguri-g \cite{treubrammerdiego16}. We elected to include, for one of our parallel estimates of $H_0$, only Oguri-a, because the team that produced it stated that they preferred that model, in a publication prior to our unblinding \cite{kawamataoguriishigaki16}. 

As plotted in Fig.~\ref{fig:ReappearanceLocation}, the  Oguri-a  prediction for the position of the reappearance SX was in $>1\sigma$ tension with the reappearance's measured location.
Because the time delay can depend sensitively on the source-plane position, we updated the model fitting to obtain a robust initial constraint on $H_0$.
We used the same pre-reappearance model with only the addition of SX's position, and reported a shift of $\sim 30$\,days (6\,km\,s$^{-1}$\,Mpc$^{-1}$).  
Our blinding process disclosed that the value had shifted by $\sim$1 sigma, but not its direction until after the measurements were unblinded.
We use the model's original prediction for SX's location when computing the weight of each model. 

The updated  Oguri-a*  prediction is $354_{-17}^{+21}$\,days, while the original prediction was $336_{-21}^{+21}$\,days. The updated Oguri-g* prediction is $331_{-17}^{+22}$\,days, while the original prediction was $311_{-24}^{+24}$\,days. The updated values of the  Oguri-a*  and Oguri-g* predictions were unblinded at the same time that we unblinded the time delays, and analysis choices were made in advance. %Even though we use the  Oguri-a*  model updated using SX's observed position to compute the time delay and magnification, 

After unblinding, we made an additional inclusive estimate of the value of $H_0$ using the full set of eight pre-reappearance models.
As shown in Table~\ref{tab:hfflensmodels}, after revision, the simply-parameterized models' predictions for the SX--S1 delay, consisting of Grillo-g, Jauzac15.2, Oguri-a*, Oguri-g*, Sharon-a*, and Sharon-g, agree with each within their 1$\sigma$ uncertainties.

\paragraph*{Likelihood calculations using Gaussian mixture model:}
The companion paper \cite{paperone} produced one-thousand sets of simulated light curves of images S1--SX of SN Refsdal, and used four separate light-curve fitting algorithms to recover the input values used to construct the simulations.
Given a relative time delay $\Delta t_{1,j}$ or magnification ratio $\mu_j / \mu_i$ between image S1 and the $j$th image estimated using fitting algorithm $k$, we can compute the residual of 
$\mathcal{O}_{1,j}^{\rm fit, k, i}$ from the input, true value $\mathcal{O}_{1,j}^{\rm tr, i}$ used for each simulation,
\begin{equation}
\delta(\mathcal{O}_{1,j}^{\rm k, i}) = \mathcal{O}_{1,j}^{\rm fit, k, i} - \mathcal{O}_{1,j}^{\rm tr, i}.
\end{equation}
\noindent

We remove $>5\sigma$ fitting outliers, where we define $\sigma = (p_{84} - p_{16}) / 2$, and $p_{16}$ and $p_{84}$ respectively correspond to the 16th and 84th percentiles of the residual distribution $\delta(\mathcal{O}_{1,j}^{\rm k, i})$. 
We require a full set of measurements from all four light-curve fitting algorithms for each simulated light curve, and 5$\sigma$ rejection leaves 897 of 1000 simulated light curves.
Our constraints on $H_0$ are robust to the rejection of outliers: the most probable value of $H_0$ and its $p_{16}$ and $p_{84}$ confidence intervals shift by $<0.3$\,km\,s$^{-1}$\,Mpc$^{-1}$ when instead only a single simulated light curve, corresponding to a 902\,day outlier for the SX time delay, is removed. 

We correct the single-point estimates for the observables inferred from the observed data for each light-curve fitting algorithm for the average biases we measure from the simulated light curves, $\mathcal{O}_{1,j}^{\rm corr,k} = \mathcal{O}_{1,j}^{\rm fit,k} - \langle \delta(\mathcal{O}_{1,j}^{\rm k, i}) \rangle$. These offsets are listed in Table~\ref{tab:offsets}. Because each observable has a distinct set of weights, for each value of $H_0$ we construct 1000 covariance matrices from which we select randomly, for each observable, a single light-curve fitting result with probability proportional to the fitting algorithm's weight $w_{k,\mathcal{O}_{1,j}}$ determined in the companion paper \cite{paperone}. For each likelihood calculation, we average the likelihood values computed using the 1000 covariance matrices. The resulting Gaussian mixture model assigns weights to each algorithm in proportion to those derived in the companion paper, while also accounting for the covariance among the measurements.

For each covariance matrix, we use Monte Carlo integration to compute the integral in Eq.~\ref{eq:probabilty} of the product of the likelihood for a skew-normal representation of the model predictions $P( \mathcal{O}_{1,j} \mid M_l; H_0)$ and the covariance likelihood $P( \mathcal{O}_{1,j} \mid {\rm LC} )$.  If the model predictions have evenly spaced confidence intervals, we instead include the model prediction uncertainty additional terms in the covariance matrix, which yields an exact calculation of the joint likelihood. The probability of the angular offset of the measured position of SX, given the RMS model uncertainty, is a final independent term:
\begin{equation}
P(\mathcal{O} | {\rm LC}; \Sigma) = \frac{\exp\left(-\frac 1 2 (\mathcal{O}^{\rm corr}_{1,j} -\mathcal{O}^{\rm }_{1,j})^\mathrm{T}{\Sigma}^{-1}(\mathcal{O}^{\rm corr}_{1,j} -\mathcal{O}^{\rm }_{1,j})\right)}{\sqrt{(2\pi)^n|\Sigma|}} P(\alpha_{\rm X},\delta_{\rm X} ; M_l)
\end{equation}
where $n$ is the number of observables in $\mathcal{O}_{1,j}$, $\Sigma$ is a covariance matrix. %and $\mathcal{O}^{\rm pred}_{1,j}$ is the predicted value of the . 

In summary, the likelihood function is, 
\begin{equation}
P(H_0 | {\rm LC}) \propto P(H_0)\,\sum_{l}\,\left[ P(M_l) \frac{1}{N} \sum_{r=1}^N \int P(\mathcal{O} | {\rm LC}; \Sigma_r) P(\mathcal{O} | M_l; H_0) d\mathcal{O}_1 ... d\mathcal{O}_n \right] ,
\end{equation}
where $N$ is the total number of covariance matrices $\Sigma_r$.

For reference, we also combine the observables  $\mathcal{O}_{1,j}^{\rm corr,k}$ by the four light-curve fitters as a weighted average $\sum_{k=1}^4 w_{k,\mathcal{O}_{1,j}} \mathcal{O}_{1,j}^{\rm corr,k}$, and compute the covariance matrix, shown in Table~\ref{tab:covariance}, for this combination. 
Because the weights maximize the maximum likelihood of the input values used to construct the simulations, they were not intended to be used to in the context of calculating a weighted average. Nonetheless, this approach yields an almost identical constraint on the value of $H_0$, and requires only a single covariance matrix.

\paragraph*{Comparison with Hera simulation:}

The Frontier Field Lens Modeling Comparison project \cite{meneghettinatarajancoe17} has produced two sets of simulated data to assess the accuracy of current lens modeling codes.  The first was a simulation of a cluster, called Ares, by assembling parameterized halos, and the second was a cluster, named Hera, from an $N$-body simulation. We sought to use these simulations to determine whether a 5\% measurement of the Hubble constant is possible using current lens reconstruction codes specifically the Oguri {\tt GLAFIC} code \cite{oguri10}. 
 
There are sharp discontinuities in the residual maps  of the Ares simulation [\citenum{meneghettinatarajancoe17}, their figure 8]. Inspection of the projected mass-density maps shows that the simulated parametric model halos used to construct Ares have a discontinuity where the halo truncates.  Such behavior is not expected for actual halos, and simply-parameterized modeling codes employ halos without such discontinuities.  Consequently, we decided not to use the Ares simulation.

The Hera simulation was constructed instead using an $N$-body simulation. We identified three difficulties in using the Hera simulations for our test, but are able to quantify the error in the expected time delay arising from each of them. First, when multiple images from the same source are projected using the lens equation, then their positions disagree by up to $0.5''$.  %These residuals appear to exhibit correlation with the mass distribution.  

For each system of $N$ images of a single strongly lensed galaxy, we compute the source-plane position $\vec{\beta_k}$ for each image $k$ using the lens equation and deflection map from the Hera simulation. We next compute the time delay $\Delta t_{ij} (\vec{\beta_k})$ between each pair of images $i$ and $j$ in the system given each of the $N$ source plane positions.
Consequently, for each pair of images, we have a set of $N$ estimates of $\Delta t_{ij} (\vec{\beta_k})$. Considering delays shorter than 6000\,days, we find that the average value of the standard deviation among the $N$ time delays $\Delta t_{ij} (\vec{\beta_k})$ is 377\,days and the  median value of the standard deviation is 309\,days.

For a fraction of systems of multiple images, the number of images $N$ is fewer than four, and the standard deviation of a small number of values $\Delta t_{ij} (\vec{\beta_k})$ can yield an underestimate of the actual uncertainty.  Therefore, when assigning an uncertainty, we use either the standard deviation of the time delays $\Delta t_{ij} (\vec{\beta_k})$ or the median value of 309\,days across all image pairs, whichever is greater. 

Second, the lensing potential of Hera is not available, so we reconstruct the lensing potential from the deflection field which requires numerical integration. Because line integrals through gradient fields should be path independent, we can use the disagreement between multiple paths to assess the error in the reconstructed potential. We first integrate along the RA axis and then the Dec axis, and then reverse the order to compute the potential.  These calculations show disagreement of $\sim 150$\,days in the strong lensing region for a source at $z=2$. 

The contributions of the inconsistency in the inferred source plane positions within each system of multiple images and in the lensing potential reconstruction from the deflection field combine to yield uncertainties of $\sim 300$\,days, on average for each time delay, although the uncertainty for each system associated with the source-plane position can exceed 500\,days. 

The softening length of the $N$-body simulations used for the Hera cluster is large, $2.3\,h^{-1}$\,kpc, which should result in less cuspy halos than is expected from high-resolution simulations. Motivated by higher-resolution dark-matter simulations, simply-parameterized modelers also used model halos with more cuspy halos than the Hera $N$-body simulation can yield.
If high-resolution dark-matter simulations indeed reproduce the true properties of cluster halos, the mismatch between the halos used by modelers and the Hera simulation may mean that the modeling codes may yield more preise constraints on $H_0$ than our comparison indicates.

\paragraph*{Astrometric error propagation:}
\label{sec:error_prop}

To first order, the fractional change in $H_0$ depends on the residual of the predicted from observed image $\delta \vec{\theta}_A$ \cite{birrertreu19}:

\begin{equation}
\frac{\delta H_0}{H_0} =  \frac{D_{\Delta t}}{c} (\vec{\theta}_B - \vec{\theta}_A) \cdot \frac{\partial \vec{\beta}}{\partial \vec{\theta}_A} \delta \vec{\theta}_A,
\end{equation}
\noindent
where $D_{\Delta t}$ is the time-delay distance, $c$ is the speed of light, $\vec{\theta}_B$ and $\vec{\theta}_A$ are the image-plane positions of images $A$ and $B$ (respectively), and $\vec{\beta}$ denotes the source-plane position. For each realization of the Markov chain from the Oguri-a* model, we compute the fractional change $\delta H_0 / H_0$ expected given the realization's predicted position for images S1 and SX. Computing the standard deviation of these residuals, we find that the precision with which the Oguri-a* model reproduces the separation of SX and S1 corresponds to a 5.5\% uncertainty in $H_0$, using first-order error propagation. Analysis of the Markov chain also shows that variation in the separation of images SX and S1 accounts for 92\% of the variance of the SX--S1 time delay. These calculations are consistent with our error budget.

\paragraph*{Gravitational lensing formalism:}
\label{sec:formalism} Galaxy-cluster lens models are constrained primarily by the positions of multiply-imaged background sources, and  are constructed (according to most methods) using the locations and properties of individual galaxies in the clusters to populate the lens plane with dark-matter halos.  The masses assigned to individual galaxies by lens models must account for both dark matter and baryonic matter. These are, in general, scaled according to the galaxies' measured stellar mass or velocity dispersion.

According to Fermat's principle, images of lensed sources form at the extrema of the time-delay surface. Given a  single gravitational lens at redshift $z_{\rm l}$, the time delay of a background source at redshift $z_{\rm s}$ appearing at a position $\vec{\theta}$ in the image plane on the sky is
\begin{equation}
t(\vec{\theta}) = C + {D_{\rm ol} D_{\rm os} \over D_{\rm ls} c} \tau(\vec{\theta}),
\label{eq:delaysurf}
\end{equation}
where $C$ is a constant, $\tau(\vec{\theta})$ is the Fermat potential, $D_{\rm ol}$ is the angular-diameter distance from the observer to the lens, $D_{\rm os}$ is the angular-diameter distance from the observer to the background source, and $D_{\rm ls}$ is the angular-diameter distance between the source and foreground lens \cite{perlick90a,perlick90b}. 

In Equation~\ref{eq:delaysurf}, the Fermat potential is the sum of a first term that accounts for the geometric delay, $(\vec{\theta} - \vec{\beta})^2$ , and a second term, the lens potential $\psi(\vec{\theta})$, which represents the combined effect of the curvature of space and gravitational time dilation,

\begin{equation}
\tau(\vec{\theta}) \equiv \left[  { (\vec{\theta}-\vec{\beta})^2 \over 2} -  \psi(\vec{\theta}) \right],
\label{eq:fermatpot}
\end{equation}

\noindent where $\vec{\beta}$ specifies the angular position in the source plane (which would be identical to the source position on the sky if the foreground lens were absent).

The lens potential 

\begin{equation}
\psi(\vec{\theta}) = \frac{1}{\pi}\int d^2 \theta^{\prime} \kappa(\vec{\theta}^{\prime}) \ln \mid\vec{\theta}-\vec{\theta}^{\prime}\mid
\end{equation}

\noindent depends on 

\begin{equation}
\kappa(\vec{\theta}) = \frac{\Sigma(D_{\rm ol}\vec{\theta})}{\Sigma_{\rm cr}},
\label{eq:kappaeq}
\end{equation}

\noindent which corresponds to the distribution of the projected mass density of the lens $\Sigma(D_{\rm ol}\vec{\theta})$ divided by the critical density,

\begin{equation}
\Sigma_{\rm cr}(z_{\rm s},z_{\rm l}) = \frac{c^2 D_{\rm os}}{4\pi G D_{\rm ls}D_{\rm ol}}.
\label{eq:critden}
\end{equation}
\noindent
Substituting Equation~\ref{eq:fermatpot}, the expression for the Fermat potential, into Equation~\ref{eq:delaysurf} yields
\begin{equation}
t(\vec{\theta}) = C +  (1 + z_{\rm l}) {D_{\rm ol} D_{\rm os} \over D_{\rm ls} c} \times \\ \left[  { (\vec{\theta}-\vec{\beta})^2 \over 2} -  \frac{1}{\pi}\int d^2 \theta^{\prime} \kappa(\vec{\theta}^{\prime}) \ln \lvert\vec{\theta}-\vec{\theta}^{\prime}\rvert \right].
\end{equation}

For a time-variable source that appears in two images, we can express the relative time delay
between the two images, $\Delta t_{\rm i_1 i_2}$, as
\begin{equation}
\Delta t_{\rm i_1 i_2} = \frac{D_{\Delta t}}{c} \Delta \tau_{\rm i_1 i_2}.
\label{eq:hnaught}
\end{equation}
where $\Delta \tau_{\rm i_1 i_2}$ is the difference in the Fermat potential at the images' positions, and the time-delay distance $D_{\Delta t}(z_{\rm l}, z_{\rm s})$ is given by
\begin{equation}
D_{\Delta t}(z_{\rm l}, z_{\rm s}) = (1 + z_{\rm l}) \frac{D_{\rm ol} D_{\rm os}}{D_{\rm ls}}.
\end{equation}
\noindent
The ratio of angular-diameter distances in the time-delay distance is inversely proportional to the value of 
$H_0$ and only weakly dependent on the values of other cosmological parameters, including $\Omega_{\rm M}$ and $\Omega_{\Lambda}$.

This formalism describes the scenario where there is only a single lens along the line of sight to the background source.  In practice, the light path for any source will be affected by many additional lensing potentials (other foreground galaxies). These may be accounted for by introducing external shear $\gamma$ and convergence $\kappa$, and through multiplane lens modeling \cite{mccullykeetonwong14,chirivisuyugrillo18}.

\paragraph{Rescaling time-delay predictions:} Here we consider how 
time-delay predictions made for $H_0 = 70$\,km\,s$^{-1}$\,Mpc$^{-1}$ and $\Omega_{\rm M} = 0.3$ can
be used to infer the value of $H_0$.  
A single-plane model is entirely defined by the distribution of projected mass on the sky, $\kappa(\vec{\theta}^{\prime})$, which
in turn deflects photons and retards their passage. 
Through the ratio of angular-diameter distances, the value of the critical density $\Sigma_{\rm cr}$ for sources at all redshifts is directly proportional to the value of $H_0$.  
From the definition of $\kappa$ in Equation~\ref{eq:kappaeq}, for a given Fermat potential, the cluster's mass density $\Sigma(D_d\vec{\theta})$  
can be rescaled by a constant factor to compensate for the  change in the value of $\Sigma_{\rm cr}$ arising from a different value of $H_0$. The ratio $\Sigma_{\rm cr}(z_s = z_1, H_0) / \Sigma_{\rm cr}(z_s = z_2, H_0)$ has no dependence on the value of $H_0$. 

Consequently, the $\kappa$ distribution is independent of the value of $H_0$ used to compute angular-diameter distances. Using a different value of $H_0$ will yield identical $\kappa$ maps and therefore lensing potentials and deflection fields.
For example, consider that $H_0 = 70$\,km\,s$^{-1}$\,Mpc$^{-1}$ is used to find the 
$\kappa(\vec{\theta}^{\prime})$ that best reproduces the positions of background images.
For $H_0 = 65$\,km\,s$^{-1}$\,Mpc$^{-1}$, $\Sigma_{\rm cr}(z_s,z_d)$ becomes $\sim 7.9$\% 
smaller, and $\Sigma(D_d\vec{\theta})$ can also be scaled by 7.9\%
to obtain a $\kappa(\vec{\theta}^{\prime})$ distribution and image positions that are identical.

Therefore, the $\kappa(\vec{\theta}^{\prime})$ distribution that yields any given set of image positions is, under the assumptions listed above, independent of the value of $H_0$ used to construct the model. An implication is that, for the model predictions given an assumed $H_0 = 70$\,km\,s$^{-1}$\,Mpc$^{-1}$, % collected by \cite{treubrammerdiego16}, 
the only term in Equation~\ref{eq:hnaught} that has a dependence on $H_0$ is the time-delay distance. It is therefore possible, in principle, to infer $H_0$ using a measured time delay by simply rescaling the predicted delays by 1/$H_0$. For the models we consider, this factor is (70\,km\,s$^{-1}$\,Mpc$^{-1}$)/$H_0$, as described in Equation~\ref{eq:rescale}.  % \cite{treubrammerdiego16}

Line-of-sight structure has been taken into account for a small set of analyses of galaxy-scale lenses, with small image separations, when its has been detected in HST imaging \cite{wongsuyuauger17,shajibbirrertreu20}. The deep Hubble Frontier Field imaging of this cluster
reveals no concentration of galaxies aside from the cluster up to the source plane at $z = 1.49$. A group or cluster with mass even a small fraction of the mass of the MACS1149 cluster [($1.4 \pm 0.3) \times 10^{15}$\,M$_{\odot}$; \cite{vdlallen14,kellyvonderlinden14,applegatevdl14}] would be easily detected. Even if such a structure existed, a predicted value of a time delay can be rescaled inversely with the Hubble constant given multiple planes. All distances will be rescaled by 1/$H_0$, and the mass densities at each redshift can be scaled by $H_0$ to preserve the convergence maps and therefore the reduced deflection angles.

The published uncertainty estimates of the pre-reappearance models did not all consider the possibility of a constant mass sheet at the cluster redshift, or scatter in the cluster-member scaling relation, which introduces a small amount of additional uncertainty \cite{grillorosatisuyu20}.
Moreover, including multiple relative time delays as constraints on the lens model could, in principle, affect the potential and therefore the value of $H_0$.
In the case of SN Refsdal, however, only the delay of SX relative to S1--S4 is constrained with high fractional precision.

This conclusion pertains to the published model predictions made for SN Refsdal that we consider in this paper, but more generally the $\kappa(\vec{\theta}^{\prime})$ distribution can, in the cases of other models constructed differently, have a dependence on the value of $H_0$. For example, if there is an independent constraint on $\Sigma(D_d\vec{\theta})$ from stellar kinematics without a free multiplicative factor that is a parameter of the fit, then $\kappa(\vec{\theta}^{\prime})$ cannot be rescaled as described above. 
We have shown, using simple mathematical arguments, that the predicted time delay can be used to infer $H_0$ with sufficient precision under the assumption that $\Omega_{\rm M}=0.3$ (or is sufficiently close), and multiplane lensing effects are not appreciable.

\paragraph*{Measured time delay does not affect inferred Fermat potential:} %We note that the inclusion of the single constraining relative time delay SX--S1 cannot be expected to affect the lens models' Fermat potential. Specifically, 
A probability distribution for the difference in the Fermat potential between two images, $\Delta \tau_{\rm A,B}$, can be computed using the image positions of strongly lensed sources alone. For a given value of $\Delta \tau_{\rm A,B}$, the predicted relative time delay $\Delta t_{\rm A,B}$ depends on the value of $H_0$. Consequently, if no assumption is made about the value of $H_0$, and $\Omega_{\rm M}$ and $\Omega_{\Lambda}$ are held fixed, a single meaningfully constrained time delay, such as the SX--S1 delay in the case of SN Refsdal, does not provide a constraint on the cluster's Fermat potential by itself.

\paragraph*{Effects of varying $\Omega_{\rm M}$ and $\Omega_{\Lambda}$ simultaneously with $H_0$:}
Next, we consider the effect of uncertainty in $\Omega_{\rm M}$ and $\Omega_{\Lambda}$ on the inference about $H_0$ using SN Refsdal.
Changing $\Omega_{\rm M}$ and $\Omega_{\Lambda}$ within the range allowed by other cosmological probes causes a small change in the ratios of the critical densities known as family ratios,
and therefore changes the shape of the Fermat surface.
Because $H_0$ factors out of the family ratio, any covariance between $H_0$ and the parameters $\Omega_{\rm M}$ and $\Omega_{\Lambda}$ is likely to be small for small changes in these parameters [\cite{grillorosatisuyu18}, their figure 3]. 

The Grillo models of the MACS\,J1149 galaxy cluster lens used the positions of multiply-imaged galaxies and a series of multiple possible values of the still-blind SX--S1 time delay spanning 315--375\,days \cite{grillorosatisuyu20}. In their calculations, they allowed both $H_0$ and $\Omega_{\rm M}$ to vary simultaneously, and found residuals of only 0.7\,km\,s$^{-1}$\,Mpc$^{-1}$ from linear rescaling with $H_0$ given a slope of $-1$ [\cite{grillorosatisuyu20}, their figure 1].
Their analysis, unlike the model predictions we consider in this paper, additionally included pre-reappearance time-delay estimates within the quad (S2--S1, S3--S1, and S4--S1) \cite{rodneystrolgerkelly16}, which we expect to generate nonlinearity in the relationship between the measured time delay and the inferred value of $H_0$. Consequently,  $\sim 0.7$\,km\,s$^{-1}$\,Mpc$^{-1}$ is likely to be an upper limit in the uncertainty resulting from holding $\Omega_{\rm M}=0.3$ fixed.

\paragraph*{Computation of time delays from published online models:} 
In Tables~\ref{tab:mod_time_delays} and \ref{tab:mod_mag_ratios}, we specify whether the relative time delays and magnifications for each model were calculated using the predicted or observed locations of the images S1--SX of SN Refsdal. 
To investigate the effect of using predicted or observed image positions, we identified publicly available models of MACS\,J1149 for which either (a)  high-resolution maps were available (pixel sizes smaller than $\sim 0.05''$, consisting of Grillo, Keeton, Oguri GLAFIC v3, Sharon v4cor), or (b) for which we could reconstruct the exact model from a list of model parameters (Jauzac Clusters As TelescopeS (CATS) v4.1).  

The Oguri GLAFIC v3 model \cite{hfflensmodels} that we use for this exercise corresponds to the pre-reappearance Oguri-a  model, where SX's position has not yet been included as an additional constraint. The Jauzac CATS v4.1 model differs from that used for the Jauzac15.2 predictions adopted in this paper, because the v4.1 model uses an updated cluster-member catalog. The Keeton* model was constructed after SN Refsdal's reappearance, and adopted a weak prior on the SX--S1 time delay of $345 \pm 45$\,days. The Sharon v4cor model is also similar to, but not identical to, that used for the pre-reappearance predictions, and it includes an additional spectroscopic redshift for strongly lensed galaxy labelled source 9 \cite{jauzacrichardlimousin16}.  The Grillo lens model corresponds to that used for the Grillo-g predictions.

In the case of the Jauzac CATS V4.1 model, we construct the lensing potential using the {\tt LENSTOOL} parameters available \cite{hfflensmodels}, and we compute the deflection maps numerically. For the Grillo model, we use the deflection maps \cite{grillomodel} to derive an approximation of the lensing potential. We use bilinear interpolation to estimate values from pixelized maps.

For this set of five models, we identified the source-plane position for the SN that minimized the residuals of the predicted image-plane positions and the observed image-plane positions. While our calculated time delays are within $\sim 2$\% of the predictions computed by the modelers where those are available, we do not expect them to agree exactly, because our calculations require interpolation across a pixel grid and reconstruction of the deflection field or potential, which will result in small errors. Nonetheless, the maps and model parameters allow us to approximate closely the exact models, and enable us to evaluate the use of observed as opposed to predicted image positions.

To identify the source-plane position that minimizes the residuals between the predicted and observed images, we use a naive Monte Carlo search algorithm within a circumscribed region of interest. The size of the region of interest, by design, decreases with each iteration until a source-plane point that minimizes the image-plane root-mean-square residuals is identified.

Table~\ref{tab:td_pos_choice} shows that, for the Grillo and Oguri GLAFIC v3 models, the use of the observed as opposed to predicted image position results in a change in the SX--S1 time delay of two days or less. Consequently, we conclude that the choice of using the predicted image by those teams is negligible. The only model that results in a larger shift (6.4\,days) is the Jauzac CATS v4.1 model, but the CATS models of MACS\,J1149 have larger RMS image-plane scatter than is the case for the Grillo or Oguri models.
All computed time delays for all models are smaller than 360\,days.

In Table~\ref{tab:mag_pos_choice}, we show that the choice of predicted versus observed image positions has a greater effect on the inferred magnification.  The agreement among models' values for the magnification, however, improves substantially when the predicted position is used. Using predicted positions provides a self-consistent method for computing time delays and magnifications.

\paragraph*{Groups of parameters contributing to error budget:}
In Fig.~\ref{fig:errorbudget} and Table~\ref{tab:error_budget}, we show the error budget for our measurement of $H_0$. 
The error budget includes the contribution of the relative time delay measurement to the error budget. As listed in Table~\ref{tab:modweights}, the Oguri models receive the greatest weight in our estimates of $H_0$, so we use the MCMC chain for the Oguri-a* model to quantify the contribution of the lens model to the error budget. 

The parameter group listed as ``SN Position'' in Table~\ref{tab:error_budget} includes the source-plane coordinates of the SN.
``Model Perturbations'' includes both external shear ($\gamma$, $\theta_{\gamma}$) and a multipole Taylor perturbation ($\epsilon$, $\theta_{\epsilon}$, $m$, $z$) of the form $\phi = (C/m) r^z \cos(\theta - \theta_{\epsilon})$ centered at the brightest cluster galaxy (BCG).
``Early-type Galaxy'' includes parameters associated with the cluster member responsible for the Einstein-cross images S1--S4, which are velocity dispersion $\sigma$ and truncation radius $r_{\rm trun}$.  
``Cluster Members'' includes the three parameters ($\sigma_{\star}$, $r_{\rm trun}$, and $\eta$) which describe both the scaling relation, $\sigma / \sigma_{\star} \propto L^{1/4}$, between cluster galaxies' velocity dispersions $\sigma$ and {\rm F814W} luminosities $L$ \cite{hfflensmodels}, as well as that between $\sigma$ and the halo truncation radius ($r_{\rm trun} / r_{{\rm trun}, \star} \propto L^{\eta}$). ``Cluster Halos'' comprises the mass, ellipticity $e$, position angle $\theta_e$, concentration $c$, and $x$ and $y$ position coordinates. The category ``Photometric Redshifts'' consists of redshifts of multiple images for which no spectroscopic redshift is available, and which were therefore optimized together with model parameters with photometric redshift prior distributions. 
The cluster lens model parameters are described in more detail elsewhere [\cite{kawamataoguriishigaki16}, their table 5].

\paragraph*{Contributions to variance of SX--S1 time delay:}
We use the the Oguri-a* model's MCMC chain to evaluate each parameter group's contribution to the variance of the model's predicted SX--S1 time delay $\Delta t_{\rm X,1}$, as follows:
\begin{equation}
\operatorname{Var}(\Delta t_{\rm X,1}) = \frac{1}{n_{\rm V}} \sum_{i=1}^{n_{\rm V}} (\Delta (t_{\rm X,1})_i - \mu)^2,
\end{equation}
\noindent
where the variable $i$ indexes the $n_{\rm V}$ samples in the Oguri-a* MCMC chain, and $\mu$ is the mean value of $\Delta (t_{\rm X,1})_i$ samples,

\begin{equation}
\mu = \frac{1}{n_{\rm V}} \sum_{i=1}^{n_{\rm V}} \Delta (t_{\rm X,1})_i. 
\end{equation}
\noindent
The sum of the squares of the residuals of $(\Delta t_{\rm X,1})_i$ from its mean value $\mu$ across all of the MCMC chains is 
\begin{equation}
\begin{aligned}
{\rm SS}_{\rm total} & = n_{\rm V} \times \operatorname{Var}(\Delta t_{\rm X,1}) \\
& = \sum_{i=1}^{n_{\rm V}} ((\Delta t_{\rm X,1})_i - \mu)^2.
\end{aligned}
\end{equation}

We next describe our model of $(\Delta t_{\rm X,1})_i$ as a function of lens model parameters.
We first ``center'' the values $p_{l,i}$ of each model parameter $p_l$ by subtracting the mean value $\bar{p_l}$: 
\begin{equation}
x_{l,i} = (p_{l,i} - \bar{p}_l).
\end{equation}
\noindent
We model $\Delta t_{\rm X,1}$ using a second-order multiple polynomial regression without an intercept,
\begin{equation}(
\Delta t_{\rm X,1})_i^{\rm pred} = \sum_{l=1}^{m} \eta_{l} \times x_{l,i}  + \zeta_{l} \times x_{l,i}^2 ,
\label{eq:modvar}
\end{equation}
where $\eta = (\eta_1,\ldots,\eta_m)$ and $\zeta = (\zeta_1,\ldots,\zeta_m)$ are vectors of coefficients that correspond to the set of $m$ lens model parameters. Through linear optimization, we identify the sets of coefficients $\eta$ and $\zeta$ that minimize the sum of the square of the residuals,
\begin{equation}
\underset{\vec{\eta},\vec{\zeta}}{\mathrm{arg~min}} \sum_{i=1}^{n_{\rm V}} \left(\vec{\eta} \cdot \vec{x_i} + \vec{\zeta} \cdot \vec{x_i^2} - (\Delta t_{\rm X,1})_i\right)^2.
\end{equation}

We next compute the residuals $\delta_i$ of the predicted time-delay values from those in the MCMC chain,
\begin{equation}
\delta_i = \vec{\eta} \cdot \vec{x_i} + \vec{\zeta} \cdot \vec{x_i^2} - (\Delta t_{\rm X,1})_i ,
\end{equation}
and calculate the sum of the square of the residuals,
\begin{equation}
{\rm SS}_{\rm model} = \sum_{i=1}^{n_{\rm V}} \delta_i^2.
\end{equation}
\noindent
For any given set of parameters, we compute the fractional reduction in variance, 
\begin{equation}
r = 1 - \frac{{\rm SS}_{\rm model}}{{\rm SS}_{\rm total}}.
\end{equation}

Our goal is to compute the reduction in the sum of the squares of residuals that occurs when we include each group of parameters. We have groups of parameters indexed by $p$, so after we add each new set of parameters, we record the change in $r$, which we designate $\delta r_p$. Owing to covariance among the parameters, the value of $\delta r_p$ depends on the order in which we incorporate the parameter groups. %affects the value of $\delta r_p$. 
Consider, for example, parameters A and B which are perfectly correlated (or even identical).
Let us, hypothetically, consider that we add parameter A first, and it reduces the variance by 20\%. In this case, subsequently adding parameter B would yield no reduction of the variance. However, if we switch the order in which we include parameters A and B, then B would account for 20\% and A would account for none. 

Consequently, we identify all possible permutations of the groups of parameters. We next compute the fractional change in the variance, $\delta r_{p,g}$, for each group of parameters and each permutation $g$ of groups. We then calculate the mean value of $\delta r_{p,g}$ across all permutations. This average value corresponds to a representative estimate of the variance due to each parameter, which also accounts for the covariance among parameters. The multiple polynomial regression is able to account for 99.3\% of the total variance, so all major contributions to the error budget.

\paragraph*{\large Supplementary Text}

\paragraph*{Position of SX in updated Oguri models:}
\label{sec:sx_positions}
From the MCMC chains, we find J2000 coordinates for image SX of $\alpha = 11^h 49^{\rm m} 36.0091^{\rm s}$ and $\delta = 22^{\circ} 23' 48.15''$ for the Oguri-a* model with uncertainty of 0.065$''$ for each coordinate, and $\alpha = 11^h 49^{\rm m} 36.006^{\rm s}$ and $\delta = 22^{\circ} 23' 48.24''$ for the Oguri-g* model with uncertainty of 0.069$''$ and 0.081$''$ for the respective coordinates. These are not used to weight models, but we include them for comparison. When weighting models, we use the pre-reappearance positions predicted by the unrevised Oguri-a and Oguri-g models.

\paragraph*{Evidence for models given $H_0$ priors:}
\label{sec:evidence_given_H0}
We next use existing measurements of $H_0$ to compare model predictions for SN Refsdal with observations. Given the Hubble tension, we choose two possible priors for $H_0$ bracketing the range of values: (i) the one from the CMB as measured by {\it Planck} \cite{planckhubble18} in a $\Lambda$CDM cosmology, and (ii) the one from the local distance ladder as measured by the SH0ES team \cite{riessyuanmacri22}.

In Table~\ref{tab:posteriorprobs}, we list the posterior probabilities for the models without a prior. In Table~\ref{tab:posteriorprobswithpriors}, with the {\it Planck} prior and the SH0ES prior, and 
Table~\ref{tab:modweightswithpriors} shows these same posterior probabilities normalized such that their weight is equal to 1. 
Together, the Oguri-a*, Oguri-g*, and Grillo-g models, all of which are simply-parameterized models, account for $> 99$\% of the posterior probability for all assumptions.

The Oguri models are penalized by the $>1\sigma$ tension between the predicted and actual position of SX.  Fig.~\ref{fig:delay_combined} compares the time-delay and magnification predictions with our constraints. These show that the Oguri predictions for images S1--S4 have, in general, substantially less estimated uncertainty than the Grillo-g predictions. The predictions of both models are consistent with the constraints that we find, with the exception of S4 which is inferred to have experienced an extreme microlensing event \cite{paperone}. 
 
The posterior probabilities for the models that are not simply-parameterized, which include the Diego-a free-form model and Zitrin-c LTM model, receive little weight. While the 2016 Zitrin-c model \cite{treubrammerdiego16} was affected by a problem affecting the time-delay surface, a corrected version called Zitrin2020-ltm \cite{zitrinrevised21} also has a posterior %probabilities listed in Table~\ref{tab:posteriorprobsrecent} that are 
much smaller than those of the simply-parameterized models.
These results suggest that the simply-parameterized models provide the most observationally favored representation of the distribution of dark matter in the galaxy cluster.

\paragraph*{Other lens models that did make a prediction:}
As described in Table~\ref{tab:hfflensmodels}, 
three additional models were available \cite{hfflensmodels} prior to the appearance of SX but were not used to make a prediction or post-reappearance calculation. 
For the Bradac {\tt SWUnited} code \cite{bradactreuapplegate09}, matter distributions can contain discontinuities and also yield numerically imprecise solutions, and such discontinuities could affect the calculation of the Fermat potential. 
As listed in Table~\ref{tab:pixscale_models} and shown in Fig.~\ref{fig:merten_resolution}, the Merten {\tt SaWLens} model has a resolution of $8.3''$ per pixel, which is similar to the SX--S1 separation and precludes an accurate time-delay calculation.

The post-reappearance Keeton* model used the early and approximate measurement of the SX--S1 delay \cite{kellyrodneytreu16} as a constraint. 
The SX--S1 time delay we calculate from the Keeton* model \cite{hfflensmodels} is 340.0 days, within
the 1$\sigma$ uncertainties of the simply-parameterized models given in Table~\ref{tab:mod_time_delays}. We list all of the relative time delays and magnifications that we calculate in Tables~\ref{tab:td_pos_choice} and \ref{tab:mag_pos_choice}. We cannot compute an uncertainty from the maps \cite{hfflensmodels}, since they do not include the potential or deflection field. We instead used an average of the Grillo and Oguri models' uncertainties, when we include the Keeton* predictions to obtain the constraints shown in Fig.~\ref{fig:empirical_prior_positions_through_2020}. %using available models through the present. %When we include the ``Keeton" model, we obtain $H_0 = 65.0_{-4.5}^{+3.5}$\,km\,s$^{-1}$\,Mpc$^{-1}$.

The Williams* {\tt Grale} model \cite{williamsliesenborgs19}, which employs only image positions as constraints, was used to make a calculation of the time delay in 2019.
This model does not reproduce the Einstein cross and has no connection between the light and mass which leads to extremely large uncertainties \cite{williamsliesenborgs19}.

\paragraph*{Other models:} Table~\ref{tab:posteriorprobsrecent} also lists the posterior probabilities with a prior on $H_0$ for Chen2020 \cite{chenkellywilliams20} and Zitrin2020-p, a grid-based simply-parameterized model \cite{zitrinrevised21}.
Table~\ref{tab:posteriorprobsrecentwithpriors} lists the posterior probabilities for these models with the CMB and SH0ES $H_0$ priors.
These models were not produced before the appearance of SX and their mass maps have not been posted with the HFF models \cite{hfflensmodels}, so were not used in our analysis. As described in the main text, the Chen2020 model implements a very different approach to lens modeling, yet predicts an SX--S1 time delay consistent with those of the simply parameterized models (Table~\ref{tab:mod_time_delays}).

\paragraph*{Revision to Zitrin model:}
The original, LTM pre-reappearance prediction, dubbed Zitrin-g, for the  relative SX--S1 time delay was $224_{-262}^{+262}$\,days for $H_0 = 70$\,km\,s$^{-1}$\,Mpc$^{-1}$ \cite{treubrammerdiego16}. A revised calculation, Zitrin-c*, recomputed after the reappearance but without addition of information about the reappearance of $37_{-11}^{+122}$\,days 
was made after setting the mass of the early-type lens galaxy to be a free parameter in the model fitting, which allowed the critical curve to pass among the four images S1--S4 of SN Refsdal detected in 2014. However, subsequent analysis has found that the calculations of the Zitrin-g and Zitrin-c* time
delays were affected by a computational problem which became apparent after inspection of the
time-delay surface. After correction, the time delay became
 $267.4_{-4.3}^{+37.9}$\,days \cite{zitrinrevised21}.

\setlength\tabcolsep{3pt}

\begin{figure*}%[htp!]
\includegraphics[width=6.5in]{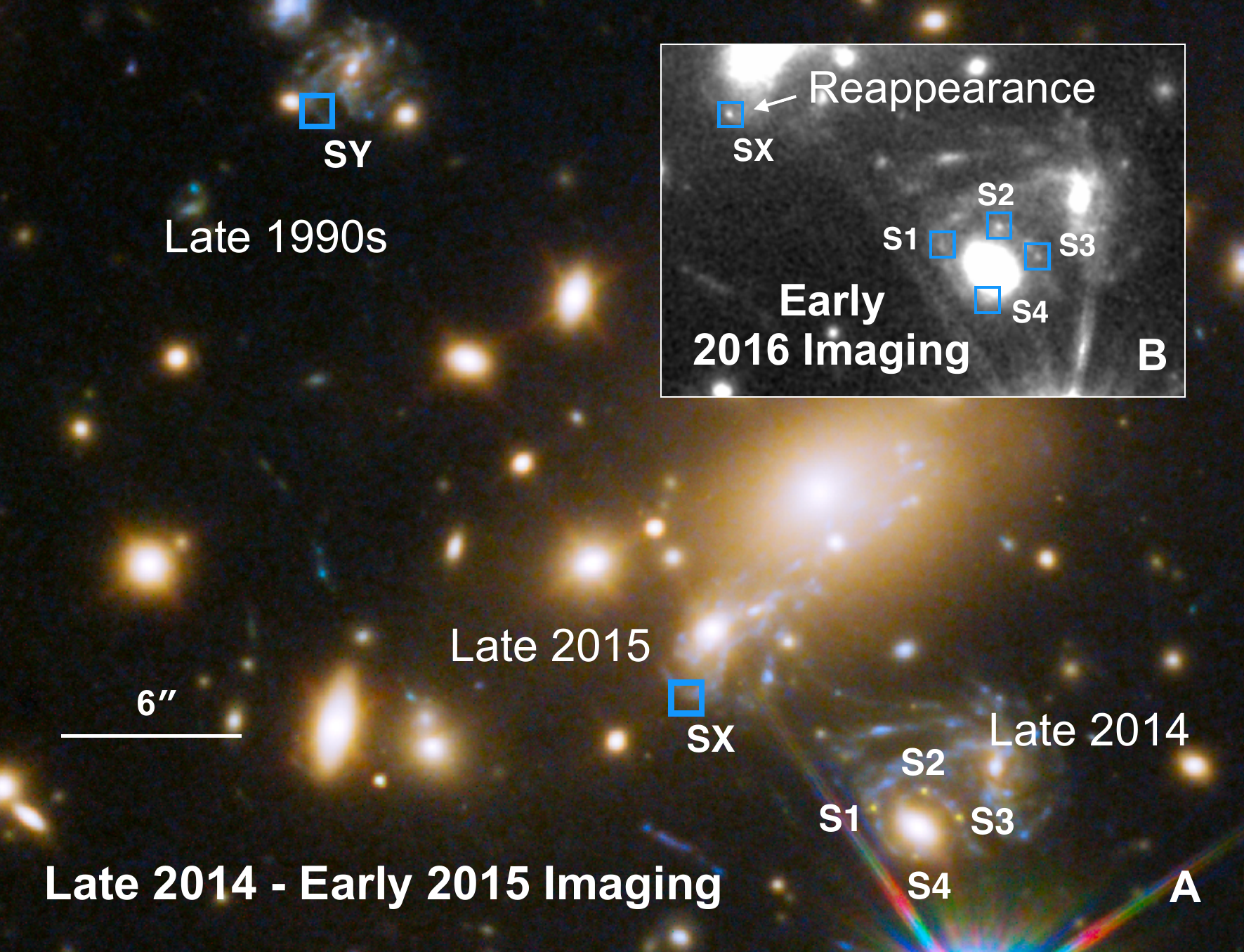}
\caption{ {\bf Configuration and arrival times of multiple images of SN Refsdal in the MACS\,J1149 galaxy-cluster field.} (A) shows the three complete images of SN Refsdal's host galaxy adjacent to labels ``Late 1990's,'' ``Late 2014,'' and ``Late 2015'' produced by the foreground MACS\,J1149 galaxy-cluster lens.
According to models, the SN first appeared as image SY at the upper left in the late 1990s, but deep imaging of the field was not acquired at that time. In November 2014,  images S1--S4 were observed in an Einstein-cross configuration \cite{kellyrodneytreu15}.
Modeling teams predicted that the SN would reappear for a final time in image SX, which was detected in late 2015 \cite{kellyrodneytreu16}.
The red, blue, and green channels in the pseudocolor image are assigned HST imaging of the field as follows: coaddition of F606W ({\it V}) and F814W ({\it I}) imaging (blue), F105W ({\it Y}) and F125W ({\it J}) (green), and F140W ({\it JH}) and F160W ({\it H}) (red).
(B) shows a coadded F125W image of the field showing the reappearance of the SN in image SX. }
\label{fig:mosaic}
\end{figure*}

\begin{figure*}%[htp!]
        \includegraphics[width=6.5in]{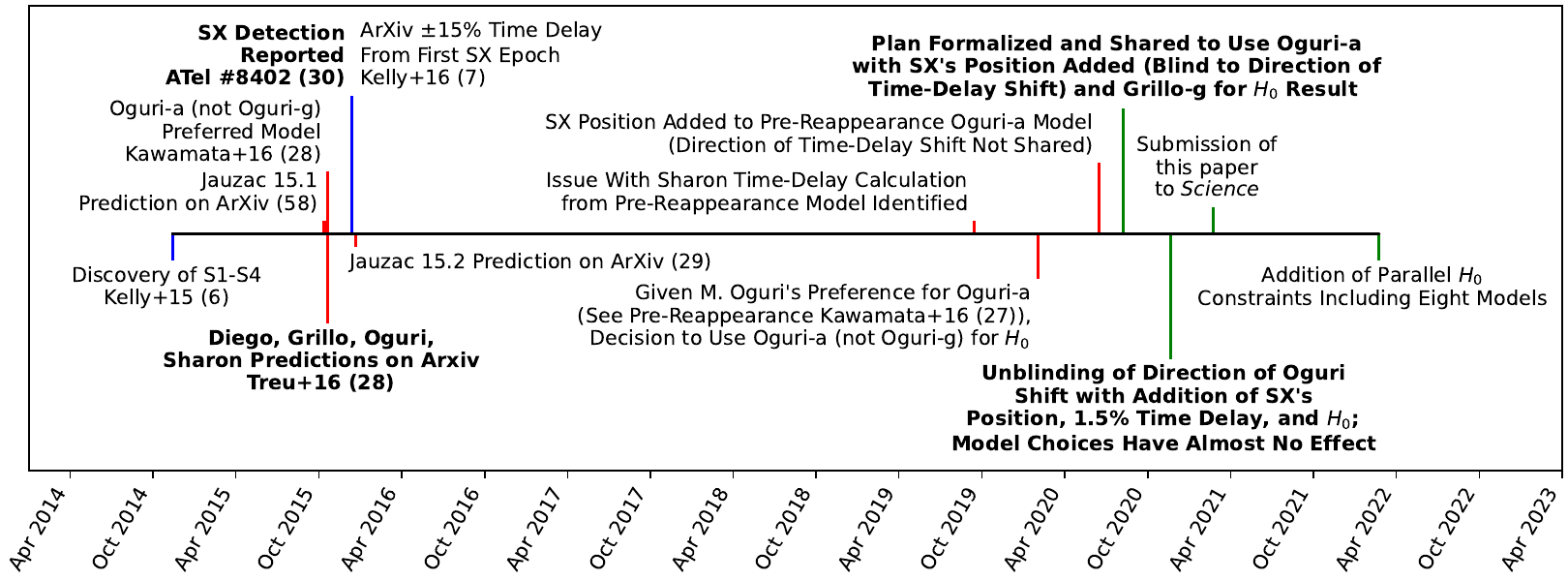}
\caption{ {\bf Timeline of model predictions and revisions, observations, and unblinding of the time delay measurement.} We do not include the Zitrin model updates in this figure, as the Zitrin predictions receive minimal weight in our $H_0$ constraints.
}
\label{fig:timeline}
\end{figure*}

\begin{figure}
\centering
\includegraphics[width=6.25in]{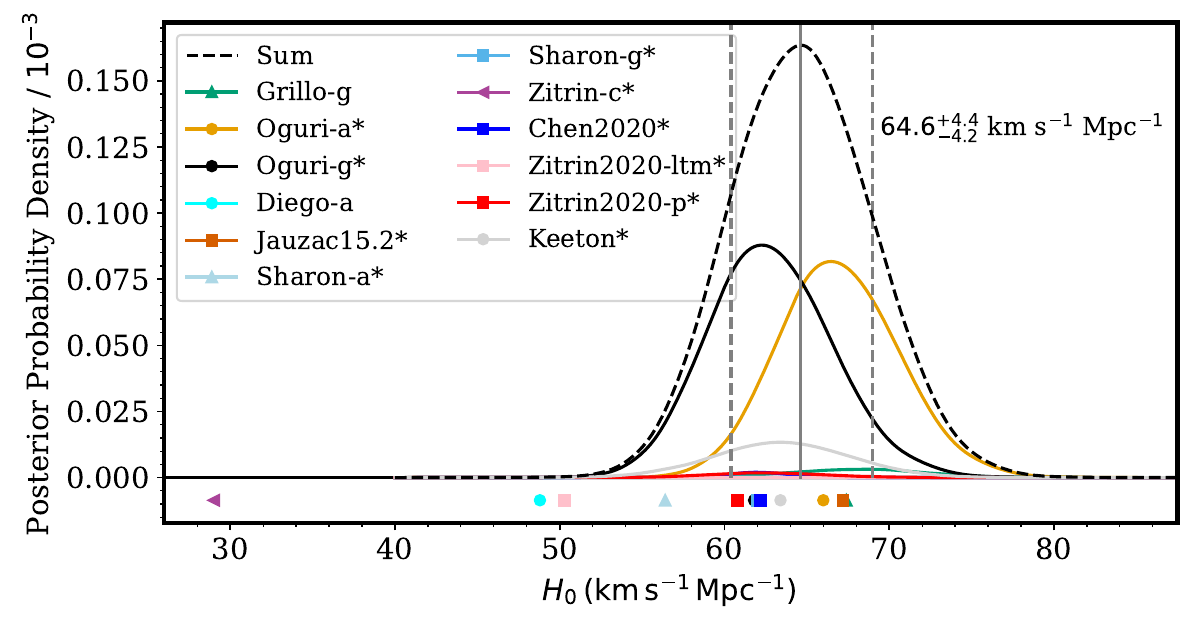}
\caption{{\bf Same as Fig.~\ref{fig:empirical_prior_positions_qualonly}, but including additional models.}}
\label{fig:empirical_prior_positions_through_2020}
\end{figure}

\begin{figure*}%[htp!]
\centering
\includegraphics[width=3.5in]{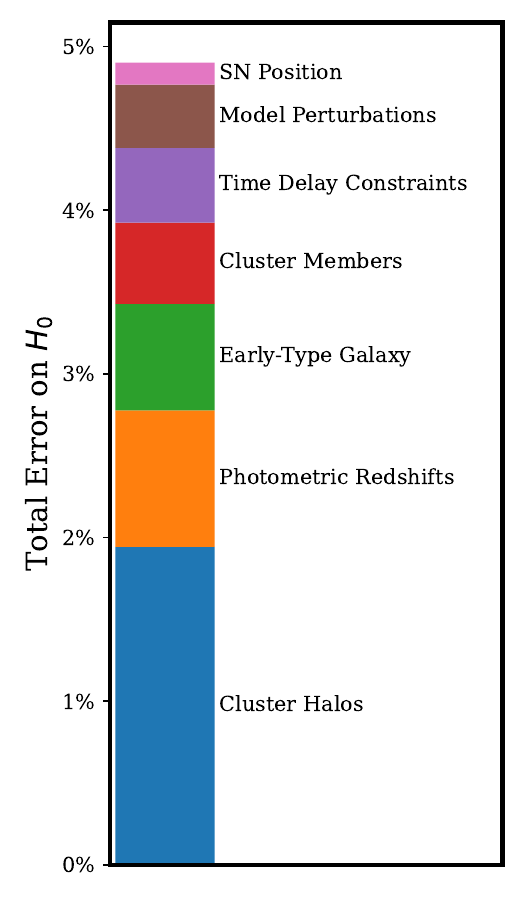}
\caption{ {\bf Error budget on constraints on $H_0$ using the  Oguri-a*  model, which receives the greatest weight in our combined estimate.} The fraction attributed to each term is proportional to its contribution to the total variance. All uncertainties listed, with the exception of the time-delay measurement, are associated with the lens model. Parameters in the lens model can have covariance with other parameters. Terms in the error budget are listed in Table~\ref{tab:error_budget}, see text for details. The uncertainties associated with the time-delay constraints include those arising from millilensing and microlensing.
\label{fig:errorbudget}}
\end{figure*}

\begin{figure*}%[htp!]
\centering
\includegraphics[width=5.5in]{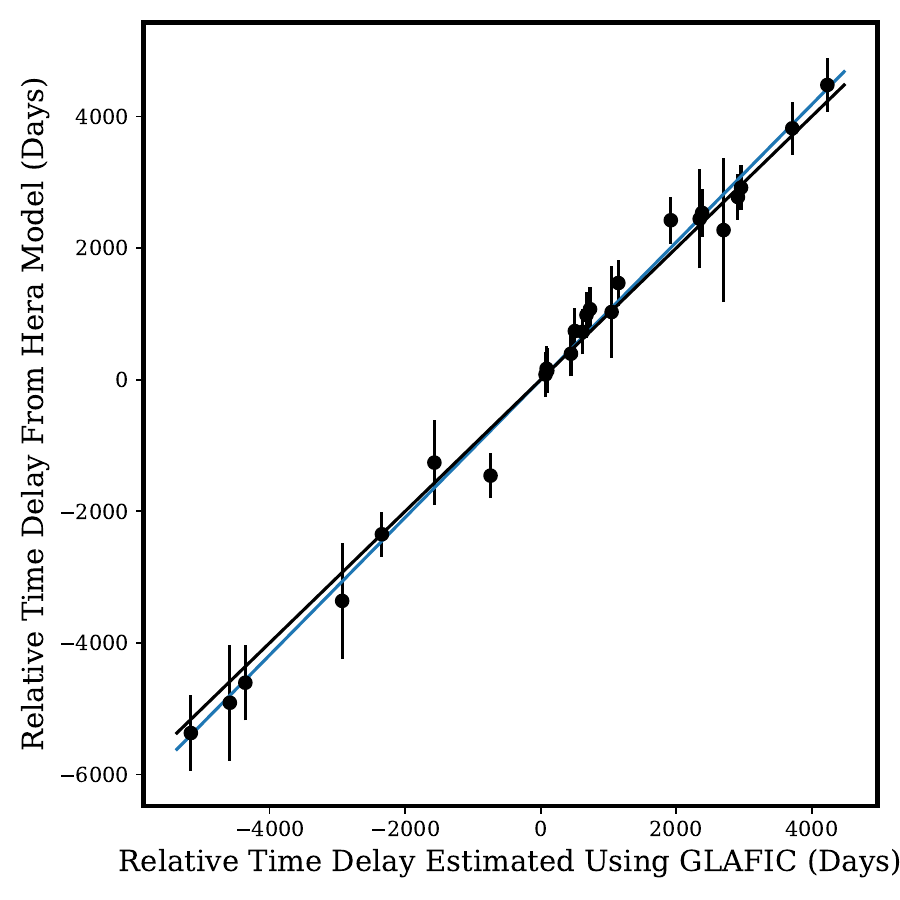}
\caption{ {\bf Comparison between the relative time delays of multiply-imaged galaxies estimated from the Hera simulation \cite{meneghettinatarajancoe17} and those recovered using the Oguri {\tt GLAFIC} lens-modeling code.} 
The vertical error bars on the circular points show the uncertainty arising from the disagreement source-plane positions recovered from images of the same galaxy in the Hera simulation. The blue line plots a one-to-one relation, and the black line shows the best-fitting linear model, which has a slope of $1.046 \pm 0.021$. This implies that the {\tt GLAFIC} Oguri code is able to constrain the value of $H_0$, which is inversely proportional to the time delay, with at least $\sim 5$\% accuracy, consistent with the error budget we estimate for $H_0$. Error bars indicate 1$\sigma$ uncertainties arising from inconsistencies in the inferred source-plane position, and from the reconstruction of the lensing potential.
\label{fig:hera_glafic_comparison}}
\end{figure*}

\begin{figure*}%[htp!]
\centering
\includegraphics[width=6.0in]{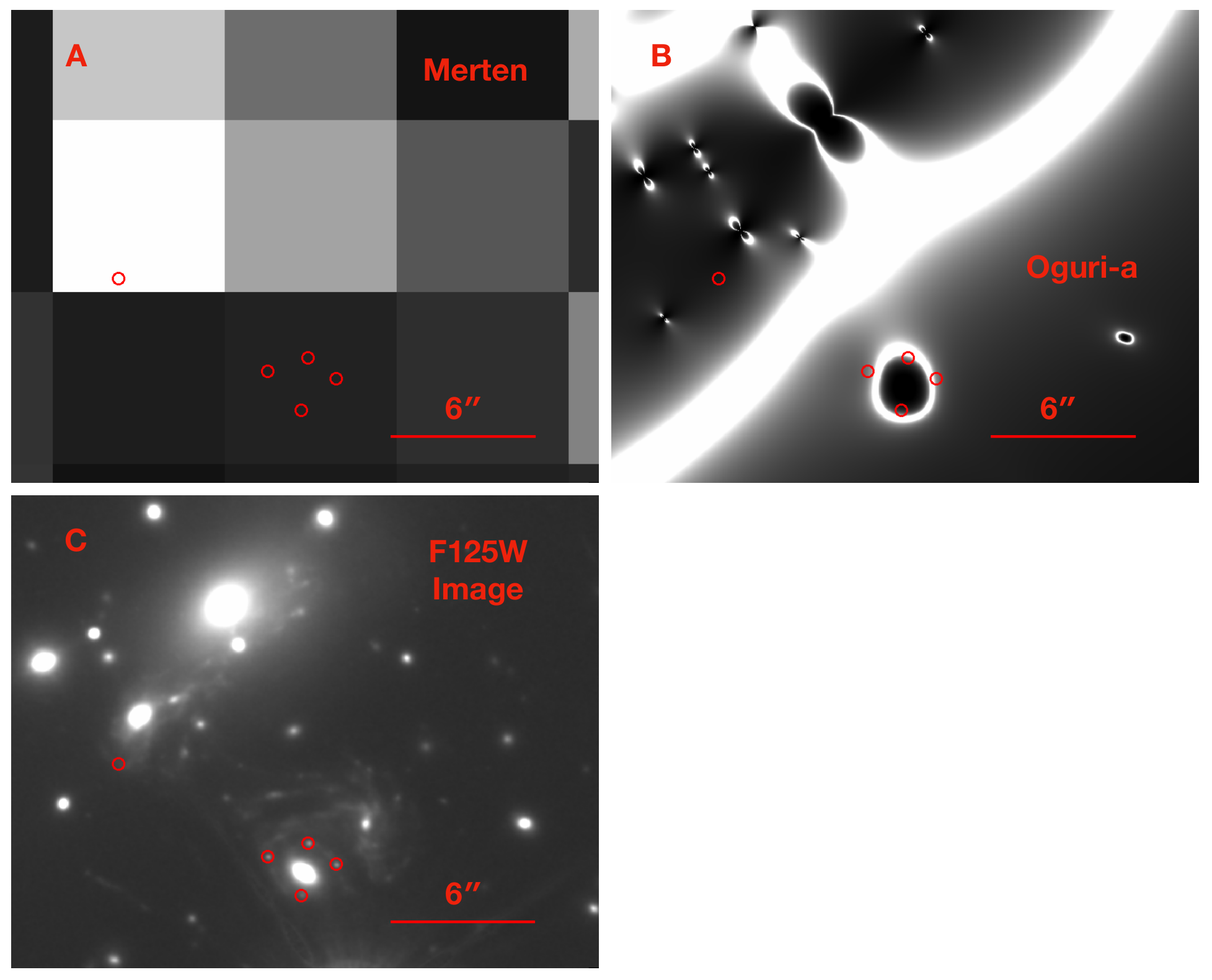}
\caption{ {\bf Resolution of the Merten model of the MACS\,J1149 cluster \cite{hfflensmodels}.} (A) and (B) show the magnification maps computed from the Merten and Oguri-a models, respectively. (C) plots the coadded F125W imaging of the galaxy cluster field. The Merten model, as listed in Table~\ref{tab:pixscale_models}, has a resolution of $8.3''$ per pixel. For comparison, the Oguri-a model has a resolution of $0.03''$ per pixel, which is a factor of 276 greater. The low resolution of the Merton model precludes its use in calculation of time delays.   
\label{fig:merten_resolution}}
\end{figure*}

\pagebreak

\begin{table*}[tpb]
\centering
  \caption{ {\bf Coordinates of the five detected images of SN Refsdal.} The coordinates are in the astrometric system of the coadded imaging from the Cluster Lensing And Supernova survey with Hubble (CLASH) program \cite{clashimages,postmancoebenitez12}. We have measured the positions from coadditions of F125W imaging acquired in 2015 and 2016.}
  %\tablecomments{Coordinates of the images are in the astrometric system used for the CLASH, GLASS, and HFF images and catalogs \url{http://www.stsci.edu/hst/campaigns/frontier-fields/}.}
  \begin{tabular}{c|cc}
  Image & R.A. & Decl.\\
 \hline
  S1 & 11$^{\rm h}$49$^{\rm m}$35.575$^{\rm s}$ & 22$^{\circ}$23$'$44.25$''$ \\
  S2 & 11$^{\rm h}$49$^{\rm m}$35.454$^{\rm s}$ & 22$^{\circ}$23$'$44.82$''$ \\
  S3 & 11$^{\rm h}$49$^{\rm m}$35.370$^{\rm s}$ & 22$^{\circ}$23$'$43.92$''$ \\
  S4 & 11$^{\rm h}$49$^{\rm m}$35.474$^{\rm s}$ & 22$^{\circ}$23$'$42.63$''$ \\
  SX & \rasx & \decsx \\
  \end{tabular}
\label{tab:positions}
\end{table*}

\begin{table}
\small
\caption{\label{tab:hfflensmodels} {\bf Models of the MACS\,J1149 galaxy cluster and predictions and revisions for the relative time delay of the reappearance of SN Refsdal.} We did use the Williams model which made a highly uncertain prediction, those which were not used to make predictions, or those which were first produced (as opposed to revised) after SX's reappearance.
We list all models posted to the HFF Lens Model website \cite{hfflensmodels} whose authors did not publish a prediction, or made only an extremely approximate prediction, given minimal model assumptions \cite{williamsliesenborgs19}. The ``Methods'' column describes whether models are, in broad terms, simply-parameterized (SP), free-form (FF), or light-traces-mass (LTM).}
\begin{tabular}{cccccc}
\hline
Used for & Model & Group & Method & SX & Post-Reappearance  \\
$H_0$ Inference &  &       &        & Prediction &   \\
\hline
\multicolumn{6}{c}{HFF Lens Models with Published Predictions} \\
\hline
Yes &  Diego-a & {\tt WSLAP+} & FF & \cite{treubrammerdiego16} & ... \\
Yes & Grillo-g & {\tt GLEE} & SP & \cite{treubrammerdiego16} & ... \\
\multirow{2}{*}{Yes} & \multirow{2}{*}{Oguri-g*} & \multirow{2}{*}{\tt GLAFIC} & \multirow{2}{*}{SP} & \multirow{2}{*}{\cite{treubrammerdiego16}} & This Paper; SX Position Added; \\
 & &  &  &  & Blind to Shift's Direction \\
\multirow{2}{*}{Yes} & \multirow{2}{*}{Oguri-a*} & \multirow{2}{*}{\tt GLAFIC} & \multirow{2}{*}{SP} & \multirow{2}{*}{\cite{treubrammerdiego16}} & This Paper; SX Position Added; \\
 & &  &  &  & Blind to Shift's Direction \\
Yes & Zitrin-g & {\tt LTM} & LTM & \cite{treubrammerdiego16} & ... \\
Yes & Zitrin-c* & {\tt LTM} & LTM & ... & \cite{treubrammerdiego16} \\
Yes & Jauzac15.1 & CATS & SP & \cite{jauzacrichardlimousin16} & ...  \\
Yes & Jauzac15.2* & CATS & SP & ... & \cite{jauzacrichardlimousin16}; Delays Recalculated \\
Yes & Sharon-a* & {\tt Lenstool} & SP & \cite{treubrammerdiego16} & This Paper; Delays Recalculated \\
Yes & Sharon-g* & {\tt Lenstool} & SP & \cite{treubrammerdiego16} & This Paper; Delays Recalculated \\
\hline
\multicolumn{6}{c}{Lens Model on HFF Site \cite{hfflensmodels} with Highly Uncertain Post-Reappearance SX--S1 Calculation} \\
\hline
\multirow{2}{*}{No} & \multirow{2}{*}{Williams*} & \multirow{2}{*}{\tt Grale} &
\multirow{2}{*}{FF} & \multirow{2}{*}{No} & \cite{williamsliesenborgs19} (2019); Highly Uncertain \\
 & &  &  &  & Uses Only Image Positions \\
\hline
\multicolumn{6}{c}{Lens Models on HFF Site \cite{hfflensmodels} with No Predictions} \\
\hline
No &  Bradac & {\tt SWUnited} & FF & No & ... \\ %\citenum{bradactreuapplegate09} \\
No &  Keeton* & ... & SP & No & ... \\ % \citenum{keeton10} \\ % (posted in 4/17) 
No & Merten & {\tt SaWLens} & FF & No & ...  \\
\hline
\multicolumn{6}{c}{Additional Calculations from 2020 (Not Used for Primary $H_0$ Constraints)}\\
\hline
No & Chen2020* & {\tt Zoe} & H & No & \cite{chenkellywilliams20}; 2020 \\
No & Zitrin2020-ltm* & {\tt LTM} & LTM & No & \cite{zitrinrevised21}; 2020 \\
No & Zitrin2020-p* & ... & P & LTM & \cite{zitrinrevised21}; 2020 \\
\hline
\end{tabular}
\end{table}

\begin{table}
\small
\caption{\label{tab:measurements} {\bf Constraints on the relative time delays and magnification ratios from companion paper \cite{paperone} used in this analysis.} Each measurement is reported as the measured (maximum likelihood) value, followed in parentheses by the 16th, 50th, and 84th percentiles of the probability distribution.}
\centering
\begin{tabular}{ccc}
    %\tablecolumns{4}
\hline
Image Pair & Time Delay ($\Delta t$; days) & Magnif. Ratio ($\mu/\mu_1$) \\
\hline
S2--S1 & \corrsimsdelaystwosoneCombinedchabriermedian   & \corrsimsmustwosoneCombinedchabriermedian \\
S3--S1 & \corrsimsdelaysthreesoneCombinedchabriermedian   &
\corrsimsmusthreesoneCombinedchabriermedian \\
S4--S1 & \corrsimsdelaysfoursoneCombinedchabriermedian & \corrsimsmusfoursoneCombinedchabriermedian \\
SX--S1 & \corrsimsdelaysxsoneCombinedchabriermedian & \corrsimsmusxsoneCombinedchabriermedian \\
\hline
\end{tabular}
\end{table}

\begin{table}
\centering
\caption{\label{tab:mod_time_delays} {\bf The relative time delays in days for each model adopted in our $H_0$ measurement.} An asterisk denotes model calculations that were revised or created after the reappearance was reported. After correction of the time delay calculation from the mass model, the predictions of the parametric models for the time delay of SX relative to S1 are within $\sim 1\sigma$ agreement of each other. Jauzac15.1 was superseded by Jauzac15.2*, which adopted an improved method of computing the position of SN Refsdal in the source plane. The last column specifies whether the calculations of S1--S4's time delays used the observed (``Obs.'') or predicted (``Pred.'') position. Uncertainties correspond to 1$\sigma$ confidence intervals around the most probable value.}
\small
\begin{tabular}{ccccccc}
\hline
Model&$\Delta t_{21}$&$\Delta t_{31}$&$\Delta t_{41}$&$\Delta t_{{\rm X}1}$&$\Delta t_{{\rm Y}1}$&Position\\
\hline
\hline
\multicolumn{7}{c}{Simply-Parameterized Models}\\
\hline
Grillo-g&$10.6^{+6.2}_{-3.0}$&$4.8^{+3.2}_{-1.8}$&$25.9^{+8.1}_{-4.3}$&$361^{+19}_{-27}$&$-6183^{+160}_{-145}$&Pred.\\
Jauzac15.1&$90\pm17$&$30\pm35$&$-60\pm41$&$449\pm45$&$-4654\pm358$&Obs.\\
Jauzac15.2*&$-0.8\pm1.6$&$8.1\pm1.6$&$0.2\pm0.4$&$361\pm42$&$-5332\pm357$&Obs.\\
Oguri-g*&$9.5^{+0.6}_{-0.8}$&$5.6\pm0.4$&$20.2^{+1.9}_{-1.4}$&$330.7^{+21.9}_{-17.4}$&$-5861.6^{+207.4}_{-249.1}$&Pred.\\
Oguri-a*&$9.6^{+0.8}_{-0.6}$&$5.8^{+0.3}_{-0.5}$&$22.1\pm1.8$&$353.8^{+20.7}_{-17.3}$&$-6208.2^{+197.6}_{-265.2}$&Pred.\\
Sharon-a*&$10^{+9}_{-7}$&$6^{+12}_{-9}$&$22^{+7}_{-6}$&$298^{+58}_{-17}$&$-6220^{+161}_{-391}$&Pred.\\
Sharon-g*&$8^{+11}_{-6}$&$4^{+13}_{-10}$&$21^{+7}_{-6}$&$331^{+30}_{-35}$&$-6371\pm203$&Pred.\\
\hline
\multicolumn{6}{c}{Free-Form Model}\\
\hline
Diego-a&$-17\pm19$&$-4.0\pm27$&$74\pm43$&$262\pm55$&$-4521\pm524$&Obs.\\
\hline
\multicolumn{6}{c}{Light-Traces-Matter Model}\\
\hline
Zitrin-c*&$-55^{+111}_{-0}$&$8.7^{+77}_{-27}$&$151^{+45.6}_{-72}$&$36.74^{+122}_{-11}$&$-7048^{+719}_{-0}$&Obs.\\
\hline
\multicolumn{6}{c}{Additional Calculations (Not Used for Primary $H_0$ Estimate)}\\
\hline
Keeton*& $5.57_{-1.78}^{+3.4}$ & $3.29_{-1.1}^{+1.8}$ & $13.6_{-2.9}^{+5.0}$ & $340.01_{-22.2}^{+20.5}$ & $-5861.6_{-249.1}^{+207.4}$ & Pred.\\
Chen2020*&$9.13\pm2.63$&$5.58\pm3.78$&$18.98\pm2.55$&$332.35\pm9.32$&...&Pred.\\
Zitrin2020-ltm*&$2.3^{+2.9}_{-1.5}$&$-10^{+7.2}_{-1.5}$&$12.9^{+6.6}_{-2.1}$&$267.4^{+37.9}_{-4.3}$&$-6335^{+252}_{-144}$&Obs.\\
Zitrin2020-p*&$9.8^{+2.0}_{-0.2}$&$2.0^{+1.0}_{-0.1}$&$24.0^{+1.8}_{-0.4}$&$325.8^{+19.8}_{-14.2}$&$-6418^{+396}_{-60}$&Obs.\\
\hline
\end{tabular}
\end{table}

\begin{table}
\centering
\caption{\label{tab:mod_mag_ratios} {\bf Same as Table \ref{tab:mod_time_delays}, but for the magnification ratios.}}
\small
\begin{tabular}{ccccccc}
\hline
Model&$\mu(2) / \mu(1)$&$\mu(3) / \mu(1)$&$\mu(4) / \mu(1)$&$\mu({\rm X}) / \mu(1)$&$\mu({\rm Y}) / \mu(1)$ & Position\\
\hline 
\hline
\multicolumn{7}{c}{Simply-Parameterized Models}\\
\hline
Grillo-g&$0.92^{+0.43}_{-0.52}$&$0.99^{+0.52}_{-0.33}$&$0.42^{+0.19}_{-0.20}$&$0.36^{+0.11}_{-0.09}$&$0.30^{+0.09}_{-0.07}$&Pred.\\
Jauzac15.1&$0.84\pm0.13$&$0.88\pm0.11$&$0.41\pm0.05$&$0.19\pm0.02$&$0.16\pm0.015$&Obs.\\
Jauzac15.2*&$0.84\pm0.13$&$0.88\pm0.11$&$0.41\pm0.05$&$0.19\pm0.02$&$0.16\pm0.015$&Obs.\\
Oguri-g*&$1.11\pm0.04$&$1.22^{+0.04}_{-0.03}$&$0.62^{+0.05}_{-0.06}$&$0.29^{+0.03}_{-0.02}$&$0.23\pm0.02$&Pred.\\
Oguri-a*&$1.14^{+0.03}_{-0.04}$&$1.20\pm0.03$&$0.62^{+0.04}_{-0.05}$&$0.28^{+0.04}_{-0.02}$&$0.24^{+0.04}_{-0.01}$&Pred.\\
Sharon-a*&$0.84^{+0.20}_{-0.19}$&$1.46^{+0.07}_{-0.49}$&$0.44^{+0.05}_{-0.10}$&$0.19^{+0.01}_{-0.04}$&$0.17^{+0.02}_{-0.03}$&Pred.\\
Sharon-g*&$0.84^{+0.18}_{-0.06}$&$1.68^{+0.55}_{-0.21}$&$0.57^{+0.11}_{-0.04}$&$0.25^{+0.05}_{-0.02}$&$0.19^{+0.03}_{-0.01}$&Pred.\\
\hline
\multicolumn{6}{c}{Free-Form Model}\\
\hline
Diego-a&$1.89\pm0.79$&$0.64\pm0.19$&$0.35\pm0.11$&$0.31\pm0.10$&$0.41\pm0.11$&Obs.\\
\hline
\multicolumn{6}{c}{Light-Traces-Matter Model}\\
\hline
Zitrin-c*&$1.37^{+0.27}_{-0.61}$&$0.97^{+0.10}_{-0.03}$&$0.70^{+0.13}_{-0.26}$&$0.2921^{+0.0557}_{0.023}$&...&Obs.\\
\hline
\multicolumn{6}{c}{Additional Calculations (Not Used for Primary $H_0$ Estimate)}\\
\hline
Keeton* & $1.28_{-0.26}^{+0.23}$ & $1.22_{-0.18}^{+0.26}$ & $0.51_{-0.24}^{+0.13}$ & $0.28_{-0.05}^{+0.07}$ & $0.23_{-0.02}^{+0.02}$ & Pred. \\
Chen2020*&$1.45\pm0.38$&$1.31\pm0.44$&$0.67\pm0.09$&$0.21\pm0.02$&...&Pred.\\
Zitrin2020-ltm*&$0.86^{+0.48}_{-0.17}$&$0.94\pm0.06$&$0.23^{+0.09}_{-0.06}$&$0.21^{+0.02}_{-0.03}$&$0.15\pm0.02$&Obs.\\
Zitrin2020-p*&$1.17^{+0.065}_{-0.33}$&$1.13^{+0.13}_{0.005}$&$0.66^{+0.03}_{-0.14}$&$0.25^{+0.005}_{-0.01}$&$0.24^{+0.005}_{-0.02}$&Obs.\\
\hline
\end{tabular}
\end{table}

\begin{table*}[tpb]
\centering
\caption{ {\bf Mean residuals of recovered light-curve parameters from input values used to construct simulations.} The residuals listed are the weighted sum of the offsets from the four light curves, where the weights are those derived in the companion paper \cite{paperone}. We subtract these residuals from our measurements to remove bias.}
\begin{tabular}{c|ccccccc}
$\mathcal{O}_{j,1}$ &
$\Delta t_{\rm 2,1}$ &
$\Delta t_{\rm 3,1}$ &
$\Delta t_{\rm 4,1}$ & 
$\Delta t_{\rm X,1}$ &
$\mu_{2}/\mu_{1}$ &
$\mu_{3}/\mu_{1}$ & 
$\mu_{X}/\mu_{1}$ \\
\hline
$\Delta_{\mathcal{O}_{j,1}}$ & -0.2956 & -0.3979 & 0.1095 & -0.3215 & 0.0008 & -0.0542 & -0.0007
\end{tabular}
\label{tab:offsets}
\end{table*}

\begin{table}
\centering
\caption{\label{tab:covariance} {\bf Covariance matrix for relative time delays and magnification ratio measurements calculated by estimating input parameters from simulated light curves that include microlensing, millilensing, and photometric noise.} }
\begin{tabular}{c|ccccccc}
&
$\Delta t_{\rm 2,1}$ &
$\Delta t_{\rm 3,1}$ &
$\Delta t_{\rm 4,1}$ & 
$\Delta t_{\rm X,1}$ &
$\mu_{2}/\mu_{1}$ &
$\mu_{3}/\mu_{1}$ & 
$\mu_{X}/\mu_{1}$ \\
\hline
$\Delta t_{\rm 2,1}$ & 10.0894 & 3.7988 & 4.6658 & 3.5722 & -0.0774 & 0.0510 & 0.0019 \\
$\Delta t_{\rm 3,1}$ & 3.7988 & 9.8915 & 4.4020 & 3.4122 & -0.0155 & 0.0319 & 0.0038 \\
$\Delta t_{\rm 4,1}$ & 4.6658 & 4.4020 & 29.4357 & 2.3142 & 0.0216 & 0.0300 & 0.0114 \\
$\Delta t_{\rm X,1}$ & 3.5722 & 3.4122 & 2.3142 & 26.8131 & -0.0179 & -0.0431 & 0.0010 \\
$\mu_{2}/\mu_{1}$ & -0.0774 & -0.0155 & 0.0216 & -0.0179 & 0.0378 & -0.0056 & 0.0017 \\
$\mu_{3}/\mu_{1}$ & 0.0510 & 0.0319 & 0.0300 & -0.0431 & -0.0056 & 0.0853 & 0.0087 \\
$\mu_{X}/\mu_{1}$ & 0.0019 & 0.0038 & 0.0114 & 0.0010 & 0.0017 & 0.0087 & 0.0019 \\
\hline
\end{tabular}
\end{table}

\begin{table*}
\centering
\caption{\label{tab:error_budget} {\bf Terms in error budget.} Most terms are associated with the lens model, except for the time-delay measurement. For each set of cluster-model parameters, we compute the reduction in the variance of the time delay after fitting a second-order model for the time delay in terms of subsets of the full set of model parameters. After permuting the order with which we add each set of parameters, we compute the mean and maximum reduction in variance. The error budget is dominated by the lens model, and the uncertainty associated with the time-delay measurement represents only a small contribution. The parameters associated with the cluster halos account for the greatest percentage of the total variance. 
The parameters included in each set are described in detail in the Supplementary Materials.}
\begin{tabular}{lcc}
\hline
Model Parameters & Mean Percentage & Maximum Percentage \\
 & of Total Variance & of Total Variance \\
\hline
Cluster Halos & 39.4 & 67.1 \\ 
Photometric Redshifts & 16.9 & 49.2 \\ 
Early-Type Galaxy & 13.2 & 34.9 \\ 
Cluster Members & 10.1 & 30.0 \\ 
Time Delay Measurement & 9.2 & 9.2 \\ 
Model Perturbations & 7.8 & 19.7 \\ 
SN Position & 2.7 & 6.7 \\ 
\hline
\end{tabular}
\end{table*}

\begin{table}
\footnotesize
\caption{\label{tab:posteriorprobswithpriors} {\bf Posterior probability for each set of lens-model predictions given SN and CMB priors on $H_0$.} The ``$H_0$ Prior'' column describes whether a prior on the value of $H_0$ was applied.} % $^{\dagger}$We note that the Zitrin-c* model presented here contains an uncorrected problem, subsequently revised in Zitrin2020-ltm (see Table~\ref{tab:posteriorprobsrecent}).}
\begin{tabular}{c|cccccccc}
\hline
 $H_0$ Prior & Diego-a & Grillo-g & Jauzac15.2* & Oguri-g* & Oguri-a* & Sharon-a* & Sharon-g* & Zitrin-c* \\  
\hline
 SN & $\sumposthybridsimsDiegoaempiricalSHohESpositionsnoqual$ & $\sumposthybridsimsGrillogempiricalSHohESpositionsnoqual$ &
$\sumposthybridsimsJauzaconefivedottwoempiricalSHohESpositionsnoqual$ &
$\sumposthybridsimsOguriaempiricalSHohESpositionsnoqual$ & $\sumposthybridsimsOgurigempiricalSHohESpositionsnoqual$ &
$\sumposthybridsimsSharonaempiricalSHohESpositionsnoqual$ &
$\sumposthybridsimsSharongempiricalSHohESpositionsnoqual$ &
$\sumposthybridsimsZitrincempiricalSHohESpositionsnoqual$ \\
 CMB & $\sumposthybridsimsDiegoaempiricalPlanckpositionsnoqual$ & $\sumposthybridsimsGrillogempiricalPlanckpositionsnoqual$ &
$\sumposthybridsimsJauzaconefivedottwoempiricalPlanckpositionsnoqual$ &
$\sumposthybridsimsOguriaempiricalPlanckpositionsnoqual$ & $\sumposthybridsimsOgurigempiricalPlanckpositionsnoqual$ &
$\sumposthybridsimsSharonaempiricalPlanckpositionsnoqual$ &
$\sumposthybridsimsSharongempiricalPlanckpositionsnoqual$ &
$\sumposthybridsimsZitrincempiricalPlanckpositionsnoqual$   \\
\hline
\end{tabular}
\end{table}

\begin{table}
\footnotesize
\centering
\caption{\normalsize \label{tab:modweightswithpriors} {\bf $H_0$ weights determined by posterior probabilities for each set of lens-model predictions given SN and CMB priors on $H_0$.} First column lists the most probable value of $H_0$ and our 16th, 84th, and 99.7th percentile confidence levels.
Other columns lists the weights (out of 1.0) we calculated using the constraints from SN Refsdal. 
The first and second rows list model weights when applying different priors from previous SN and CMB measurements of $H_0$, respectively.
The relative weights are proportional to the models' posterior probabilities. The ``$H_0$ Prior'' column describes whether a prior on the value of $H_0$ was applied.} % $^{\dagger}$We note that the Zitrin-c* problem presented here was affected by a calculation problem and was subsequently corrected in Zitrin2020-ltm (see Table~\ref{tab:posteriorprobsrecent}).}
\begin{tabular}{c|cccccccc}
  \hline
$H_0$ Prior & Diego-a & Grillo-g & Jauzac15.2* & Oguri-g* & Oguri-a* & Sharon-a* & Sharon-g* & Zitrin-c* \\  
\hline
SN & $\sumfracposthybridsimsDiegoaempiricalSHohESpositionsnoqual$ & $\sumfracposthybridsimsGrillogempiricalSHohESpositionsnoqual$ &
$\sumfracposthybridsimsJauzaconefivedottwoempiricalSHohESpositionsnoqual$ &
$\sumfracposthybridsimsOguriaempiricalSHohESpositionsnoqual$ & $\sumfracposthybridsimsOgurigempiricalSHohESpositionsnoqual$ &
$\sumfracposthybridsimsSharonaempiricalSHohESpositionsnoqual$ &
$\sumfracposthybridsimsSharongempiricalSHohESpositionsnoqual$ &
$\sumfracposthybridsimsZitrincempiricalSHohESpositionsnoqual$ \\
CMB & $\sumfracposthybridsimsDiegoaempiricalPlanckpositionsnoqual$ & $\sumfracposthybridsimsGrillogempiricalPlanckpositionsnoqual$ &
$\sumfracposthybridsimsJauzaconefivedottwoempiricalPlanckpositionsnoqual$ &
$\sumfracposthybridsimsOguriaempiricalPlanckpositionsnoqual$ & $\sumfracposthybridsimsOgurigempiricalPlanckpositionsnoqual$ &
$\sumfracposthybridsimsSharonaempiricalPlanckpositionsnoqual$ &
$\sumfracposthybridsimsSharongempiricalPlanckpositionsnoqual$ &
$\sumfracposthybridsimsZitrincempiricalPlanckpositionsnoqual$ \\
\hline 
\end{tabular}

\end{table}

\begin{table}
\small
\caption{\label{tab:pixscale_models} {\bf Pixel scale of models available on the HFF Lens Model website \cite{hfflensmodels}.} The Merten model has low resolution. The ``Principal'' column corresponds to the main cluster model provided by each modeling team. The ``Range'' column lists the resolution of the set of models corresponding, in most cases, to individual MCMC realizations used to estimate uncertainties.}
\centering
\begin{tabular}{ccc}
\hline
Model Version & Pixel Size (Main) & Pixel Size (Range) \\
\hline
\multicolumn{3}{c}{Included in $H_0$ Estimate} \\
\hline
Jauzac CATS v4.0 & 0.3$''$ & 0.5$''$ \\ 
Jauzac CATS v4.1 & 0.3$''$ & 0.5$''$  \\ 
Diego v4.1 & 0.42$''$ & 0.42$''$ \\
Oguri {\tt GLAFIC} v3 & 0.03$''$ & 0.3$''$  \\
Sharon v4cor & 0.05$''$ & 0.2$''$ \\
\hline
\multicolumn{3}{c}{Not Included in $H_0$ Estimate} \\
\hline
Keeton v4 & 0.04$''$ & 0.19$''$  \\
Williams v4 & 0.12$''$ & 0.12$''$ \\
Bradac v1 & 0.04$''$ & 0.04$''$  \\
Merten v1 & 8.3$''$ & 8.3$''$ \\
\hline
\end{tabular}
\end{table}

\begin{table}
\centering
\caption{\label{tab:td_pos_choice} {\bf Comparison of SX--S1 time delays using the predicted or observed image position for selected models.} We determine a source-plane position by minimizing the residuals of the predicted positions with the observed images. Our calculations using the Grillo-g and  Oguri-a*  pixelized best-fitting maps show that using the observed versus predicted image positions changes the SX--S1 time delay by less than one day.  The Oguri-a GLAFIC v3 model corresponds to the version of the model before the addition of SX as an additional constraint. We do not expect perfect agreement with the modeler's calculations, because we are using the pixeled model. For the Jauzac CATS v4.1 model, we used the lens parameters to construct model. 
Models were only included if they have resolution of $\lesssim 0.05''$ per pixel (Table~\ref{tab:pixscale_models}). }
\begin{tabular}{c|cc|cc|cc|cc}
\hline
Model& \multicolumn{2}{c|}{S2--S1} & \multicolumn{2}{c|}{S3--S1} & \multicolumn{2}{c|}{S4--S1} & \multicolumn{2}{c}{SX--S1} \\
  & Pred. & Obs. & Pred. & Obs. & Pred. & Obs. & Pred. & Obs.  \\
\hline
Grillo-g & 6.3 & 6.2 & 3.4 & 3.2  & 26.5 & 26.0 & 354.6 & 354.1  \\
Jauzac CATS v4.1 & 11.7 & 8.8 & 4.5 & 4.2 & 28.8 & 25.6 & 315.3 & 321.7 \\
Keeton* & 5.6 & 5.7 & 3.3 & 3.5 & 13.6 & 12.6 & 340.0 & 338.0 \\
Oguri-a GLAFIC v3 & 9.2 & 9.2 & 5.9 & 6.1 & 19.7 & 20.3 & 342.8 & 344.1 \\
Sharon v4cor & 6.6 & 6.5 & 0.8 & 0.8 & 19.0 & 19.0 & 351.8 & 351.8 \\
\hline
\end{tabular}
\end{table}

\begin{table}
\centering
\caption{\label{tab:mag_pos_choice} {\bf Same as Table~\ref{tab:td_pos_choice}, but for the magnification ratios.}}
\begin{tabular}{c|cc|cc|cc|cc|cc}
\hline
Model & \multicolumn{2}{c|}{S1} & \multicolumn{2}{c|}{S2} & \multicolumn{2}{c|}{S3} & \multicolumn{2}{c|}{S4} & \multicolumn{2}{c}{SX} \\
  & Pred. & Obs. & Pred. & Obs. & Pred. & Obs. & Pred. & Obs. & Pred. & Obs.  \\
\hline
Grillo-g & 13.9 & 13.0 & 15.2 & 16.6 & 12.6 & 12.6 & 7.5 & 7.8 & 4.7 & 5.0  \\
Jauzac CATS v4.1 & 18.1 & 26.9 & 15.2 & 34.4 & 16.3 & 15.6 & 6.4 & 11.0 & 4.9 & 4.8 \\
Keeton* & 20.5 & 16.4 & 26.4 & 122.5 & 25.0 & 14.3 & 10.5 & 23.7 & 5.8 & 5.4 \\
Oguri-a GLAFIC v3 & 15.3 & 15.1 & 17.9 & 22.5 & 18.8 & 15.5 & 9.7 & 5.8 & 4.5 & 4.4 \\
Sharon v4cor & 19.9 & 18.9 & 23.3 & 25.4 & 21.4 & 22.6 & 9.3 & 8.8 & 4.3 & 4.3  \\
\hline
\end{tabular}
\end{table}

\begin{table}
\centering
\small
\caption{\label{tab:posteriorprobsrecent} {\bf Same as Table~\ref{tab:posteriorprobs}, but for the Zitrin-ltm \cite{zitrinrevised21}, Zitrin-p \cite{zitrinrevised21}, and Chen2020* \cite{chenkellywilliams20} models and without SX--S1 separation included in the likelihood calculation}. These models were produced after the appearance of SX but before the time delay was unblinded. Grillo-g and Oguri-a* are included for comparison. Neither Zitrin model uses the position of SX as a constraint. The ``$H_0$ Prior'' column describes whether a prior on the value of $H_0$ was applied.}
\begin{tabular}{ccccc}
 \multicolumn{5}{c}{Probability for Each Model Given Measurements} \\  
\hline
 Zitrin2020-ltm* & Zitrin2020-p* & Chen2020* & Grillo-g & Oguri-a* \\  
\hline
$\sumposthybridsimsZitrintwoohtwoohltmempiricalnopriornopositionsnoqualthroughtwoohtwooh$ & $\sumposthybridsimsZitrintwoohtwoohpempiricalnopriornopositionsnoqualthroughtwoohtwooh$ &
$\sumposthybridsimsChentwoohtwoohempiricalnopriornopositionsnoqualthroughtwoohtwooh$ & $\sumposthybridsimsGrillogempiricalnopriornopositionsnoqualthroughtwoohtwooh$  & $\sumposthybridsimsOguriaempiricalnopriornopositionsnoqualthroughtwoohtwooh$  \\
\hline
\end{tabular}
\end{table}

\begin{table}
\centering
\small
\caption{\label{tab:posteriorprobsrecentwithpriors} {\bf Same as Table~\ref{tab:posteriorprobswithpriors}, but for the Zitrin2020-ltm* \cite{zitrinrevised21}, Zitrin2020-p* \cite{zitrinrevised21}, and Chen2020* \cite{chenkellywilliams20} models and without SX--S1 separation included in the likelihood calculation.} Grillo-g and Oguri-a* are included for comparison. Neither Zitrin model uses the position of SX as a constraint. The ``$H_0$ Prior'' column describes which prior on the value of $H_0$ was applied.}
\begin{tabular}{c|ccccc}
\multicolumn{1}{c|}{Included} & \multicolumn{5}{c}{Probability for Each Model Given Measurements} \\  
\multicolumn{1}{c|}{Constraints} & \multicolumn{5}{c}{ } \\
\hline
    $H_0$ Prior & Zitrin2020-ltm* & Zitrin2020-p* & Chen2020* & Grillo-g & Oguri-a* \\  
 %Prior &  &  &   \\
\hline
 SN & $\sumposthybridsimsZitrintwoohtwoohltmempiricalSHohESnopositionsnoqualthroughtwoohtwooh$ & $\sumposthybridsimsZitrintwoohtwoohpempiricalSHohESnopositionsnoqualthroughtwoohtwooh$ & $\sumposthybridsimsChentwoohtwoohempiricalSHohESnopositionsnoqualthroughtwoohtwooh$& $\sumposthybridsimsGrillogempiricalSHohESnopositionsnoqualthroughtwoohtwooh$& $\sumposthybridsimsOguriaempiricalSHohESnopositionsnoqualthroughtwoohtwooh$  \\
 CMB & $\sumposthybridsimsZitrintwoohtwoohltmempiricalPlancknopositionsnoqualthroughtwoohtwooh$ & $\sumposthybridsimsZitrintwoohtwoohpempiricalPlancknopositionsnoqualthroughtwoohtwooh$ & $\sumposthybridsimsChentwoohtwoohempiricalPlancknopositionsnoqualthroughtwoohtwooh$ & $\sumposthybridsimsGrillogempiricalPlancknopositionsnoqualthroughtwoohtwooh$  & $\sumposthybridsimsOguriaempiricalPlancknopositionsnoqualthroughtwoohtwooh$   \\
\hline
\end{tabular}
\end{table}

\end{document}